\documentclass[11pt]{article}
\pdfoutput=1  
\usepackage{jcappubmod}
\usepackage{graphicx}
\usepackage{color}
\usepackage{amssymb}
\usepackage{amsmath}
\usepackage{mathabx}
\usepackage{stmaryrd}
\usepackage{multirow}
\usepackage{amscd}
\usepackage{mfpic}
\usepackage{verbatim}
\usepackage{upgreek}

\definecolor{cream}{rgb}{.97, .95, .88}
\definecolor{darkcream}{rgb}{1., .88, .5}
\definecolor{lightpink}{rgb}{0.98, 0.88, 0.87}
\definecolor{lightwhite}{rgb}{1., 0.98, 0.95}
\definecolor{lightsalmon}{rgb}{1., 0.95, 0.90}
\definecolor{lightviolet}{rgb}{0.9, 0.8, 0.9}
\definecolor{lightgray}{rgb}{.96, .96, .96}  
\definecolor{lgray}{rgb}{.75, .75, .75}
\definecolor{LemonChiffon}{rgb}{0.95, 1., 0.7}
\definecolor{lightolivegreen}{rgb}{0.84, 0.89, 0.25}
\definecolor{lightgreen}{rgb}{.664, 1., .52}
\definecolor{llgreen}{rgb}{.900, .983, .960}
\definecolor{tristle}{rgb}{0.87, 0.67, 0.77} 
\definecolor{pink}{rgb}{0.95, 0.45, 0.75}
\definecolor{magenta}{rgb}{1., 0, 1.}
\definecolor{violet}{rgb}{0.9, 0.20, 0.85}
\definecolor{darkolivegreen}{rgb}{0.55, 0.65, 0.35}
\definecolor{maroon}{rgb}{0.7, 0.26, 0.56}
\definecolor{lightmaroon}{rgb}{0.85, 0.38, 0.58}
\definecolor{darkmaroon}{rgb}{0.604, 0.169, 0.451}
\definecolor{ddarkmaroon}{rgb}{0.2, 0.03125, 0.150}
\definecolor{mediumorchid}{rgb}{0.8, 0.33, 0.83}
\definecolor{mediumorchidd}{rgb}{1., 0.33, 0.63}
\definecolor{darkgreen}{rgb}{0.1, 0.6, 0.13}
\definecolor{lightyellow}{rgb}{1., 1., 0.82}
\definecolor{turquoise}{rgb}{0.042, 0.586, 0.512}
\definecolor{turquoisel}{rgb}{0.66, 0.94, 0.83}
\definecolor{darkturquoise}{rgb}{0.21, 0.55, 0.50}
\definecolor{coral}{rgb}{1., 0.6, 0.21}
\definecolor{lightorange}{rgb}{1., 0.88, 0.75}
\definecolor{orangered}{rgb}{1., 0.5, 0.}
\definecolor{orange}{rgb}{1., 0.65, 0.1}
\definecolor{orangel}{rgb}{1., .85, .3}
\definecolor{darkorange}{rgb}{0.875, 0.4, 0.204}
\definecolor{ddarkorange}{rgb}{.675, .218, .05}
\definecolor{bluesky}{rgb}{0.48, 0.53, 1.}
\definecolor{gold}{rgb}{1., 0.85, 0.25}
\definecolor{goldd}{rgb}{0.95, 0.75, 0.05}
\definecolor{darkviolet}{rgb}{0.54, 0.04, 0.84}
\definecolor{ddarkviolet}{rgb}{.382, .063, .657}
\definecolor{lightblue}{rgb}{0.30, 0.86, 0.89}
\definecolor{LightBlue}{rgb}{0.68, 0.85, 0.9}
\definecolor{lblue}{rgb}{0.78, 0.90, 0.95}
\definecolor{darkblue}{rgb}{.105, .308, .707}
\definecolor{lightmaroon}{rgb}{0.85, 0.38, 0.58}
\definecolor{darkmaroon}{rgb}{0.604, 0.169, 0.451}
\definecolor{darkpink}{rgb}{0.879, 0.020, 0.766}
\definecolor{ddarkpink}{rgb}{0.738, 0.195, 0.406}
\definecolor{grey}{rgb}{0.717, 0.717, 0.717}
\definecolor{lightgrey}{rgb}{0.800, 0.800, 0.800}
\definecolor{brown}{rgb}{0.740, 0.323, 0.182}
\definecolor{redbrown}{rgb}{.575, .158, .05}
\definecolor{darkbrown}{rgb}{0.34, 0.25, 0.05}
\definecolor{orangebrown}{rgb}{0.433, 0.262, 0.06}
\definecolor{pinkl}{rgb}{1., 0.788, 0.918}
\definecolor{salmon}{rgb}{1., 0.66, 0.5}
\definecolor{lightbrown}{rgb}{0.703, 0.508, 0.121}

\def\etal{{\it et al.}}

\def\Journal#1#2#3#4{{#1} {\bf #2}, (#3) #4}

\def\AA{\em A.\& A.}

\def\AIP{\em AIP Conf.Proc.}
\def\AMT{\em Acta. Math.}

\def\APH{\em Annals Phys.}

\def\CPC{\em Comp. Phys. Com.}

\def\CMP{\em Commun. Math. Phys.}
\def\CPC{\em Comp. Phys. Com.}

\def\CRP{\em Compt. Rendu. Phys.}

\def\GRG{\em Gen. Rel. Grav}

\def\IMD{{\em Int. J. Mod. Phys.} D}

\def\JCA{\em J. Cosmol. Astrop. Phys.}

\def\JHE{\em J. High Ener. Phys.}
\def\JMP{\em J. Math. Phys.}
\def\JPC{\em J. Phys.: Conf. Series}

\def\LNP{\em Lect. Notes Phys.}

\def\MPL{{\em Mod. Phys. Lett.} A}
\def\MRA{\em MNRAS}

\def\NPB{{\em Nucl. Phys.} B}

\def\PDU{\em Phys.Dark Univ.}

\def\PLB{{\em Phys. Lett.} B}

\def\PRD{{\em Phys. Rev.} D}
\def\PREE{{\em Phys. Rev.} E}
\def\PRL{\em Phys. Rev. Lett.}
\def\PRV{\em Phys. Rev.}
\def\PRE{\em Phys. Rep.}
\def\PRLA{{\em Proc.Roy.Soc.Lond.} A}

\def\RPP{\em Rept. Prog. Phys.}
\def\SCI{\em Science}

\def\ZPH{\em Z. Phys.}

\def\be{\begin{equation}}
\def\ee{\end{equation}}
\def\bea{\begin{eqnarray}}
\def\eea{\end{eqnarray}}
\def\bes{\begin{equation*}}
\def\ees{\end{equation*}}
\def\beas{\begin{eqnarray*}}
\def\eeas{\end{eqnarray*}}

\def\tr{\text{tr}}
\def\mg{\mathsf g}

\def\um{\mathcal U}

\def\hm{\mathcal H}
\def\gm{\mathcal G}

\begin{document}
\title{Non-equilibrium evolution of quantum fields during inflation and late accelerating expansion}

\author[a,b]{Houri~Ziaeepour}
\affiliation [a]{Institut UTINAM, CNRS UMR 6213, Observatoire de Besan\c{c}on, Universit\'e de Franche Compt\'e, 41 bis ave. de l'Observatoire, BP 1615, 25010 Besan\c{c}on, France}
\affiliation [b]{Mullard Space Science Laboratory, University College London, Holmbury St. Mary, 
GU5 6NT, Dorking, UK}

\emailAdd{houriziaeepour@gmail.com}

\abstract
{To understand mechanisms leading to inflation and late acceleration of the Universe it is important 
to see how one or a set of quantum fields may evolve such that the classical energy-momentum 
tensor behave similar to a cosmological constant. Phenomenological models assume that condensation 
of a scalar field dominating other constituents is responsible for the onset of inflation and 
dark energy. However, conditions for formation of such a condensate and whether it is a necessary 
ingredient for generation of inflation and late acceleration are not clear. In this work we consider 
a toy model including 3 scalar fields with very different masses to study the formation of a light 
axion-like condensate, presumed to be responsible for inflation and/or late accelerating expansion 
of the Universe. Despite its simplicity, this model reflects hierarchy of masses and couplings of 
the Standard Model and its candidate extensions. The investigation is performed in the framework of 
non-equilibrium quantum field theory in a consistently evolved FLRW geometry. We discuss in details 
how the initial conditions for such a model must be defined in a fully quantum setup and show that 
in a multi-component model interactions reduce the number of independent initial 
degrees of freedom. Numerical simulation of this model shows that it can be fully consistent with 
present cosmological observations. For the chosen range of parameters we find that quantum 
interactions rather than effective potential of a condensate is the dominant contributor in the 
energy density of the Universe and triggers both inflation and late accelerating expansion. 
Nonetheless, despite its small contribution in the energy density, the light scalar field - in both 
condensate and quasi free particle forms - has a crucial role in controlling the trend of 
heavier fields. Furthermore, up to precision of our simulations we do not find any IR singularity 
during inflation. These findings highlight uncertainties in attempts to extract information about 
physics of the early Universe by naively comparing predictions of local effective classical 
models with cosmological observations, neglecting inherently non-local nature of quantum processes. 
}

\maketitle

\section{Introduction} \label{sec:intro}
Cosmological observations have demonstrated that at least during two epochs the Universe has gone 
through accelerating expansion. The first era, usually called {\it inflation}~\cite{infrev} 
occurred at or close to the birth of the Universe. The second epoch has begun at around redshift 
0.5 - roughly half of the age of the Universe - and is ongoing now. Its unknown cause is given the 
generic name of {\it dark energy}~\cite{derev,modgrrev} (reviews) and at present it is the dominant 
contributor in the average energy density of the Universe. If dark energy is not an elusive 
Cosmological Constant (CC), its origin may be a modification of Einstein theory of gravity or a new 
field in the matter side of the Einstein equation~\cite{quin}. It is possible that inflation and 
dark energy be the manifestation of the same phenomenon at different epochs~\cite{quininfunify} 
(review). Homogeneity of present expansion rate~\cite{hubbleanisoconstr} and properties of inflation 
concluded from observations of the Cosmic Microwave Background (CMB) anisotropies~\cite{infplanck} 
indicate a slowly varying - in both space and time - energy density for dark energy and inflaton. 
This requirement can be phenomenologically formulated with one or multiple light quantum scalar 
fields, which their effective flat potential dominates the energy density of the Universe during 
the epochs of accelerating expansion.

In two ways a quantum field may generate an effective flat energy-momentum tensor: either through 
non-zero 1-point Green's function - also called condensate, mean, or background field - of a scalar 
(or vector) field $\varphi \equiv \langle \Phi \rangle \neq 0$ and its close to flat potential; or 
through quantum interactions producing an approximately constant average energy density and small 
quantum fluctuations around it. Even in the latter case, which corresponds to a subdominant 
contribution of condensate in the energy density of the Universe and in the expansion rate, the 
condensate may have a crucial role in cosmological phenomena through Higgs mechanism and breaking 
of symmetries~\cite{higgssymmbrrev} (reviews). Thus, motivations for investigating formation and 
evolution of condensates in cosmology go beyond their role in inflation and late accelerating 
expansion.

Evolution of quantum scalar fields in curved spacetimes are extensively studied, specially in de 
Sitter geometry as a good approximation for the geometry of the Universe during inflation and 
reheating~\cite{reheatrev}. For instance, authors of~\cite{renormadiab0} study reheating and 
evolution of mean field (condensate) and their backreaction on the metric for an $O(N)$ symmetric 
multi-scalar field model. However, their formulation and simulations include only local quantum 
corrections to effective mass. In \cite{renormadiab1} the evolution of a pre-existing condensate 
during preheating for the same model as previous and thermalization of quantum fluctuations are 
studied. Their simulations consistently evolve geometry and take into account non-local quantum 
corrections to effective potential. But, they are performed for unrealistically large couplings. 
Models with more diverse field content are also studied~\cite{infsecular,infsecularir}, mostly in 
de Sitter space and without backreaction on the geometry. The common aim of practically all these 
and many other works in the literature have been the study of particle production during and after 
inflation. Estimation of non-Gaussianity generated by quantum processes is another topic related 
to inflation which is extensively investigated~\cite{infqftnongauss}. However, by the nature of 
this subject, the concentration has been on the quantum correlation of fluctuations rather than 
effective secular component driving inflation itself. In fact, the onset and evolution of quantum 
field(s) leading to a shallow slope effective potential for the condensate or all components and 
onset of an inflationary era from a pre-inflationary epoch is not extensively studied.

A controversial issue about inflation, which is not yet completely settled, is the stability of IR 
modes. Instability of these modes, and thereby the de Sitter geometry, is first concluded 
in~\cite{desitterinstab}. Its effect on the evolution of inflation and de Sitter geometry is studied 
in~\cite{desitcorrmass}. The particular case of massless scalar fields is investigated 
in~\cite{desitterirmassless,desitterirmassless0} (using parametric representation of path integrals, 
and adiabatic vacuum subtraction renormalization, respectively). Breaking of symmetries by 
condensation of light scalars, acquisition of mass by some fields, and generation of massless 
Goldstone modes for others are investigated in~\cite{desittersymmbreak} (using Winger-Weisskopf 
method~\cite{wwmethod}). In addition, analogy between particle creation and vacuum instability in 
a constant electric field and de Sitter space vacuum is used to study the IR instability of the 
latter models in~\cite{desitterinstab}. It is also shown that subhorizon and superhorizon modes 
become entangled when a transition from fast roll to slow roll occurs. This convoys the effect of 
non-observable IR singularities to observable subhorizon fluctuations. Furthermore, sudden 
variations of inflaton field(s) lead to particle production, suppression of dynamical mass, and 
anomalous decay of inflaton~\cite{desitterir}. Quantum IR modes and ultra light particles are also 
suggested as the origin of dark energy~\cite{deultralightir}.

Furthermore, a fully non-equilibrium quantum field theoretical calculation of IR modes in de 
Sitter space~\cite{infdesiter2pi} shows that due to dynamical acquisition of mass by a massless 
scalar, these modes are naturally regulated and no singularity arises. Other works using the same 
method~\cite{inf2piremormcurv} confirm these results for inflaton alone, but find that IR quantum 
corrections become large and non-perturbative in curvaton models, which include spectator scalar 
field(s). Other perturbative and non-perturbative methods are also used to investigate the issue 
of IR modes. For instance, in~\cite{desitterinstab1} parametric representation of path integrals 
are used to show that de Sitter space is instability-free in presence of massive fields, in 
contrast to the case of massless fields studied with the same method by the same authors. 
Non-perturbative renormalization group technique is used by~\cite{rgdesitter} to find quantum 
corrections to classical potential of $O(N)$ model in de Sitter space. They conclude that due to 
large IR fluctuations symmetries are radiatively restored, i.e. no condensate is formed. However, 
their calculation includes only local quantum corrections and solutions of free field equation are 
used to estimate the evolution of condensate. These approximations do not seem reasonable when the 
issue of large distant correlations is studied. Stochastic approach to 
inflation~\cite{infstochas,infstochas0} is used by~\cite{infqftstoch} to take into account, in a 
non-perturbative manner, quantum corrections to the same model as previous works cited here. 
The equivalence of stochastic and Schwinger-Keldysh 2 Particle Irreducible (2PI) method for 
IR modes is shown in~\cite{infstochasirequiv}. They conclude that no condensate is formed and 
symmetry of the $O(N)$ model is preserved. 

In what concerns the estimation of long distance correlations and formation of a condensate, which 
may lead to symmetry breaking, one of the main shortcoming of works reviewed above is neglecting the 
backreaction of quantum effects on the evolution of geometry. This issue is an addition to other 
approximations which had to be considered to make models analytically tractable. Moreover, these 
studies have been mostly concentrated on a single field or $O(N)$ symmetry inflaton, and exceptionally 
on models with additional fields possessing mass and coupling hierarchies. Additionally, the issue 
of condensate formation and symmetry breaking needs more accurate calculation than what have been 
done in previous works, because analytical approximations may have important and misleading impact 
on conclusions - we will discuss an example of such problems later in this work. Although formal 
description of perturbative expansion and Feynman diagrams contributing in the evolution of 
condensates are worked out in details in~\cite{cond2piexpand}, an analytical approach including 
consistent evolution of geometry is not available.

Studies of dark energy models in full quantum field theoretical setups are very rare. Example of 
exceptions are~\cite{deultralightir,quinvactunnel,houricond,deeft2pi}. However, even these works 
miss some of the most important features which a simple but realistic model should cover, namely: 
taking into account both local and nonlocal quantum corrections, at least at lowest order; 
backreaction of matter on the geometry; mass hierarchy; proper calculation of formation and 
evolution of condensate, etc. Indeed many dark energy models are simply phenomenological and do 
not have a well defined and renormalizable quantum formulation. In particular, in many modified 
gravity models - as an alternative to a cosmological constant - the dilaton scalar field has 
a non-standard and non-renormalizable Lagrangian. These models must be considered as effective 
theories and their quantization is either meaningless or must be restricted to lowest order to avoid 
renormalization issues. By contrast, the class of models generally called {\it interacting 
quintessence} include cases with quantum mechanically well defined and renormalizable interactions, 
such as a monomial/polynomial $\phi^n$-type self-interaction~\cite{houridmquin} with $n > 0$ or a 
gauge field~\cite{quinqft0}. We should remind that despite various definitions and classification 
procedures in the literature~\cite{dediscrimscale,houridephono,dediscrim}, there is not a general 
consensus about how a dark energy model should be classified as modified gravity or quintessence. 
Here we use the definition of~\cite{houridephono}: if the scalar field responsible for accelerating 
expansion has the same coupling to all fields, including itself, the model is considered to be a 
modified gravity; otherwise it is called (interacting)-quintessence.

As we mentioned above little work has been performed on the formation of a condensate. Specifically, 
studies of inflation, reheating, and dark energy models usually assume a pre-existing condensate as 
initial condition. {\bf Therefore, our aim is to understand how a condensate is formed from an initial state where it is absent. Another goal is to understand whether and under which conditions it may become energetically dominant, as it is usually assumed in classical approaches to inflation and dark energy.} For this purpose, in a previous work~\cite{houricond} we studied formation and evolution of 
the condensate of a light scalar produced by the decay of a massive particle in FLRW geometry. 
The model includes 3 fields which present the three important mass scales, namely a heavy field 
presenting sub-Planckian/GUT physics, an axion-like light field as inflaton or quintessence field, 
and an intermediate mass field presenting Standard Model particles. The calculation takes into 
account the lowest order quantum corrections to effective potential of condensate, but not quantum 
corrections to propagators. Nonetheless, this model stands out from those reviewed above by 
including fields with very different masses and in this respect it is a better representation of 
what we see in particle physics. This study showed that during radiation domination the amplitude 
of the condensate builds up very quickly - indeed similar to parametric resonance and particle 
production during preheating. But in matter domination era all but the longest modes decay. Moreover, 
only for self-interaction potentials of order $\lesssim 4$ long (IR) modes survive the faster 
expansion of the Universe. This result is consistent with conclusions of~\cite{infsecular} about 
the evolution of inflation condensate. Furthermore, it is shown that only by taking into account 
quantum corrections the condensate may survive. Therefore, if dark energy is not due to an 
alternative gravity model, it may be a large scale non-local quantum phenomenon, which could not 
exist in the realm of a classical expanding universe. However, approximate analytical approach used 
in~\cite{houricond} works for a fixed geometry and backreaction of matter evolution on the geometry 
cannot be followed.

The goal of present work is to improve the investigation performed in~\cite{houricond} by using 
full 2PI formalism in a consistently evolved FLRW geometry according to a semi-classical Einstein 
equation. The toy model of~\cite{houricond} can be considered as an inflation model, interacting 
quintessence or both, because only initial conditions discriminate between these epochs. To go 
further than~\cite{houricond}, it is important to properly evolve different components, specially 
the condensate, and investigate their role in the process of accelerating expansion and 
formation of anisotropies. Unfortunately, these goals cannot be accomplished analytically. Numerical 
simulations are necessary, and they have their own difficulties and imprecisions. Nonetheless, similar 
to other hard problems in theoretical physics, such as strong coupling regime of QCD and evolution of 
Large Scale Structures (LSS) of the Universe, the hope is that the quality of such simulations would 
be gradually improved.

In Sec. \ref {sec:model} we present the model. To fix notations a brief review of 2PI formulation is 
given in Sec. \ref {sec:2pirev} and applied to the model in Sec. \ref {sec:2pievol}. Renormalization 
is discussed in Sec. \ref {sec:renorm}. The semi-classical energy-momentum tensor and Einstein 
equation are obtained in Sec. \ref {sec:effenermom}. As the model includes 3 fields with very 
different masses, and apriori it can have a non-zero initial condensate, the initial state and 
initial conditions for solving dynamical equations must be chosen in a consistent way. These topics 
are discussed in Sec. \ref {sec:initcond}. The issue of consistently defining and taking into 
account the contribution of non-vacuum initial state in the 2PI is not 
trivial~\cite{inin,2piinitcond}. In the literature the case of a thermal initial condition is 
extensively studied~\cite{2piinitcond}. But, in inflation and dark energy physics an initial 
coherent condensate state alone or along with a non-condensate is physically plausible and an 
interesting case to consider. In subsection \ref {sec:densitymatrix} we discuss interesting initial 
states for the model. In particular, we determine the density matrix of a generalized coherent 
state and discuss its contribution in the generating functional of 2PI formalism. In 
Sec. \ref {sec:initevol} initial conditions for evolution equations of propagators and condensate 
are discussed. Initial conditions for the semi-classical Einstein equation is described in 
Sec. \ref {sec:geowave}. We will show that they also fix the normalization of the wave-functions 
of constituents. Numerical simulations of the model and their results are presented in 
Sec. \ref {sec:simulation}. Physical implications of the results of simulations are discussed 
in Sec. \ref {sec:discuss}. Sec. \ref {sec:conclusion} summarizes the outlines of this work. 

In Appendix \ref{app:classpot} we calculate extrema of classical potential of the model. 
Appendix \ref {app:prop} reminds the definition of various propagators and Appendix \ref{app:propag} 
presents description of free propagators with respect to solutions of evolution equations for 
a given initial state. In Appendix \ref {app:initparticle} we present the general description of 
initial state and density matrix. In Appendix \ref {app:distphia} we obtain momentum distribution 
of remnants of a decaying heavy particle. Appendix \ref {app:connect} presents Christoffel 
coefficients for the linearized metric gauge used here. Appendix \ref {app:solution} reviews 
solutions of free evolution equation for cases of radiation and matter dominated homogeneous FLRW 
metric, and WKB approximation for other geometries and for renormalization of the model. Initial 
conditions described in Sec. \ref {sec:initcond} give a unique solution for integration constants 
of renormalized initial propagators and condensate. They are determined in Appendix 
\ref {app:gensolution}. 

\section {Model} \label{sec:model}
We consider a phenomenological model with 3 scalar fields which their masses are in 3 physically 
interesting and relevant ranges: a heavy particle $X$ with a mass a few orders of magnitude less 
than Planck scale - presumably in GUT scale; a scalar field $A$ with an intermediate mass of order 
the of electroweak mass scale, that is in GeV-TeV range; and finally a vary light axion-like scalar 
$\Phi$. The model can be easily extended to the case in which $X$ and $A$ are fermions. Extension 
to vector fields and a full Yang-Mills model is also straightforward, but because of their 
additional complexities, we do not consider them here. We believe that the simplest case of 
scalars without internal symmetries is generic enough for investigating properties of condensate 
and effective potential, which may affect expansion of the Universe. In particular, the extension 
of the model to the case where each scalar field has an internal $O(N)$ symmetry only modifies 
multiplicity of Feynman diagrams. Such extensions are widely used in the literature in the framework 
of large $N$ expansion technique to take into account non-perturbative effects, see 
e.g.~\cite{2pirev} for a recent review and~\cite{infnongauss} for its application in study of 
non-Gaussianity in cosmological models.

A similar model has been studied as an alternative to simple quintessence models for dark energy, 
classically in~\cite{houriustate,houridmquin} and with lowest quantum corrections in~\cite{houricond}, 
and its extension to inflationary epoch may provide a unified theory for both phenomena. In addition, 
interaction between massive and light fields is known to influence the evolution of 
fluctuations~\cite{infsecular}, and thereby IR modes~\cite{infsecularir,infmassivecmb} of both the 
condensate and quantum fluctuations. The third field with intermediate mass may be considered as a 
prototype of an average mass dark matter or Standard Model fields, if the heavy field is considered 
as a meta-stable dark matter.

Considering the simplest interactions between the 3 constituents of the model, the classical 
Lagrangian can be written as the following:
\bea
{\mathcal L} &=& {\mathcal L}_{\Phi} + {\mathcal L}_{X} + {\mathcal L}_{A} + {\mathcal L}_{int} 
\label{lagrange}\\
{\mathcal L}_{\Phi} &=& \int d^4 x \sqrt{-g} \biggl [\frac{1}{2} g^{\mu\nu}
{\partial}_{\mu}\Phi {\partial}_{\mu}\Phi - \frac{1}{2}m_{\Phi}^2 {\Phi}^2 - 
\frac{\lambda}{n!}{\Phi}^n \biggr ] \label{lagrangphi} \\
{\mathcal L}_{X} &=& \int d^4 x \sqrt{-g} \biggl [\frac{1}{2} g^{\mu\nu}
{\partial}_{\mu}X {\partial}_{\mu}X - \frac{1}{2}m_X^2 X^2 \biggr ] 
\label{lagrangx}\\
{\mathcal L}_{A} &=& \int d^4 x \sqrt{-g} \biggl [\frac{1}{2} g^{\mu\nu}
{\partial}_{\mu}A {\partial}_{\mu}A - \frac{1}{2}m_A^2 A^2 - 
\frac{\lambda'}{n'!}A^{n'} \biggr ] \label{lagranga} \\
{\mathcal L}_{int} &=& \int d^4 x \sqrt{-g} \begin{cases} 
\mg \Phi X A, & \quad \quad \text{(a)} \\
\mg \Phi X A^2, & \quad \quad \text{(b)} \\ 
\mg {\Phi}^2 XA, & \quad \quad \text{(c)} \end{cases}  \label{lagrangint}
\eea
Model (a) is the simplest interaction and in presence of an internal symmetry either $X$ is in the 
same representation as one of the other fields and the third one is a singlet, or it is singlet 
and $A$ and $\Phi$ are in conjugate representations. Other cases in (\ref{lagrangint}) can have 
more diverse symmetry properties. In this work we only consider the model (a). Moreover, we assume 
that only $\Phi$ has a self interaction, thus $\lambda' = 0$\footnote{This assumption is for 
simplifying the problem in hand and simulations described in Sec. \ref{sec:simul}. Indeed, as a  
representative of SM fields, $A$ must have self-interaction.}. It is well known that quantum 
corrections increases the dynamical mass of $\Phi$. For this reason, usually a shift symmetry i.e. 
a periodic potential is assumed for light scalar fields~\cite{shiftsymm}. But such potentials are 
not renormalizable perturbatively. Moreover, the assumption of small self-coupling and coupling to 
other fields ensures the suppression of high order corrections very quickly. Indeed numerical 
simulations discussed in Sec. \ref{sec:simuldense} show that at the onset of inflation the 
effective mass of the scalar falls off at approaches its initial value. Although we are only interested 
in fully quantum treatment of the model, it is useful to know the classical behaviour of the system 
presented by Lagrangian (\ref{lagrange}), in particular extrema of its classical potential. They are 
calculated in Appendix \ref{app:classpot}.

In~\cite{houricond} we found that the amplitude of the condensate decreases very rapidly with the 
mass of $\Phi$. This observation can understood  as the following: Assume that condensates are 
coherent states as defined in Sec. \ref{sec:condmatrix}. Because these states are quantum 
superposition of many particle states, heavier the field smaller is the probability of the 
production of a large number of particles. Therefore, it is expected that condensate component of 
$X$ and $A$ be subdominant. For this reason we ignore them to simplify the model and its numerical 
simulation\footnote{Condensates of $X$ and $A$ may be important for UV scale phenomena. For 
instance, $A$ may be identified with Higgs. In this case, although the cosmological contribution 
of its condensate would be negligible with respect to $\Phi$, it would have important role in 
symmetry breaking and induction of a dynamical mass for other fields.}. 

\subsection {2PI formalism} \label {sec:2pirev}
The method of effective action~\cite{2piintro} - also called 2 Particle Irreducible (2PI) 
formalism - is closely related to Schwinger-Keldysh~\cite{schwingerkeldyshpath} and 
Kadanoff-Baym~\cite{kadanoffbaym} equations, which generalize Boltzmann equation - more exactly 
BBGKY hierarchy - to describe non-equilibrium systems in the framework of quantum field theory. 
The advantage of 2PI is in the fact that all 1PI corrections are included in the propagators, and 
owing to integration over higher order corrections, better precision for amplitudes of processes 
can be achieved at a lower order of perturbative expansion, see for instance~\cite{2pirev} for a 
recent review and example of applications. The 2PI formalism is also extended to curved 
spacetimes~\cite{heatkernel,2picurved}. 

The effective action depends on both 1-point $\varphi \equiv \langle \hat{\Phi} (x) \rangle$ 
and 2-point expectation values:
\bea
&& G (x,y) \equiv \langle \Psi |T\hat{\Phi} (x) \hat{\Phi} (y)| \Psi \rangle - 
\varphi (x) \varphi (y) = \langle \Psi |T\hat{\phi} (x) \hat{\phi} (y)| \Psi \rangle = 
\tr (\hat{\rho} \hat{\phi} (x) \hat{\phi} (y)) \label{green2} \\ 
&& \hat{\phi} \equiv \hat{\Phi} - I \varphi, \quad \quad \hat{\rho} \equiv 
| \Psi \rangle \langle \Psi | \label{densitymatrixdef}
\eea
where in Heisenberg picture the density matrix $\hat{\rho}$ is independent of time. Note that in 
the definition (\ref{green2}) we have omitted internal indices of fields. Indeed 2-point Green's 
functions can be defined for two fields with different indices if the model has e.g. an $O(N)$ 
symmetry. We call these 2-point expectation values {\it mixed propagators}. 

Using the definition of perturbatively free states in Appendix \ref{app:initparticle}, it is evident 
that a state with $\varphi \neq 0$ cannot contain finite number of particles. We call such a state a 
{\it condensate}. A general state $\rho$ can be a superposition of condensates and perturbatively 
free particles. The condensate component has its own fluctuations, which manifest themselves in time 
and position dependence of the classical field $\varphi$.

The density operator $\hat{\rho}$ can be a vacuum state or otherwise. In Minkowski spacetime vacuum 
state is defined as the state annihilated by number operator 
$\hat{N}_\alpha \equiv a_\alpha^{\dagger} a_\alpha$ for any mode $\alpha$. However, in curved spacetimes 
this definition is frame dependent, and under a general coordinate transformation such a state becomes 
a state with infinite number of particles~\cite{qmcurve}. An alternative definition of vacuum is a 
superposition of condensate states such that the amplitude of all components approaches 
zero~\cite{hourivacuum}. It is shown that this vacuum state is annihilated by number operator in any 
frame\footnote{Any superposition of states with finite number of particles or modes and non-vanishing 
amplitudes by definition is not vacuum in a discrete manner, i.e. one of $N < \infty$ superposition 
states must have a close to 1 amplitude. Only if the number of states in the superposition, and 
thereby the number of particles or modes, goes to infinity, their amplitudes can asymptotically 
approaches zero to make a vacuum. In this limit states with finite number of particles and vanishingly 
small amplitude can be added to the superposition without changing its expectation value. Therefore, 
at this limit case any state of many particle bosons can be considered as a superposition of coherent 
states with vanishing amplitudes. \label{footvac}}. Through this work {\it vacuum} refers to such a 
state.

In 2PI formalism the effective action can be decomposed as~\cite{2piintro,heatkernel,2pirev}:
\be
\Gamma [\{\varphi_\alpha\}, \{G_{\alpha\beta}\}] = S [\{\varphi_\alpha\}] + \frac{i}{2} 
\biggl (tr [\ln G_{\alpha\beta}^{-1}] + tr [\mathcal{G}_{\alpha\beta}^{-1} G_{\alpha\beta}] \biggr ) + 
\Gamma_2 [\{\varphi_\alpha\}, \{G_{\alpha\beta}\}] + const \label{effaction}
\ee
From left to right the terms in the r.h.s. of (\ref{effaction}) are classical action for condensates 
of quantum fields $\{\Phi_\alpha\}$, 1PI contribution, and 2PI contribution to the effective action 
$\Gamma [\{\varphi_\alpha\}, \{G_{\alpha\beta}\}]$. Propagators $G_{\alpha\beta}$ 
are the 2-point Green's functions defined in (\ref{green2}) for quantum fields $\hat{\Phi}_\alpha$ and 
$\hat{\Phi}_\beta$. The trace is taken over both flavor indices $\alpha$ and spacetime. The free 
propagator $\mathcal{G}_{\alpha\beta}$ is the second functional derivative of the classical 
action\footnote{Through this work we use $(+, -, -, -)$ signature for the metric. Space components 
of position vectors are presented with bold characters.}:
\bea
&& \mathcal{G}_{\alpha\beta}^{-1} (x,y) = i\delta_{\alpha\beta} \biggl (\frac{1}{\sqrt{-g}}{\partial}_{\mu}
(\sqrt{-g} g^{\mu\nu}{\partial}_{\nu}) + V''(\varphi_\alpha) \biggr ) \delta^4 (x,y) 
\label{freepropag} \\
&& \int d^4x \sqrt {-g}~\delta^4 (x,y) f(x) \equiv f (y), \quad \forall f \label{deltafunc}
\eea
where $V$ is the interaction potential in the Lagrangian of the model. The propagator 
$G_{\alpha\beta}$ is assumed to contain all orders of perturbative quantum corrections and in this 
sens it is exact. 

To fix notations and 2PI equations that we will apply to the model studied here, we briefly review 
how (\ref{effaction}) is obtained. In the framework of Schwinger-Keldysh Closed Time Path integral 
(CTP)~\cite{schwingerkeldyshpath,kadanoffbaym} (also called in-in formalism) the generating functional 
$\mathcal{Z}[\varphi,G]$ of Feynman diagrams can be expanded as~\footnote{In addition to flavor 
indices, in CTP integrals fields and propagators have path indices $+$ or $-$. For the sake of 
simplicity of notation here we show either species or CTP indices, depending on which one is more 
relevant for the discussion. The other indices are assumed to be implicit.}:
\bea
\mathcal{Z} (J_a,K_{ab};\varrho) & \equiv & e^{i W[J_a,K_{ab}]} = \int \mathcal{D} \Phi^a \mathcal{D} 
\Phi^b \exp \biggl [i S(\Phi^a) + i\int d^4x \sqrt {-g} J_a(x) \Phi^a(x) + \nonumber \\ 
&& \frac{i}{2} \int d^4x d^4y \sqrt {-g (x)} \sqrt {-g (y)} \Phi^a (x) K_{ab} (x,y) \Phi^b (y) \biggr] 
\langle \Phi^a | \hat{\varrho} | \Phi^b \rangle \label{genpathint}
\eea
where indices $a,b \in \{+,-\}$ indicate two opposite time branches. They are contracted by the 
diagonal tensor $c_{ab} \equiv \text{diag} (1, -1)$. Apriori the spacetime metric must be also defined 
separately on two time paths, but we follow~\cite{heatkernel,2picurved} and assume 
$g_{\mu\nu}^a = g_{\mu\nu}~\forall a \in \{+,-\}$. This is a good approximation when matter distribution 
is close to uniform and gravitational effect of energy density fluctuations is much smaller than their 
quantum effects and propagation of fields along {\it in} and {\it out} paths cannot be felt by the 
local classical field $g^{\mu\nu}$.

States $|\Phi^a \rangle$ consist of an orthonormal basis of eigen vectors of quantum 
field $\hat{\Phi}$. Their eigen values are identified with field configurations $\Phi^a$. The 
density matrix $\hat{\rho}$ can be pure or mixed. Here we only consider the case of pure 
states\footnote{The configuration field $\Phi$ and thereby density operator should be considered 
to present infinite number of particles. States with finite number of particles can be assumed as 
special cases where only a measure zero subset of configurations have non-zero amplitude.}. The last 
factor in (\ref{genpathint}) is expected to be a functional of $\Phi^a$:
\be 
\langle \Phi^a | \hat{\varrho} | \Phi^b \rangle = \exp (iF[\Phi^a,\Phi^b]) \label{rhomatrix}
\ee
and its contribution can be added to other terms in square brackets in (\ref{genpathint}) as a 
functional which is non-zero only at initial time $t_0$~\cite{inin,2piinitcond}. Notably, terms up to 
order 2 in the Taylor expansion of $F [\Phi]$ can be added to $J$ and $K$ currents and will be 
absorbed in the initial condition of 1-point and 2-point Green's functions. We first consider this 
simplest - Gaussian - case and then discuss more general cases, in which $F[\Phi]$ depends on higher 
orders of $\Phi$.

Ignoring both flavor and path integral branch indices, the functionals $J$ and $K$ are defined such 
that:
\be
\frac{\partial W[J,K]}{\partial J(x)} = \varphi (x), \quad \quad \frac{\partial W[J,K]}
{\partial K(x,y)} = \frac{1}{2}\biggl (G (x,y) + \varphi (x) \varphi (y) \biggr ) \label{jkcondition}
\ee
The effective action $\Gamma [\{\varphi_\alpha\}, \{G_{\alpha\beta}\}]$ must be independent of auxiliary 
functionals $J$ and $K$, and is defined by a double Legendre transformation:
\be
\Gamma [\varphi, G] = W[J,K] - \int d^4x \sqrt {-g} J(x) \varphi(x) - \frac{1}{2} \int d^4x d^4y 
\sqrt {-g (x)} \sqrt {-g (y)} K (x,y) [G (x,y) + \varphi (x)\varphi (y)] 
\label{effactlegendre}
\ee
Derivatives of (\ref{effactlegendre}) with respect to $\varphi$ and $G(x,y)$ are:
\bea
&& \frac {\partial\Gamma [\varphi, G]}{\partial \varphi (x)} = -J (x) - 
\int d^4y \sqrt {-g (y)} K(x,y) \varphi (y) \label {gammavarphi} \\
&& \frac {\partial\Gamma [\varphi, G]}{\partial G (x,y)} = - \frac{1}{2} K(x,y) \label {gammag}
\eea
After eliminating auxiliary functional $J$ and $K$ by adding the last term in (\ref{gammag}) to the 
action and performing again a Legendre transformation, one obtains (\ref{effaction}) up to an 
irrelevant constant which can be included to normalization of fields and ignored. 
The 2PI effective Lagrangian $\Gamma_2 [\varphi,G]$ includes terms which are not included in the 
modified 1IP effective action and consists of 2PI Feynman diagrams without external lines. 

The effective action can be treated as a classical action depending on fields $\varphi$ and $G$. 
Their evolution equations satisfy usual variational principle:
\bea
&&\frac{\partial \Gamma [\varphi,G]}{\partial \varphi} = \frac{\partial S[\varphi]}
{\partial \varphi} - \frac{i}{2} \biggl (tr [G^{-1} \frac{\partial G}{\partial \varphi}] - 
tr [\mathcal{G}^{-1}\frac{\partial G}{\partial \varphi}] \biggr ) + 
\frac{\partial \Gamma_2 [\varphi,G]}{\partial \varphi} = 0. \label{phievol} \\
&&\frac{\partial \Gamma [\varphi,G]}{\partial G} = \frac{i}{2} tr [\mathcal{G}^{-1} - G^{-1}] + 
\frac{\partial \Gamma_2 [\varphi,G]}{\partial G} = 0. \label{propagevol}
\eea
The last term in (\ref{propagevol}) is proportional to self-energy defined as:
\be
\Pi (\varphi,G) \equiv 2i \frac{\partial \Gamma_2 [\varphi,G]}{\partial G} \label{2piselfener}
\ee
In presence of internal symmetry among fields the effective Lagrangian depends on both {\it pure} 
and mixed propagators, and evolution equation (\ref{propagevol}) also applies to the both types.

\subsubsection {Non-Gaussian states} \label{sec:density}
Equation (\ref{rhomatrix}) defines the elements of density matrix with respect to eigen vectors of 
field operator. In Appendix \ref{app:initparticle} we show that any initial density matrix can be 
expanded as:
\be
F [\Phi] = \sum_{n=0}^\infty \int d^3\mathbf {x}_1 \ldots d^3\mathbf {x}_n \alpha ~(\mathbf{x}_1,\ldots,
\mathbf{x}_n) \Phi (\mathbf {x}_1) \ldots \Phi (\mathbf {x}_n) \label{fexpan}
\ee
where non-local n-point coefficients $\alpha ~(\mathbf{x}_1,\ldots,\mathbf{x}_n)$ include non-local 
correlation and entanglement in the initial state. Equation (\ref{fexpan}) can be also considered as 
the definition of initial state without relating it to a state in the Fock space of a physical system. 
This interpretation is specially useful for systems in a mixed state. Initial correlation and mixing 
can be induced, for instance by factoring out high energy physics~\cite{infinitcond0} or by 
interaction with an external system such as a thermal bath~\cite{oneparticledens}.

After replacing the density matrix components in (\ref{genpathint}) with (\ref{rhomatrix}) the 
classical action can be redefine as~\cite{2piinitcond}: 
\be
S[\Phi] \rightarrow \tilde{S}[\Phi] = S[\Phi] + F[\Phi] \label{classactionmod}
\ee
As we discussed earlier, in 2PI formalism 1-point and 2-point terms in $F[\Phi]$ can be included in 
auxiliary currents $J$ and $K$ and do not induce additional Feynman diagrams to the perturbative 
expansion. Nonetheless, they contribute to the initial conditions for the solution of evolution 
equations (\ref{phievol}) and (\ref{propagevol}). In nPI formalism, which can be constructed by 
repetition of Legendre transformation and inclusion of n-point Green's functions in the effective 
action, $\alpha$ coefficients up to n-point can be included in the auxiliary fields analogous to 
$J$ and $K$.

The 2PI effective action for $\tilde {S}[\Phi]$ is:
\bea
&& \tilde{\Gamma} [\varphi,G] = \tilde {S} [\varphi] + \frac{i}{2} (tr [\ln G^{-1}] + 
tr [\tilde{\mathcal{G}}^{-1} G]) + \tilde{\Gamma}_2 [\varphi, G] - \frac{i}{2}\tr I
\label{effactionconst} \\
&& \tilde{\mathcal{G}}^{-1} (x,y) = \mathcal{G}^{-1} (x.y) + i\frac{\partial^2 F [\varphi]}{\partial
\varphi (x) \partial \varphi (y)} \label{tildfreepropag}
\eea
where $\tilde{\Gamma} [\varphi,G]$ is determined with a vacuum initial condition. Evolution 
equations (\ref{phievol}-\ref{propagevol}) must be also written for $\tilde{\Gamma}$ and 
$\tilde{\mathcal{G}}^{-1}$. 
Non-local terms in $\tilde{S}[\Phi]$ and $\tilde{\Gamma}_2$ 
induce non-local interaction vertices in the effective action, which similar to local interactions, 
can be perturbatively expanded. They also interference with local interactions in the classical 
Lagrangian, but only at initial time. It is proved~\cite{oneparticledens} that in theories with a Wick 
decomposition, also called Gaussian, $n-$point Green's functions for $n > 2$ can be expanded with 
respect to 1 and 2-point Green's functions. Examples of such models are free thermal systems and their 
extension where each energy mode has a different temperature. For these initial states $F[\varphi]$ 
has the form of an Euclidean action and one has to add an imaginary time branch to the closed time 
path integral, see e.g.~\cite{inin,2piinitcond}.

For the model studied here and its simulations it is important to take into account the effect 
of a non-vacuum initial state, including a condensate. The reason is that it is very difficult to 
use a single and continuous simulation beginning with a vacuum state for the light field before 
inflation and ending at present epoch, where it dominates energy density. If numerical simulations 
are broken to multiple epochs, the initial condition of intermediate eras would not be vacuum and 
we must consistently include initial correlations in the evolution of condensates and propagators. 
In Sec. \ref{sec:initcond} we calculate density matrix of physically realistic condensate states 
and determine their $F[\Phi]$ functional.


\subsection {2PI evolution of condensates and propagators in the toy model} \label {sec:2pievol}
We begin this section by presenting 2PI diagrams that contribute to the effective action of the 
toy model (\ref{lagrange}). The models in (\ref{lagrangint}) have two types of vertices: 
self-interaction vertex for $\Phi$ and interaction between 3 distinct fields $X$, $A$, and $\Phi$. 
Of course, diagrams can have a combination of both vertices, but assuming that both couplings 
$\lambda$ and $\mg$ are very small, only lowest order diagrams have significant amplitudes. 
As mentioned earlier, the model (\ref{lagrangint}) can be easily extended to the case where $\Phi$ 
has a flavor presenting an $O(N)$ symmetry. In this case, in order to have a singlet potential, 
the self-interaction order $n$ must be even\footnote{It is possible to construct singlet odd-order 
interaction potentials by using forms of the internal symmetry space. The best example is a 
Chern-Simon interaction. But these models do not have $N = 1$ limit, which for the time being is the 
only case implemented in our simulation code. For this reason, we do not consider them in this work.}. 

\begin{figure}
{\Large \color{darkblue}
\begin{tabular}{p{6cm}p{6cm}p{6cm}}
$D_1$ & $D_2$ & $D_3$ \\
\includegraphics[width=6cm]{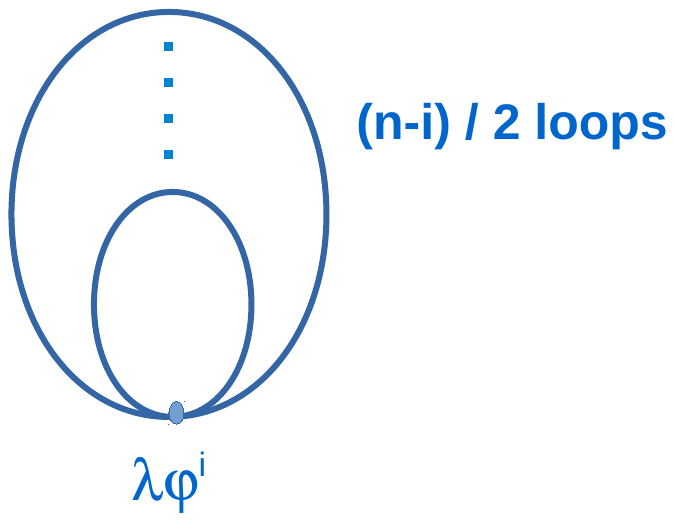} & 
\includegraphics[width=6cm]{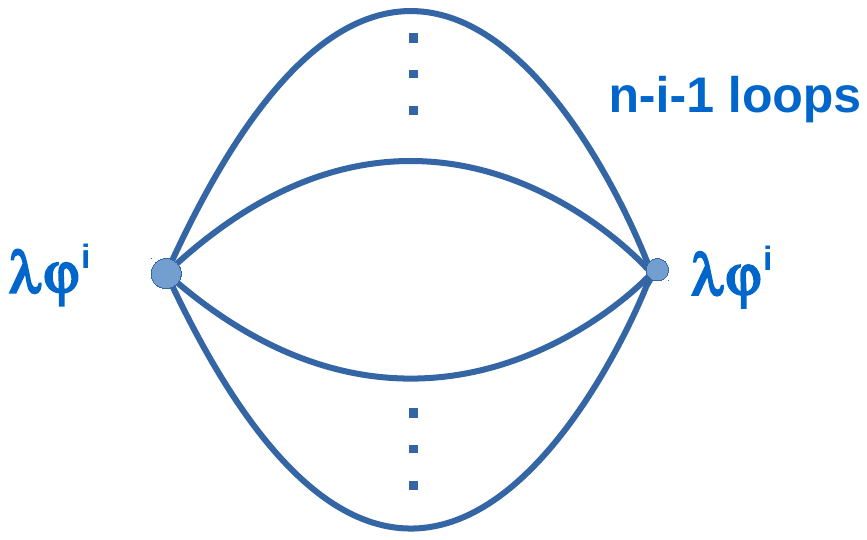} & 
\includegraphics[width=6cm]{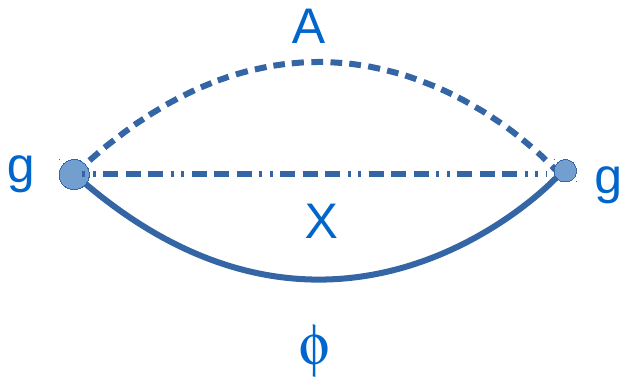} \\

\multicolumn {3}{p{18cm}}{\vspace{-5cm}$\Gamma_2 = \sum\limits_{i=0, n-i=2k}^n N_1 D_1 + 
\sum\limits_{i=0}^{n-3} N_2 D_2 + \mg^2 D_3 + \ldots $} \\
\multicolumn {3}{p{18cm}}{\vspace{-4cm} $N_1 = \frac{\lambda}{n!} C_i^n C_2^{n-i},~
N_2 = (\frac{\lambda}{n!})^2 \frac{(n-i)!}{2} (C_i^n)^2$} 
\end{tabular}
}
\vspace{-3.5cm}
\caption{Diagrams contributing to $\Gamma_2 (\varphi, G)$ up to $\lambda^2$ and $\mg^2$ order of 
model (\ref{lagrange}). If self-interaction of 
$\Phi$ is not monomial, similar diagrams with different values of $n$ weighted by the amplitude of 
monomial terms in the potential must be added to $\Gamma_2 (\varphi, G)$. \label{fig:gammaeffdiag}}
\end{figure}

Fig. \ref{fig:gammaeffdiag} shows the lowest order 2PI diagrams contributing to $\Gamma_2 [\varphi,
{G_{\alpha\beta}}],~\alpha,~\beta \in {X,~A,~\Phi}$ for a vacuum initial condition. 
Derivatives of these diagrams with respect to $\varphi$ and ${G_{\alpha\beta}}$ determine their 
contribution to equations (\ref{phievol}) and (\ref{propagevol}), respectively. 

\subsubsection{Condensates} \label{sec:condevol}
For the condensate field $\varphi$ of model (a) in (\ref{lagrangint}) the evolution equation 
(\ref{phievol}) is expanded as:
\be
\frac{1}{\sqrt{-g}}{\partial}_{\mu}(\sqrt{-g} g^{\mu\nu}{\partial}_{\nu}
\varphi) + m_{\Phi}^2 \varphi + \frac{\lambda}{n!}\sum_{i=0}^{n-1} (i+1)
C^n_{i+1}{\varphi}^i\langle{\phi}^{n-i-1}\rangle - \mg \langle XA\rangle = 0 \label {dyneffa} \\ 
\ee
We should emphasize that this equation is exact at all perturbative order and can be directly 
obtained by decomposing $\Phi = \varphi + \phi$, $\langle \phi \rangle = 0$ in the classical action 
and applying variational principle to classical field $\varphi$. To calculate in-in expectation 
values we use Closed-Time Path integral (CTP) as explained in details in~\cite{houricond}, but in 
place of using free propagators, we use exact propagators determined from equation 
(\ref{propagevol}). In this work we only take into account the contribution of the lowest order 
perturbative terms, which inevitably makes final solutions approximative.

The condensate components of $X$ and $A$ fields satisfy the same evolution as (\ref{dyneffa}) if we 
replace $\Phi$ with $X$ or $A$, respectively. Moreover, because we assumed no self-interaction for 
these fields, the corresponding terms in (\ref{dyneffa}) would be absent. 

\subsubsection{Propagators} \label{sec:propevol}
Using symmetric and antisymmetric propagators defined in Appendix \ref{app:prop} and 
equations (\ref{propagevol}), evolution equations of these 
propagators~\cite{heatkernel,2pirev,2picurved} for the three fields of the model are obtained as:
\bea
\biggl [\frac{1}{\sqrt{-g}}{\partial}_{\mu}(\sqrt{-g} g^{\mu\nu}{\partial}_{\nu}) + M_i^2(x) \biggr] ~ 
G_i^F(x,y) & = & -\int_{-\infty}^{x^0}d^4z \sqrt {-g (z)} ~ \Pi_i^\rho(x,z) ~ G_i^F(z,y) + \nonumber \\
&& \int_{-\infty}^{y^0}d^4z \sqrt {-g (z)} ~ \Pi_i^F(x,z) ~ G_i^\rho(z,y) \label{evolgf} \\
\biggl [\frac{1}{\sqrt{-g}}{\partial}_{\mu}(\sqrt{-g} g^{\mu\nu}{\partial}_{\nu}) + M_i^2(x) \biggl] ~ 
G_i^\rho(x,y) & = & -\int_{y^0}^{x^0}d^4z \sqrt {-g (z)} ~ \Pi_i^\rho(x,z) ~ G_i^\rho(z,y) \label{evolgrho}
\eea
\be
M_\Phi^2 (x) = m_\Phi^2 + \frac{\lambda}{(n-1)!} \sum_{j=0}^{[n/2]-1} C^{[n/2] - 1}_j 
\varphi^{n-2(j+1)}(x) (G^F_{\Phi} (x,x))^j, \quad M_{X, A}^2 = m_{X, A}^2 \label {local2pi}
\ee 
where $i = X,~A,~\phi$. In (\ref{local2pi}) $[n/2]$ means the integer part of $n/2$ and $C^i_j$ is 
the combinatory coefficient. Effective masses $M_i,~i=X,~A,~\phi$ include local 2PI corrections. 
However, as $X$ and $A$ are assumed not to have self-interaction, no local mass correction is 
induced to their propagators. If the fields of the models have internal symmetries, $G$'s and 
$\Pi$'s may have internal symmetry indices. In this case, eq. (\ref{evolgrho}) applies also to mixed 
propagators. Here we mostly consider the simpler case of single fields without internal symmetries 
and only briefly mention the case with internal symmetry. We also ignore species index $i$ when 
there is no risk of confusion. If we assume that all interactions are switched on at the initial 
time $t_0$, the lower limit of integrals in (\ref{evolgrho}) will shift to $t_0$. Self-energies 
$\Pi^F$ and $\Pi^\rho$ are defined in Appendix \ref{app:prop}. Symmetric and antisymmetric propagators 
are suitable for studying the evolution of a quantum system, specially numerically, because the 
r.h.s. of their evolution equations are explicitly unitary and causal~\cite{heatkernel,2pirev}.

\begin{figure}
{\Large \color{darkblue}
\begin{tabular}{p{6cm}p{6cm}p{6cm}}
$D_4$ & $D_5$ & $D_6$ \\
\includegraphics[width=6cm]{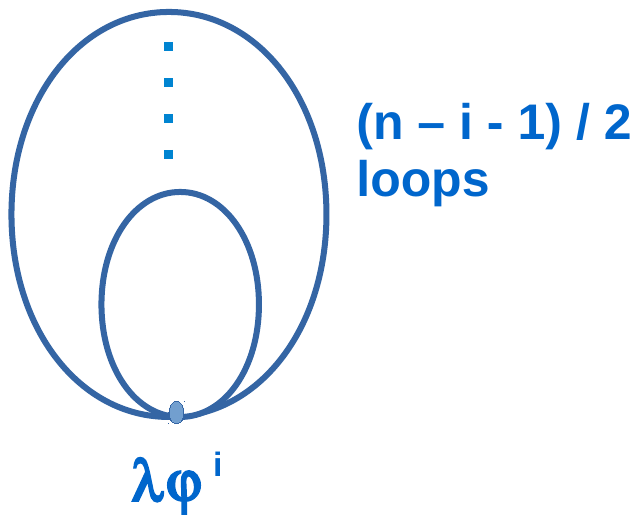} & 
\includegraphics[width=6cm]{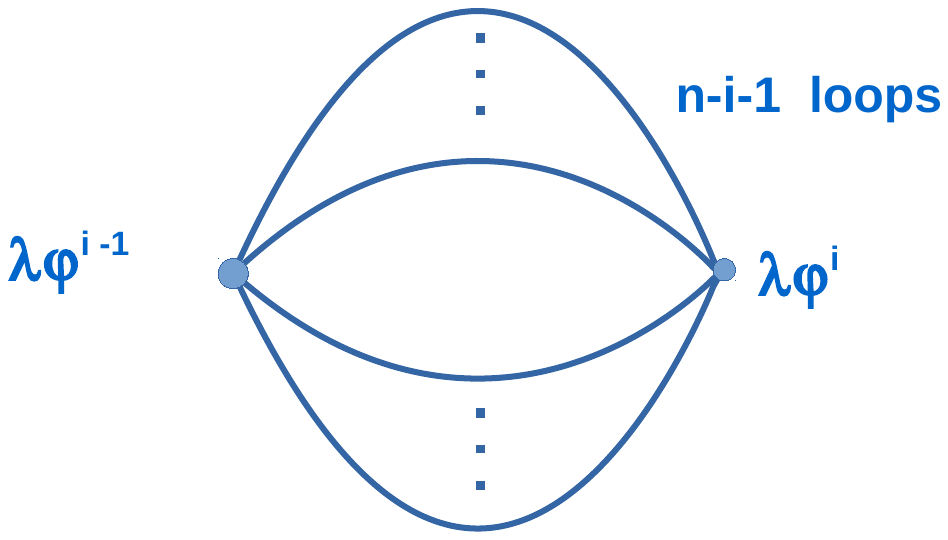} & 
\includegraphics[width=6cm]{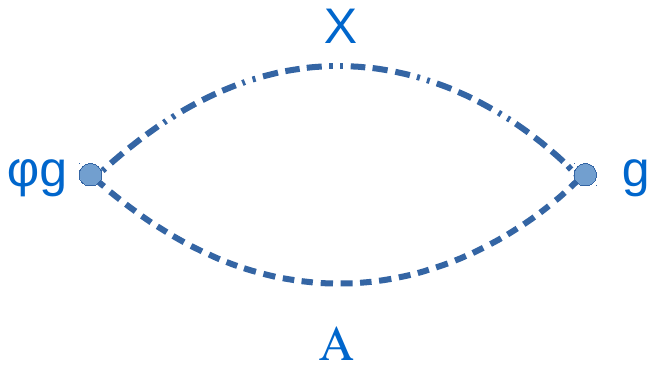} \\
\multicolumn {3}{p{18cm}}{\vspace{-5cm}$\frac{\partial\Gamma_2}{\partial \varphi} = 
\sum\limits_{i=0, n-i=2}^{n-1} i N_1 D_4 + 2 \sum\limits_{i=1}^{n-3} i N_2 D_5 + 2 \mg^2 D_6 + \ldots $}
\end{tabular}
}
\vspace{-3.5cm}
\caption{Diagrams contributing to $\partial \Gamma_2 (\varphi, G)/\partial \varphi$ up to $\lambda^2$ 
and $\mg^2$ order. They correspond to correlation functions in (\ref{dyneffa}). Coefficients $N_1$ and 
$N_2$ are defined in Fig. \ref{fig:gammaeffdiag}. Diagram $D_6$ contributes to 
$\langle X(x)A(x) \rangle$ in eq. (\ref{dyneffa}) and presents contraction of $\varphi(y)X(y)A(y)$ in 
$\mg^2$-order correction $\langle X(x)\varphi(y)X(y)A(y)A(x) \rangle$ to this correlation function. 
\label{fig:condensatediag}}
\end{figure}

\begin{figure}
{\Large \color{darkblue}
\begin{tabular}{p{6cm}p{6cm}p{6cm}}
$D_7$ & $D_8$ & $D_9$ \\
\includegraphics[width=6cm]{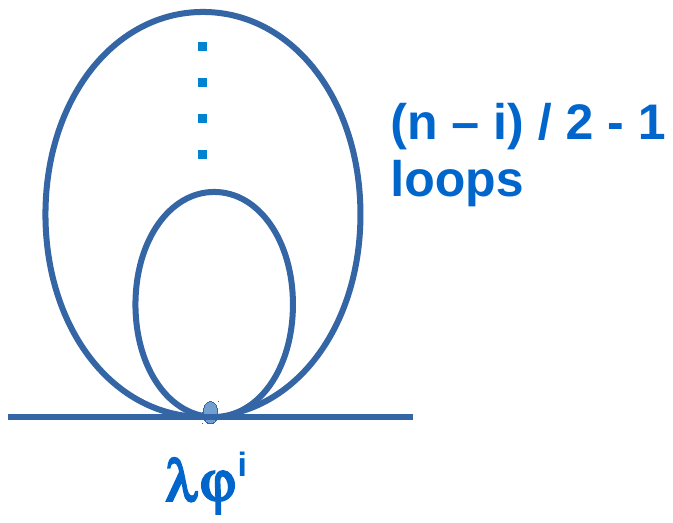} & 
\includegraphics[width=6cm]{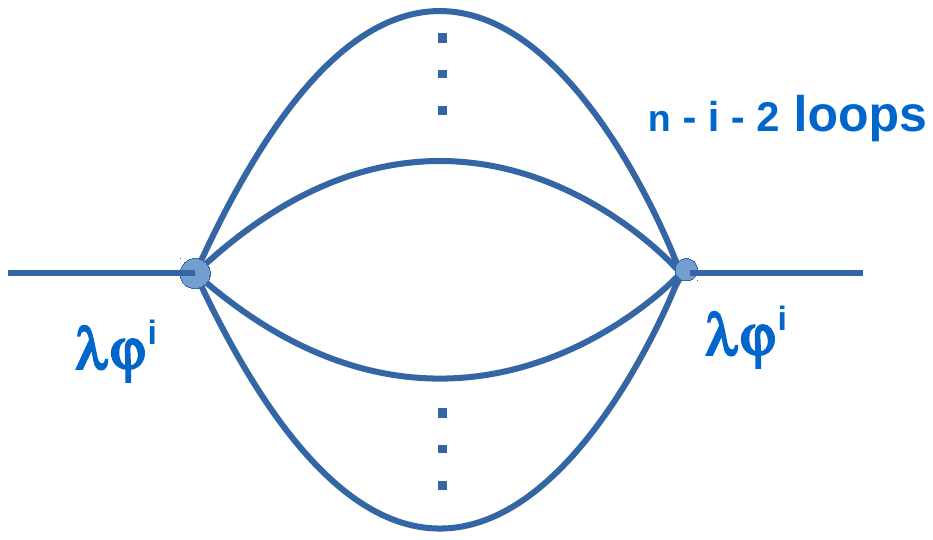} & 
\includegraphics[width=6cm]{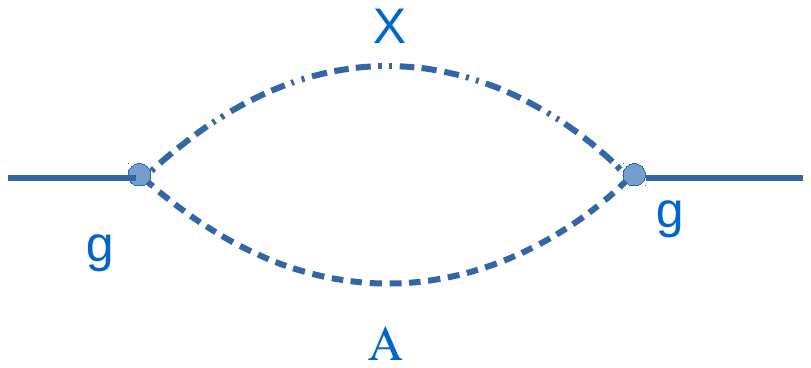} \\
\multicolumn{3}{p{18cm}}{\vspace{-5cm}$\frac{\partial\Gamma_2}{\partial G_\phi} = \sum\limits_{i=0, n-i=2k}^{n-2} 
\frac {(n - i)N_1}{2} D_7 + \sum\limits_{i=0}^{n-3} (n-i) N_2 D_8 + \mg^2 D_9 \ldots $} \\
\vspace{-3.5cm} $D_{10}$ & \vspace{-3.5cm} $D_{11}$ & \\
\vspace{-3.5cm} \includegraphics[width=6cm]{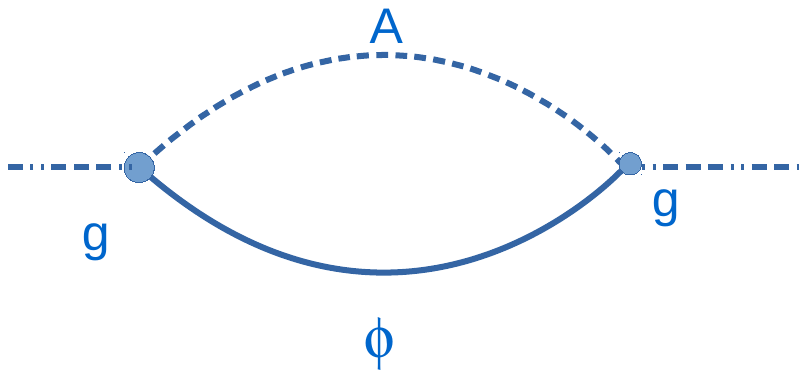} & 
\vspace{-3.5cm} \includegraphics[width=6cm]{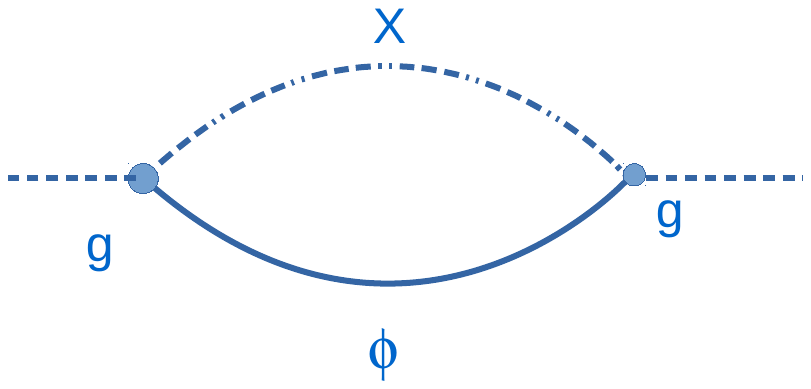} & \\
{\vspace{-6cm} $\frac{\partial\Gamma_2}{\partial G_X} = \mg^2 D_{10}$} & 
{\vspace{-6cm} $\frac{\partial\Gamma_2}{\partial G_A} = \mg^2 D_{11}$ } &
\end{tabular}
}
\vspace{-4.5cm}
\caption{Diagrams contributing to $\partial \Gamma_2 (\varphi, G)/\partial G_\phi$, $\partial 
\Gamma_2 (\varphi, G)/\partial G_X$, and $\partial \Gamma_2 (\varphi, G)/\partial G_A$, or 
in other words to self-energies. Coefficients $N_1$ and $N_2$ are defined in 
Fig. \ref{fig:gammaeffdiag}. The tadpole diagrams only contribute to effective mass term 
(\ref{local2pi}) and do not appear in r.h.s. of equations (\ref{evolgf}) and (\ref{evolgrho}). 
\label{fig:propagdiag}}
\end{figure}

To proceed with detailed construction of evolution equations, we need to specify 2PI diagrams that 
contribute to in-in expectation values in (\ref{dyneffa}) and self-energy in (\ref{evolgf}) and 
(\ref{evolgrho}). Figs. \ref {fig:condensatediag} and \ref {fig:propagdiag} show these diagrams. 
We remark that for interaction (a) in (\ref{lagrangint}) a non-zero condensate does not induce 
a local mass. By contrast it is easy to see that interactions (b) and (c) can be considered as  
effective mass for $A$ and $\Phi$, respectively. In these cases the mass matrix of fields is not 
diagonal and the model has an induced $O(2)$ symmetry when condensates are present and in addition 
to usual loop diagrams, one must consider mixed propagators $G_{AB}$, where $A$ and $B$ are different 
fields. Like their diagonal counterparts evolution of mixed propagators is ruled by 
eqs. (\ref{evolgf}) and (\ref{evolgrho}), but additional Feynman diagrams~\cite{2pirev} including 
condensate insertion contribute to these equations. However, because the amplitude of induced 
mass (insertion) is proportional to the coupling, diagrams with mixed propagators have higher 
perturbative order than their single-field counterparts.

\subsection {Renormalization} \label{sec:renorm}
Renormalization of 2PI formulation of $\Phi^n$ models in Minkowski space is studied in details 
in~\cite{renormflat}, with thermal initial state in~\cite{renormflat0}, and that of gauged models 
in~\cite{renormflat1}. Numerical simulation of 2PI renormalization using both BPHZ~\cite{renormbphz} 
counterterm method and exact renormalization group equation~\cite{nprge,rgoptimalregul} is 
described in details in~\cite{renormflatsimul}. 

Although significant development on the renormalization of quantum field theories in curved 
spacetimes is achieved, specially using the method called {\it adiabatic 
regularization}~\cite{renormadiab,qmcurve}, their application to 2PI formalism has been 
mostly in de Sitter space. For instance, heat kernel~\cite{heatkernel,deultralightir} and 
non-perturbative Renormalization Group (RG) flow are used to determine the effect of quantum 
corrections on the evolution of inflation and scalar perturbations~\cite{qftdergflow} . The exact 
renormalization group equation is also employed to determine quantum corrected effective potential 
of inflation~\cite{rgdesitter}. Moreover, the BPHZ counterterm method is used to renormalize this 
quantity as well as the energy-momentum tensor~\cite{renorminf2pi}. Aside from the importance of 
effective potential for comparison with cosmological observations, it also determines whether at 
the end of inflation symmetries broken by the inflaton condensate were 
restored~\cite{desittersymmbreak}. 

Application of the Weinberg power counting theorem shows that the model studied here is 
renormalizable for all the interaction options between $X,~A,~\Phi$ fields considered in 
(\ref{lagrangint}), and for self-interaction order $n = 3 ~ \& ~ 4$. Although all renormalization 
techniques lead to finite physical observables and their running with scale, some methods may be 
more suitable for some applications than others. Notably, adiabatic subtraction is more suitable 
and straightforward for numerical solution of evolution equations and has been used for calculation 
of nonequilibrium quantum effects during reheating after inflation~\cite{renormadiab0,renormadiab1}. 

In this method rather than renormalizing effective Lagrangian, which is performed in BPHZ and RG 
techniques, Green's functions are renormalized. For renormalizing a n-point Green's function, the 
expansion of vacuum Green's function $G_{vac}$ of the same order (number of points) with respect to 
expansion rate and its derivatives up to finite terms is subtracted, mode by mode, from bare 
Green's function\footnote{If Green's functions are computed for a non-vacuum state, free rather 
than vacuum solution must be used for the expansion. Moreover, renormalized value of mass $M$ rather 
than $m$ must be used in the solutions.}~\cite{renormadiab,qmcurve,renormadiab0}. Propagators 
are determined at desired perturbative order using the solution of equation (\ref{evolgf}) with 
r.h.s. put to zero and vacuum initial conditions - corresponding to $|\psi|^2 = 0$ in 
(\ref{propfu1part}). As no analytical solution for evolution equations with an arbitrary $a (t)$ 
is known, one has to use a WKB expansion~\cite{renormadiab0,renormadiab1}. Exact solutions of 
field equations, when they exist, and WKB approximation and its expansion with respect to $\dot{a}$ 
and its derivatives are reviewed in Appendix \ref{app:solution}.

Although the exact expressions of the solutions of evolution equations depend on the initial 
conditions, which we discuss in detail is Sec. \ref{sec:initcond}, a simple power counting of the 
integrals in (\ref{propfu1part}) shows that they are UV divergent. For $\Phi$ propagators 
these singularities are generated by the local term in self-energy, which is quadrically divergent, 
and by 2-vertex diagrams, which have logarithmic UV singularities, see Fig. \ref {fig:propagdiag}. 
Self-energy diagrams of $X$ and $A$ are only logarithmically divergent because we assumed 
$\lambda' =0$ in Lagrangian (\ref{lagrange}). Similarly, tadpole and 2-vertex expectation values 
in the evolution equation of $\Phi$ condensate $\varphi$ are 
quadrically divergent.

A theorem by Fulling, Sweeny, and Wald (FSW)~\cite{renormsingular} states that if a singular 2-point 
Green's function $G(x,y)$ at $x \rightarrow y$ can be decomposed to smooth functions $u(x,y)$, 
$v(x,y)$ and $w(x,y)$ in an open neighbourhood on a Cauchy surface such that:
\bea
&& G(x,y) = \frac{u (x,y)}{\sigma} + v (x,y) \ln \sigma + w (x,y), \quad \quad \sigma 
\equiv \frac{1}{2}|x-y|^2 \label{singulardecomp} \\
&& v (x,y) = \sum_n v_n(x,y) \sigma^n, \quad \quad w (x,y) = \sum_n w_n(x,y) \label{hadamardfuncs}
\eea
where $u$, $v_n$ and $w_n$ satisfy Hadamard recursion relation, then $G(x,y)$ has the Hadamard 
form (\ref{singulardecomp}) everywhere and evolution of Cauchy surface preserves this property. 
In curved spacetime this theorem assures that the structure of singularities of adiabatic 
vacuum propagators is preserved during the evolution of fields and geometry\footnote{In an 
expanding universe the condition for existence of an asymptotically Minkowski behaviour of mode $k$ 
is $k^2/a^2 + m^2 \gg (\dot{a}/a)^2 = H^2$, where $H$ is the Hubble function~\cite{qmcurve}. Modes 
which satisfy this condition have negligible probability to be produces by Unruh radiation due to 
the expansion. If all the modes of a quantum field satisfy this condition, its vacuum is called 
{\it adiabatic vacuum}. It is clear that if at some epoch $M < H$, IR modes will not respect 
adiabaticity condition. For this reason, vacuum subtraction must be performed for an arbitrary 
mass $m$ before applying $m \rightarrow M$~\cite{qmcurve}.}.

Power counting of singularities of the effective mass term explained above shows that their  
singularities are of the same order as those in (\ref{singulardecomp}). Therefore, $G$ and $G_{vac}$ 
have the same sort of singularities, and according to FSW theorem subtraction of their divergent 
terms should lead to a finite and renormalized theory. However, this theorem is proved for 2-point 
operator valued distributions which satisfy a wave function equation of the form:
\be
\biggl( D_\mu D^\mu + M (x) \biggr ) G (x,y) = I (x,y) \label{adiaevol}
\ee
where $I (x,y)$ is a smooth external source. Therefore, apriori it cannot be applied to exact 
propagators in 2PI, which satisfy the integro-differential equations (\ref{evolgf}) and 
(\ref{evolgrho}). On the other hand, we can heuristically and perturbatively consider the integrals 
on the r.h.s. of (\ref{evolgf}) as an small external source, which depends on second and higher 
orders of coupling constants. In this case FSW theorem would be applicable, and we can define 
renormalized propagators 
as:
\be
G_R (x,y) = G_B (x,y) - G_{vac_N} (x,y) \label{renormgreen}
\ee
where $R$ and $B$ indicate renormalized and bare quantities, respectively. The index $N$ is the 
adiabatic order in the expansion of vacuum with respect to derivatives of expansion factor. It must 
correspond to divergence order of the Green's function. More generally, renormalized expectation 
value of any operator $\mathcal {O}$ can be {\it formally} expressed as:
\be
\langle \mathcal {O} \rangle_R = \langle \mathcal {O} \rangle_B - \langle \mathcal {O} \rangle_{vac,N} 
\label{obsrenorm}
\ee
where $N$ must correspond to singularity order of $\langle \mathcal {O} \rangle_B$. The reason for 
calling this expression {\it formal} is that it does not explicitly show how subdivergences -  
divergent subdiagrams - are renormalized. Indeed, the method of adiabatic regularization was 
originally developed for regularization of expectation value of number operator on vacuum state of 
a free scalar field~\cite{adiaborgin}. Nonetheless, the technique can be applied to interacting 
models by subtracting adiabatic expansion of a vacuum solution separately for each mode in each 
loop. This hierarchical subtraction procedure is similar to the addition of counterterms to Lagrangian 
to remove subdivergences in BPHZ method. For instance, tadpole diagrams $D_4,~D_7$ in 
Figs. \ref{fig:condensatediag} and \ref{fig:propagdiag} for $n=4$ self-coupling model, 
which contribute to the effective mass of condensate and propagator, respectively, can be 
renormalized as:
\be
[D_4 \& D_7]_R ~ \propto ~ \frac{1}{(2\pi)^3} \int dk^3 e^{-ik.\mathbf{x}} a^{-3/2} [G_B (k,t) - 
|\mathcal {U}^{(2)}_k (t)|^2]
\ee
where $G_B (k,t)$ is the bare propagator evolving according to eq. (\ref{evolgf}). The function 
$\mathcal {U}^{(2)}_k$, $|\mathcal {U}^{(2)}_k|^2 \equiv G^{(2)}_{vac} (k,t)$ is the adiabatic expansion 
up to order 2 of the solution of free field equation defined in (\ref{soldecomp}). The adiabatic 
order corresponds to divergence order of $D_4$ diagram and the expansion is performed according to 
expression (\ref{wkbappoxsol}). 1-loop diagrams in $D_2$, $D_5$ and $D_6$ are only logarithmically 
divergent. Therefore, in Fourier space we have to determine subtractions of form 
$G_B (k,t) G_B (k,t) - G^{(0)}_{vac} (k,t) G^{(0)}_{vac} (k,t)$, where we have omitted species and path 
indices. Implementation of this renormalization procedure in numerical calculations is much easier 
than e.g. abstract counterterms in BPHZ method or variation of dimension in dimensional 
regularization and renormalization. In any case, diagrams in 
Figs. \ref{fig:gammaeffdiag}-\ref{fig:propagdiag} do not contain any divergent sub-diagram and 
problem of subdivergence does not arise at perturbation orders considered in this work.


Renormalized condensate $\varphi_R$ is obtained by using renormalized expectation values in its 
evolution equation (\ref{dyneffa}) and no additional renormalization would be necessary. From now 
on we assume that adiabatic renormalization procedure is applied to observables and drop the 
subscript $R$ when it is not strictly necessary.

\subsubsection{Initial conditions for renormalization} \label{sec:renormcond}
In order to fix renormalized mass, self-coupling, and coupling between $X$, $A$, and $\Phi$ 
we define the following initial conditions:
\bea
&& \frac{\delta^2 \Gamma_R (\varphi_R, G_R)}{\delta \varphi_R^2} \biggr|_{\varphi_R = 0,\mu_0} = 
-m^2_{R\Phi}, \quad \frac{\delta^n \Gamma_R (\varphi_R, G_R)}{\delta \varphi_R^n}
\biggr|_{\varphi_R = 0,\mu_0} = -\lambda_R, \quad \frac{\delta^2 \Gamma_R (\varphi_R, G_R)}{\delta 
(\partial_\mu \varphi_R) \delta (\partial_\nu \varphi_R)} \biggr|_{\varphi_R = 0,\mu_0} = 
g^{\mu\nu} \nonumber \\ 
&& \label{renormphi}\\
&& \frac{\delta^3 \Gamma_R (\varphi_R, G_R)}{\delta G_{R_\phi}(x,y) \delta G_{R_X} (x,y) \delta 
G_{R_A} (x,y)} \biggr|_{\varphi_R = 0} = \mg^2_R. \quad \label{renorm34} \\
&& \frac{\delta \Gamma_R (\varphi_R, G_R)}{\delta G_{R_X} (x,x)} \biggr|_{\varphi_R = 0,\mu_0} = 
M_{R_X}^2 (x) = m_{R_X}^2 \quad \quad 
\frac{\delta \Gamma_R (\varphi_R, G_R)}{\delta G_{R_A} (x,x)} \biggr|_{\varphi_R = 0,\mu_0} = 
M_{R_A}^2 (x) = m_{R_A}^2. \label{renormax} \\
&& \frac{\delta^3 \Gamma_R (\varphi_R, G_R)}{\delta \varphi_R(x,y) \delta G_{R_X} (x,y) \delta 
G_{R_A} (x,y)} \biggr|_{\varphi_R = 0} = \mg_R. \quad \label{renormxacond}
\eea
where a renormalization scale $\mu_0 \ll M_X \ll M_P$ is assumed. Due to interaction with the 
condensate, masses and couplings depend on the amplitude of the condensate $\varphi$ and their values 
at renormalization scale must be defined for a given value of the condensate. The choice of 
$\varphi = 0$ in (\ref{renormphi})-(\ref{renormxacond}) is motivated by the fact that we assume 
$\varphi (t_0) = 0$, where $t_0$ is the initial time in simulations discussed in 
Sec. \ref{sec:simulation}. Similar to Lagrangian renormalization techniques, a renormalization group 
equation can be written for adiabatic subtraction method with respect to adiabatic time scale 
$T \sim 1/\mu_0$, which is used for adiabatic expansion, see Appendix \ref{app:solution} 
and~\cite{renormadiab,qmcurve} for more details. Equation (\ref{renormxacond}) is a consistency 
condition for coupling of the classical field $\varphi$ with $X$ and $A$. It is not independent of 
$XA\Phi$ vertex defined in (\ref{renorm34}) and is included in the renormalization conditions for 
the sake of completeness.

We remind that the Lagrangian (\ref{lagrangint}) is not symmetric with respect to fields $X,~A$ and 
$\Phi$, and there is no mixed propagator in the model\footnote{We remind that the correlation 
$\langle X(x)A(x)\rangle$ is not a propagator and would be null if the coupling constant 
$\mg \rightarrow 0$}. However, if we consider an internal symmetry for each of the three $X,~A$ and 
$\Phi$ fields, the effective Lagrangian will depend on mixed propagators carrying 2 different internal 
indices. In this case, additional renormalization conditions for mixed propagators and interaction 
vertices, which must respect symmetries, would be necessary. As in the simulations discussed in 
Sec. \ref{sec:simulation} we only consider the simple case of fields without internal symmetry, we do 
not discuss the case with internal symmetry further. Scalar field models with $O(N)$ symmetry 
and their renormalization are extensively studied in the literature, see e.g.~\cite{renorminf2pi}. 

In a cosmological context the expansion of the Universe pushes all scales to lower energies. Thus, 
cutoffs can be considered as time-dependent and correlated with the evolution of the model. This 
induces more complications in interpretation of results, for instance whether inflation is IR stable 
and long range quantum correlations are suppressed~\cite{desitterinstab}-\cite{desitterir}. 
In de Sitter space the symmetry of space allows to write time-dependence of cutoffs as a 
factor~\cite{rgdesitter} and dependence of quantities on the cutoff can be studied in the same way 
as in Minkowski space. But in a general FLRW geometry, even in homogeneous case, such a factorization 
does not occur~\cite{houricond}. Other choices of regulator, for instance explicit dependence of 
renormalization scale to expansion factor~\cite{renormadiab0}, that is replacement of $\mu_0$ 
with $\mu = a (\eta) \mu_0$, are also suggested. However, they induce non-trivial effects at 
IR limit and only in De Sitter space the IR limit can be followed 
analytically~\cite{desitterstable,desittersymmbreak,desitterinstab,desitterinstab1}.

\subsection {Effective energy-momentum tensor and metric evolution} \label{sec:effenermom}
In semi-classical approach to gravity the effective action (\ref{effactionconst}) can be 
used~\cite{qmcurve,heatkernel} to define an effective energy-momentum tensor $T^{\mu\nu}_{eff}$, which 
is then used to evolve metric according to Einstein equations or alternatively a modified gravity 
model~\cite{modgrrev}. Here we only consider Einstein gravity\footnote{It is 
shown~\cite{qftcurvreno,qmcurve} that for renormalizing energy momentum tensor one has to add terms 
proportional to $R^2$ and $R_{\mu\nu\rho\sigma}R^{\mu\nu\rho\sigma}$ to gravitation Lagrangian. However, in 
Einstein frame these terms can be transferred to matter side and perturbatively included in 
renormalized effective energy-momentum tensor.}:
\be
G_{\mu\nu}(x) \equiv 8\pi \gm ~ T^{\mu\nu}_{eff}, \quad \quad T^{\mu\nu}_{eff} \equiv 
\langle \hat{T}^{\mu\nu} (x) \rangle_R = \frac{2}{\sqrt {-g}} 
\biggl (\frac{\partial \Gamma_R}{\partial g_{\mu\nu} (x)} \biggr ) \label{effenermom}
\ee
where $G_{\mu\nu} \equiv R_{\mu\nu} - 1/2 g_{\mu\nu} R$ is the Einstein tensor and the index $R$ means that 
for this calculation we use the renormalized effective action. From now on we drop this index where 
this does not induce any confusion. We remind that effective energy-momentum tensor $T^{\mu\nu}_{eff}$ 
is a classical quantity and as such it must be finite, if the underlying quantum theory is physically 
meaningful. Thus, no additional regularization or renormalization condition should be imposed on it. 
By contrast, the exact expression for $\hat{T}_R^{\mu\nu} (x)$ with respect to fields of the model is 
unknown and its bare version may include singularities. Assumption of energy-momentum tensor as a 
classical effective quantity is in strict contrast to usual approach, in which classical Lagrangian 
is used to define a quantum energy-momentum operator $\hat{T}^{\mu\nu} (x)$. This field has usually a 
quartic divergence and must be renormalized. By contrast, in the semi-classical approach 
(\ref{effenermom}), once quantities in the effective Lagrangian are renormalized, derived quantities 
such as $T^{\mu\nu}_{eff}$ are finite. However, initial conditions for renormalization defined in 
(\ref{renormphi}) and (\ref{renorm34}) do not fix the wave-function normalization. In 
Sec. \ref{sec:initcond} we show that the initial value of $T^{\mu\nu}_{eff}$, which is necessary for 
solving Einstein equations, fixes the wave-function renormalization and the ensemble of condensate, 
propagators, and metric evolution equations can be solved in a consistent manner.

Using (\ref{effactionconst}) the energy-momentum tensor is described as\footnote{The consistency 
of in-in formalism imposes the limit condition $\varphi^+ = \varphi^-$ at the spacetime point in 
which the expectation value of an operator depending on a single spacetime point is 
calculated~\cite{heatkernel}. The reason is similar to the case of metric, because like the latter 
$\varphi$ is a classical field.}:
\be
T^{\mu\nu}_{eff}(x)  = \frac{2}{\sqrt {-g}} \biggl \{ \frac{\partial S (\varphi)}{\partial 
g_{\mu\nu} (x)} + \frac{i}{2} \sum_{i = \Phi, X, A} \biggl [\tr \biggl (\frac{\partial \ln G_i^{-1}}{\partial 
g_{\mu\nu} (x)} \biggr ) + \tr \biggl (\frac{\partial \mathcal {G}_i^{-1} G_i}{\partial g_{\mu\nu} (x)} 
\biggr ) \biggr ] + \frac{\partial \Gamma_2}{\partial g_{\mu\nu} (x)} \biggr \} \label{tmunu}
\ee
The first term in (\ref{tmunu}) is the energy-momentum tensor $T^{\mu\nu}_{cl}(\varphi)$ of the classical 
condensate field $\varphi$:
\be
T^{\mu\nu}_{cl}(\varphi) \equiv \frac{2}{\sqrt {-g}} \frac{\partial S (\varphi)}{\partial 
g_{\mu\nu} (x)} = \partial^\mu \varphi \partial^\nu \varphi + g^{\mu\nu} V_{eff}(\varphi) - 
\frac{1}{2} g^{\mu\nu} g^{\rho\sigma} \partial_\rho \varphi \partial_\sigma \varphi \label{effenermomcond}
\ee
where $V_{eff}$ is the effective interaction potential of condensate in which the bare mass $m$ is 
replaced by quantum corrected mass $M(x)$. Other terms in (\ref{tmunu}) can be calculated separately 
as the followings (for the sake of notation simplicity we drop species index):
\bea
\frac{i}{\sqrt {-g}} \biggl (\tr \frac{\partial \ln G^{-1}}{\partial g_{\mu\nu} (x)} \biggr ) & = &
\frac{i}{\sqrt {-g}}~\frac{\partial}{\partial g_{\mu\nu} (x)}~\int d^4x \sqrt {-g (x)} 
~\int d^4y \sqrt {-g (y)} \ln G^{-1}(x,y) \delta^4 (x,y) \nonumber \\
& = & -\frac{i}{2} g_{\mu\nu}(x) tr \ln G^{-1}  \label{enermom1}
\eea
where we used the equality $\partial \sqrt {-g} / \partial g_{\mu\nu} = -g^{\mu\nu} \sqrt {-g}/2$. We 
notice that the l.h.s. of (\ref{enermom1}) contributes to Einstein equation as a cosmological 
constant and its value depends on the normalization of wave function, which we discuss in 
Sec. \ref {sec:wfrenorm}. We drop this term from $T_{eff}^{\mu\nu}$ because we show later that it can be 
included in the wave function renormalization of fields.

The next term in (\ref{effenermomcond}) can be expanded as:
\bea
\frac{i}{\sqrt {-g}} \tr \biggl (\frac{\partial \mathcal {G}^{-1} G}{\partial g_{\mu\nu} (x)} 
\biggr ) & = & \frac{-1}{\sqrt {-g}} \frac{\partial}{\partial g_{\mu\nu} (x)} \int d^4x' \sqrt {-g(x')} 
\int d^4y' \sqrt {-g(y')} \biggl [D^\rho D_\rho^{x'} + M^2 (x') \biggr ] \delta^4 (x',y') G(x',y') 
\nonumber \\
& = & \frac{-1}{\sqrt {-g (x)}} \frac{\partial}{\partial g_{\mu\nu} (x)} \int d^4x' \sqrt {-g(x')} 
\biggl [D^\rho D_\rho^{x'} + M^2 (x') \biggr ] G(x',y'=x') \nonumber \\
\label{enermom2mid}
\eea
where we have used the definition of $\mathcal {G}^{-1}$ in (\ref{freepropag}). As expected, if 
non-local 2PI quantum corrections are neglected, $G^{-1} \rightarrow \mathcal {G}^{-1}$, the integrand 
in the second line of (\ref{enermom2mid}) becomes $\delta^4 (x',y'= x')$, and the integral becomes 
a constant, which can be added to vacuum/wave function renormalization. 

Using:
\be
\biggl [\frac{\partial}{\partial g_{\mu\nu}}, D_\rho \biggr ] = 
\biggl [\frac{\partial}{\partial g_{\mu\nu}}, D^\rho \biggr ] = 0 \label{metricvarcommut}
\ee
the functional derivative in the second line of (\ref{enermom2mid}) is determined as:
\be
\frac{i}{\sqrt {-g}} \tr \biggl (\frac{\partial \mathcal {G}^{-1} G}{\partial g_{\mu\nu} (x)} 
\biggr ) = \frac{1}{2} \biggl [g_{\mu\nu} \biggl ( D^\rho D_\rho + M^2 \biggr ) G (x, x) - 
D_\mu D_\nu G (x,x) - D_\nu D_\mu G (x,x) \biggr ] \label {enermom2mid1}
\ee
The last term of (\ref{tmunu}) is the contribution of 2PI in the energy-momentum tensor and is model 
dependent. It is determined from derivatives of diagrams in Fig. \ref{fig:gammaeffdiag}, and up to 
$\lambda^2$ and $\mg^2$ order has the following explicit expression:
\bea
\frac{2i}{\sqrt {-g}} \frac{\partial \Gamma_2}{\partial g_{\mu\nu} (x)} &=& i g_{\mu\nu} 
\biggl [(\frac{-i\lambda}{n!}) \sum_{i=0}^{[n/2]} C^n_{2i} C^{2i}_2 G^i (x,x) \varphi^{n-2i} + 
\nonumber \\
&& (\frac{-i\lambda}{n!})^2 \sum_{i=0}^{n-2} (C^n_i)^2 (n-i)! \oint d^4y \sqrt {-g (y)} 
\varphi^i (x) \varphi^i (y) G^{n-i} (x,y) \biggr ] + \nonumber \\
&& (i\mg)^2 g_{\mu\nu} \oint d^4y \sqrt {-g (y)} G_\Phi (x,y)~G_X (x,y)~G_A (x,y) + \nonumber \\
&& (i\mg)^2 g_{\mu\nu} \oint d^4y \sqrt {-g (y)} \varphi (x) \varphi (y)~G_X (x,y)~G_A (x,y) + \ldots 
\label{gamma2gmunu}
\eea
where $\oint$ means closed time path and $G^>$ and $G^<$ are used on advance and reverse time 
branches, respectively. We assume equal condensates on the two branches. Thus, 
$\varphi^- = \varphi^+$\footnote{In Schwinger closed time path formalism one extends time 
coordinate to a complex space and $t \pm i\epsilon$ present two branches with different time 
directions of a path which closes at $t \rightarrow \pm\infty$. In n-point, $n > 1$ Green's 
functions opposite time directions of field operators change the ordering of field operators on 
them. Thus, in general Green's functions with different branch indices are not equal. By contrast, 
in $n = 1$ case there is only one operator. Thus, there is no time ordering and no difference 
between branches. Another way of reasoning is by using evolution equation of condensate. 
Expectation values in this equation are not sensitive to branch index of their $\varphi$ factors. 
Thus, evolution equations for $\varphi^+$ and $\varphi^-$ are the same, and if the same initial 
conditions are applied to them, their solution will be equal.}. 

Finally, the renormalized energy-momentum tensor is be explicitly written as\footnote{We have used 
the following equalities: 
$\frac{\delta g_{\lambda a}}{\delta g_{\mu \nu}} =\delta ^{\mu}_{\lambda}\delta ^{\nu}_{a}$ and
$\frac{\delta \partial_{\kappa} g_{\lambda \alpha }}{\delta g_{\mu \nu}} = 
\partial_{\kappa}(\delta^{\mu}_{\lambda}\delta ^{\nu}_{\alpha}) = 0.$
}:
\bea
T^{\mu\nu}_{eff} & = & T^{\mu\nu}_{cl}(\varphi_R) + \frac{1}{2} \sum_{i = \Phi, X, A} 
\biggl [g^{\mu\nu} \biggl (g^{\rho\sigma} D_\rho D_\sigma +  M_i^2(x) \biggr ) G^F_{Ri} (x, x) - 
\biggl (D^\mu D^\nu G^F_{Ri} (x,x) + D^\nu D^\mu G^F_{Ri} (x,x) \biggr ) \biggr ] + \nonumber \\
&& \frac{2i}{\sqrt {-g}} \frac{\partial \Gamma_2 [G_B]}{\partial g_{\mu\nu} (x)} 
\label{effenermomdet}
\eea

To get a physical insight into the terms in (\ref{effenermomdet}) we write $T_{eff}^{\mu\nu}$ as a fluid. 
The energy-momentum tensor of a classical fluid is defined as:
\be
T^{\mu\nu} = (\rho + p) u^\mu u^\nu - g^{\mu\nu} p + \Pi^{\mu\nu}, \quad \quad g_{\mu\nu} \Pi^{\mu\nu} 
\equiv 0, \quad \quad u_\mu u_\nu \Pi^{\mu\nu} \equiv 0, \quad \quad u^\mu u_\mu \equiv 1 
\label{tmunukinetic}
\ee 
It is straightforward to obtain following relations for Lorentz invariant density $\rho$, pressure 
$P$ and for shear tensor $\Pi^{\mu\nu}$:
\be
\rho = u_\mu u_\nu T^{\mu\nu} \quad \quad T \equiv g_{\mu\nu} T^{\mu\nu} = \rho -3p \label{rhop}
\ee
The unit vector $u^\mu$ is arbitrary. It defines the equal-time 3D surfaces and the only condition 
it must satisfy is $u_\mu u^\mu = 1$. In kinetic theory it is conventionally chosen in the direction 
of the movement of the fluid.

Definitions (\ref{tmunukinetic}) and (\ref{rhop}) leads to the following expressions for fluid 
description of a classical scalar field with potential $V$:
\be
\rho^{(cl)}_\varphi = \frac{1}{2}\partial_\mu \varphi \partial^\mu \varphi + V (\varphi), \quad \quad 
p^{(cl)}_\varphi = \frac{1}{2}\partial_\mu \varphi \partial^\mu \varphi - V (\varphi), \quad \quad 
\Pi^{(cl)}_\varphi = 0. \label{udef}
\ee
After decomposing the effective energy-momentum tensor (\ref{effenermomdet}) as a fluid we find 
$\rho$, $p$ and $P^{\mu\nu}$ as the followings:
\bea
\rho & = & \rho^{(cl)}_\varphi + \sum_{i = \Phi, X, A} \frac{1}{2} \biggl [(g^{\rho\sigma} D_\rho D_\sigma + 
M_i^2 (x)) G^F_i (x,x) - (u^\rho u^\sigma D_\rho D_\sigma + u^\sigma u^\rho D_\sigma D_\rho) G^F_i (x,x) 
\biggr ] + \nonumber \\
&& \frac{2i}{\sqrt{-g}} u_\rho u_\sigma \frac{\partial \Gamma_2}{\partial g_{\rho\sigma}} 
\label {rhotot} \\
p & = & p^{(cl)}_\varphi + \sum_{i = \Phi, X, A} \frac{1}{2} \biggl [\frac{1}{3} (g^{\rho\sigma} - u^\rho u^\sigma)
(D_\rho D_\sigma + D_\sigma D_\rho) G^F_i (x,x) - [g^{\rho\sigma} D_\rho D_\sigma + M_i^2 (x)] 
G^F_i (x,x) \biggr ] + \nonumber \\
&& \frac{2i}{3 \sqrt{-g}} (u_\rho u_\sigma - g_{\rho\sigma}) \frac{\partial \Gamma_2}
{\partial g_{\rho\sigma}} \label {ptot} \\
\Pi^{\mu\nu} & = & \sum_{i = \Phi, X, A} \frac{1}{2} \biggl \{(D_\rho D_\sigma + D_\sigma D_\rho) G^F_i (x,x) 
\biggl [u^\mu u^\nu (\frac{4}{3} u^\rho u^\sigma - \frac{1}{3} g^{\rho\sigma}) -\frac{g^{\mu\nu}}{3} 
(u^\rho u^\sigma - g^{\rho\sigma}) - g^{\rho\mu} g^{\sigma\nu} \biggr ] \biggr \} + \nonumber \\
&& \frac{2i}{\sqrt{-g}} \biggl \{\frac{\partial \Gamma_2}{\partial g_{\mu\nu}} - 
\biggl [u^\mu u^\nu (\frac{4}{3} u_\rho u_\sigma - \frac{1}{3} g_{\rho\sigma}) - 
\frac{g^{\mu\nu}}{3} (u_\rho u_\sigma - g_{\rho\sigma}) \biggr ]
\frac{\partial \Gamma_2}{\partial g_{\rho\sigma}} \biggr \} \label{sheartot}
\eea
where $V = V_{eff}$ is used in (\ref{udef}) which defines $\rho^{(cl)}_\varphi$ and $p^{(cl)}_\varphi$ for 
the condensate. The terms $(g^{\rho\sigma} D_\rho D_\sigma + M_i^2 (x)) G^F_i (x,x)$ in 
(\ref{rhotot}-\ref{sheartot}) can be replaced by the r.h.s. of (\ref{evolgf}). Therefore, if 2PI 
quantum corrections are neglected, these terms would be null. As expected, the shear $\Pi^{\mu\nu}$ is 
a functional of $G_i(x,x)$ and is non-zero only when quantum corrections are taken into account. 
In (\ref{sheartot}) the terms in the curly brackets are due to 1PI and 2PI quantum corrections, 
respectively.

Despite unusual appearance of the above expressions for $\rho$ and $p$ they are consistent with 
fluid formulation when 2PI corrections are neglected. To see this, consider the case of a 
relativistic fluid, that is when $M (x) \rightarrow 0$ and the condensate $\varphi = 0$. In this 
case the contribution of different fields in (\ref{rhotot}-\ref{sheartot}) can be separated and 
application of (\ref{evolgf}) to these equations shows that $w \equiv p/\rho = 1/3$ and 
$\Pi^{\mu\nu} = 0$ for each field component with $M (x) \rightarrow 0$, as expected for a 
relativistic classical fluid of particles. If $M (x) \neq 0$ in a homogeneous universe with small 
perturbations at zero order $n^\mu = (1,0,0,0)$ and contribution of the first term in (\ref{ptot}) 
is zero and we find $p \rightarrow 0$ when quantum corrections generated by interaction between 
fields are neglected.

If we neglect 2PI terms, $u^\mu$ can be different for each component. For instance, it can be chosen 
such that space components vanish in a homogeneous universe. This choice is suitable when components 
are studied or observed separately. Alternatively, the same $u^\mu$ can be used for all components. 
It is proved that in multi-field classical models of inflation such a choice leads to adiabatic 
evolution of superhorizon modes in Newtonian gauge~\cite{adabaticinf}. We notice that due to the 
interaction between fields - more precisely the term proportional to 
$\partial \Gamma_2 / \partial g_{\rho\sigma}$ - it is not possible to define density and pressure 
separately for each species, unless we neglect 2PI corrections. 

Comparison of expressions (\ref{rhotot}) and (\ref{ptot}) with $\rho_\varphi$ and $p_\varphi$ shows that 
not all the term induced by interactions can be considered as an effective potential, which contributes  
in $\rho$ and $P$ with opposite sign. Although some of 1PI terms in $\rho$ and $p$ behave similar to 
a classical potential, others - including 2PI corrections which contain integrals and are non-local - 
do not follow the rule of a classical potential. Therefore, an {\it effective classical scalar field} 
description cannot present full quantum corrections, even if we neglect the shear - the viscosity - 
term. In addition, the contribution of species without a condensate is, as expected, a functional 
of their propagators and its expression is not similar to a simple fluid with $p ~\propto~ \rho^\alpha$. 
Thus, $T_{eff}^{\mu\nu}$ cannot be even phenomenologically described by a fluid. Of course, 
we can always consider the effective action (\ref{effactionconst}) and its associated effective 
energy-momentum tensor (\ref{effenermomdet}) as a phenomenological classical model. But, such a model 
has very little similarity with bare Lagrangian of the underlying quantum model described in 
(\ref{lagrangphi}-\ref{lagrangint}). This observation highlights difficulties and challenges 
of deducing the physics of early Universe from cosmological observations, which in a large extend 
reflect only classical gravitational effect of quantum processes. Specifically, the effect of quantum 
corrections can smear contribution of the {\it classical} $\rho_\varphi$ and $p_\varphi$, which 
reflect the structure of classical Lagrangian. Therefore, 
conclusions about underlying inflation models by comparing CMB observations with predictions of models 
treated classically or with incomplete quantum corrections should be considered premature. See also 
simulations in~\cite{infqftsimul,infinitsuperpos} which show the backreaction of quantum corrections 
and their role in the formation of spinodal instabilities in natural inflation models, even when 
only local quantum corrections are considered. Nonetheless, constraints that CMB observations impose 
on the amplitude of tensor modes generated by $\Pi^{\mu\nu}$ and measurement of the power spectrum 
properties should be considered in the selection of parameters of any candidate quantum model of the 
early Universe. See also Sec. \ref{sec:simulation} for more discussion about these issues.

\subsubsection{Fixing metric gauge}
To proceed to solving evolution equations of the model, either analytically or numerically, we must 
choose an explicit description for the metric in a given gauge. We consider a homogeneous flat 
FLRW metric for the background geometry and add to it both scalar and tensor fluctuations that 
subsequently will be truncated to linear order:
\be
ds^2 = a^2(\eta) (1 + 2 {\boldsymbol \psi}) d\eta^2 - a^2(\eta) [(1 - 2 {\boldsymbol \psi}) 
\delta_{ij} + h_{ij}] dx^i dx^j, \quad \quad dt = a d\eta \label{perturbmetric}
\ee
where $t$ and $\eta$ are comoving and conformal times, respectively. Explicit expression of connection 
for this metric is given in Appendix \ref{app:connect}. This parametrization contains one redundant 
degree of freedom and does not completely fix the gauge. Nonetheless, it has the advantage of containing 
both scalar and tensor perturbations and can be easily transformed to familiar Newtonian and conformal 
gauges. The redundant degree of freedom can be removed from final results by imposing a constraint on 
$h_{ij}$ and ${\boldsymbol \psi}$. For instance, if $h_{ij} = 0$, this metric takes the familiar form of 
Newtonian gauge for scalar perturbations when anisotropic shear is null. If $h_{ij} \propto \delta_{ij}$, 
the metric gets the general form of Newtonian gauge with two scalar potentials ${\boldsymbol \psi}$ 
and $\upphi \equiv {\boldsymbol \psi} - h/6$, where $h \equiv \delta^{ij} h_{ij}$. If in addition 
${\boldsymbol \psi} = h = 0$, the metric becomes homogeneous in conformal gauge form.

For solving evolution equations either analytically - which in the case of the model described 
here is not possible - or numerically, it is preferable to scale the condensate and propagators 
such that their evolution equations (\ref{dyneffa}), (\ref{evolgf}) and (\ref{evolgrho}) depend 
only on the second derivative with respect to conformal time $\eta$. It is straightforward to 
show that for the metric (\ref{perturbmetric}) the following scaling changes the evolution 
equations of condensate and propagators to the desired from:
\bea
&& \frac{1}{\sqrt{-g}}{\partial}_{\mu} \biggl (\sqrt{-g} g^{\mu\nu}{\partial}_{\nu} \Xi (x) \biggr ) + 
M^2(x) \Xi (x) = \text{\color{blue} [interaction and quantum corrections]} \label{wavefun} \\
&& \Xi_\chi (x) \equiv a (1 - 2{\boldsymbol \psi} + \frac{h}{4})~\Xi (x) \label{chidef} \\
&& \Xi_\chi'' - \frac{1}{1 - 2{\boldsymbol \psi} + \frac{h}{4}} \partial_i \biggl [
\biggl ((1+\frac{h}{2}) \delta^{ij} + h^{ij} \biggr ) \partial_j 
\biggl (\frac{\Xi_\chi}{1 - 2{\boldsymbol \psi} + \frac{h}{4}} \biggr )\biggr ] + \nonumber \\
&& \hspace{1.5cm} \biggl [a^2 M^2 (x) (1 + 2{\boldsymbol \psi}) - \biggl (\frac{a''}{a} 
(1 - 2{\boldsymbol \psi} - \frac{h}{4}) - 4 \frac{a'}{a} ({\boldsymbol \psi}' - \frac{h'}{8}) - 
2 ({\boldsymbol \psi}'' - \frac{h''}{8})\biggr ) \biggr ] \Xi_\chi  = \nonumber \\
&& \hspace{1.5cm} a^3 (1 - \frac{h}{4}) \text{\color{blue} [interaction and quantum corrections]} 
\label{wavefunchi}
\eea
where $\Xi$ is any of propagators or the condensate with quantum corrected mass $M(x)$. From 
now on prime means derivative with respect to conformal time $\eta$. When $\Xi$ is a propagator, 
it depends on two spacetime coordinates, but differential operators are applied only to one of them. 
Thus, in (\ref{wavefun}) the dependens on coordinates of the second point is implicit. Interaction 
and quantum correction terms in the r.h.s. of (\ref{wavefunchi}) are the same as ones in 
(\ref{wavefun}) (with respect to unscaled variable $\Xi$). The last arbitrary degree of freedom 
in metric (\ref{perturbmetric}) can be chosen to simplify (\ref{wavefunchi}) without loosing the 
generality at linear order. For instance, if we choose ${\boldsymbol \psi} = h/8$ the evolution 
equation becomes:
\bea
&& \Xi_\chi'' - \partial_i \biggl [\biggl ((1+\frac{h}{2}) \delta^{ij} + h^{ij} \biggr ) \partial_j 
\Xi_\chi \biggr ] + \biggl [a^2 M^2 (x) (1 + \frac{h}{4}) - \frac{a''}{a} \biggr ] \Xi_\chi = \nonumber \\
&& \quad \quad a^3 (1 - \frac{h}{4}) \text{\color{blue} [interaction and quantum correction terms]} 
\label{wavefunchigauge}
\eea
The presentation of scaled solution of field equation for linearized Einstein equations is for 
the sake of completeness of discussions and for future use, because in the simulations presented 
in Sec. \ref{sec:simulation} we only use a homogeneous background metric.

\section {Initial conditions} \label {sec:initcond}
To solve semi-classical Einstein equation (\ref{effenermom}) we need evolution of effective 
energy-momentum tensor $T_{eff}^{\mu\nu}$, which depends on the propagators 
$G_i (x,y), i \in \Phi,~X,~A$ and the condensate field $\varphi (x)$. Evolution of these 
quantities is governed by a system of second order differentio-integral equations needing two 
initial or boundary conditions for each equation. This is in addition to the initial state density 
which appears in the generating functional $\mathcal{Z}$, because the state of the system at 
initial time $t_{-0}$ does not give any information about that of $t_{+0}$ in Schr\"odinger or 
interaction picture or evolution of operators in Heisenberg picture. In addition, initial 
conditions for evolution equations of propagators and condensate(s) in a multi-component model 
are not independent from each others and their consistency must be respected. 

The model formulated in the previous sections is independent of the cosmological epoch to which 
it may be applicable. However, for fixing initial or boundary conditions we have to take into 
account physical conditions of the Universe at the epoch in which this model and its constituents 
are supposed to be {\it switched on}. Two epochs are of special interest: (pre)-inflation; 
and epoch of the formation of the component which may play the role of dark energy at present. 
These two eras may be the same if dark energy is a leftover of inflationary epoch, otherwise 
different conditions may be necessary for each. In the following subsections we first describe 
physically interesting initial quantum states for the model. Then, we specify initial conditions 
for solutions of evolution equations. Explicit description of constraints used for the determination 
of initial conditions and their solutions are described in Appendix \ref{app:gensolution}.
 
A word is in order about the initial conditions for bare and adiabatic vacuum Green's functions, 
because they are primary rather than derived quantities which their evolution is implemented in 
the numerical simulations. Initial conditions for these functions are arbitrary and different 
conditions are equivalent to performing a Bogoliubov transformation on creation and annihilation 
operators. Only initial conditions for renormalized quantities are physically meaningful, lead to 
observable effects, and must respect observational constraints.

\subsection{Density matrix of initial state} \label {sec:densitymatrix}
Our main purpose in studying the model (\ref{lagrange}) is to learn how the light fields $\Phi$ and 
$A$ are created from the decay of the heavy field $X$ and how they evolve to induce an accelerating 
expansion. Therefore, it is natural to assume a vacuum state for $\Phi$ and $A$ at initial time $t_0$. 
The initial state of $X$ can be more diverse. Physically motivated cases are Gaussian, double Gaussian, 
and free thermal states. The only difference between the first and the second case is the choice 
of cosmological rest frame. The last case is motivated by hypothesis of a thermal early Universe 
and the assumption that interaction of $X$ with other fields is switched on at $t_0$. As we 
discussed earlier, both a Gaussian and a free thermal states are Gaussian~\cite{oneparticledens,inin}. 
We remind that as it is assumed that interactions are switched on at $t_0$, $X$ is initially a free 
field. Consequently, the contribution of its density matrix can be included in 1-point and 2-point 
correlations and no additional Feynman diagram is needed. 

Simulations discussed in Sec. \ref{sec:simulation} are performed in several steps to prevent 
exponential increase of numerical errors. The initial state of $\Phi$ and $A$ in intermediate 
simulations is not any more vacuum and due to interactions the initial state of the system may be 
non-Gaussian. However, considering the large mass and small coupling of $X$ and $A$, a Gaussian or free 
thermal initial states for both seem a good approximation. In this case, their density functional 
$F$ do not change the effective action. However, a non-zero condensate component needs special care. 
For this reason in the next subsection we calculate elements of matrix density for a condensate state. 

\subsubsection {Density matrix of coherent states} \label{sec:condmatrix}
Following the decomposition (\ref{densitymatrixdef}), the state of a scalar 
can be factorized to $|\Psi\rangle = |\Psi_C\rangle \otimes |\Psi_{NC}\rangle$ where $|\Psi_C\rangle$ 
is a condensate state and $|\Psi_{NC}\rangle$ is non-condensate consisting of quasi-free 
particles\footnote{This decomposition is virtual in the sense that condensate and non-condensate 
parts may be inseparable and entangled.}. There is no general description for a condensate state, 
but special cases are known. A physically interesting example of known condensate states, which has 
been also realized in laboratory~\cite{bec}, is a Glauber coherent state\cite{coherglauber}. 
See also~\cite{coherstaterev} for a review of other coherent states and their applications. 
The Glauber coherent state is defined as an eigen state of annihilation operator:
\bea
a_k |\Psi_C\rangle & = & C_k |\Psi_C\rangle \label{coheredef} \\ 
|\Psi_C\rangle & \equiv & e^{-|C_k|^2 / 2} e^{C a_k^{\dagger}} |0\rangle = e^{-|C|^2 / 2} \sum_{i=0}^\infty 
\frac {C_k^i}{i!}(a_k^{\dagger})^i |0\rangle
\label{condwavef}
\eea
It can be generalized to a superposition of condensates of different modes\footnote{In equation 
(\ref{condwavefsuperpos}-\ref{genpathintcoher}) a $\sqrt{-g}$ factor is included in $A_k$. See 
Appendix \ref{app:initparticle} for details.}:
\be
|\Psi_{GC}\rangle \equiv \int d^3k A_k e^{C_k a_k^{\dagger}} |0\rangle = \int d^3k A_k 
\sum_{i=0}^\infty \frac {C_k^i}{i!}(a_k^{\dagger})^i |0\rangle \label{condwavefsuperpos}
\ee
If the support of mode $k$ is discrete, the integral in (\ref{condwavefsuperpos}) 
is replaced by a sum. A condensate may be also a combination of condensates of different 
fields or modes:
\be
|\Psi_{mGC}\rangle \equiv \prod_i \int d^3k_i A_{k_i} e^{C_{k_i} a_{k_i}^{\dagger}} |0\rangle = 
\prod_i \int d^3k_i A_{k_i} \sum_{j=0}^\infty \frac {C_{k_i}^j}{j!}(a_{k_i}^{\dagger})^j |0\rangle 
\label{condwavefmulti}
\ee
where $i$ runs over the set of fields. 

It is proved that if a density operator commutes with number operator, its elements over field eigen 
states have a Gaussian form~\cite{oneparticledens}. However, coherent states are neither eigen states 
of field operator $\hat{\Phi}$ nor number operator $\hat {N}$. In fact they are explicitly a 
superposition of states with any number of particles. Elements of density matrix operator $\varrho_{GC}$ 
of the coherent state $|\Psi_{GC}\rangle$ can be expanded as:
\be
\langle \Phi'| \varrho_{GC} | \Phi \rangle = \langle \Phi'| \int d^3k' A^*_{k'} e^{C^*_{k'} u^{*-1}_{k'} 
\int d^3 y e^{-ik'y} \hat \phi^+ (y)} |0 \rangle \langle 0 | \int d^3k A_k e^{C_k u^{-1}_k 
\int d^3 x e^{ikx} \hat \phi^- (x)} | \Phi \rangle \label {densmatrixgc}
\ee
Because $\hat \phi^- (x) |0 \rangle = 0$ and $\langle 0 | \hat \phi^+ (x) = 0$, we can replace 
$\hat \phi^-$ and $\hat \phi^+$ in (\ref{densmatrixgc}) with $\hat \phi$ and apply a normal 
ordering operator $::$ to each factor. Then, using Wick theorem 
$:\hat{A}\hat{B}:~\equiv~\hat{A}\hat{B} - \langle 0 | \hat{A}\hat{B} | 0 \rangle I$, we find:
\bea
\langle \Phi'| \varrho_{GC} | \Phi \rangle & = & \phi_0 \phi^{'*}_0 \int d^3k' A^*_{k'} 
e^{C^*_{k'} u^{*-1}_{k'} \int d^3 y e^{-ik'y} \phi' (y)} \int d^3k A_k e^{C_k u^{-1}_k \int d^3 x e^{ikx} \phi (x)} - 
\nonumber \\
&& \langle 0| \int d^3k' A^*_{k'} e^{C^*_{k'} u^{*-1}_{k'} \int d^3 y e^{-ik'y} \hat \phi (y)} |0 \rangle 
\langle 0 | \int d^3k A_k e^{C_k u^{-1}_k \int d^3 x e^{ikx} \hat \phi (x)} | 0 \rangle \nonumber \\
& = & \phi_0 \phi^{'*}_0 \int d^3k'~e^{\int d^3 y F^*_{k'} e^{ik'y} \hat \phi{'*} (y)} 
\int d^3k~e^{\int d^3 x F_k e^{ikx} \hat \phi (x)} - \int d^3k |A_k|^2 \label {densmatrixgcexp} \\
F_k & \equiv & C_k u^{-1}_k \ln A_k \label {fdef}
\eea
where $\phi_0$ ($\phi'_0$) is the zero mode of the decomposition of $|\Phi\rangle~(|\Phi'\rangle)$ 
to $n$-particle states $|n\rangle~\forall~n \in \mathbb{Z}$ and $\phi_k$ is the 3D Fourier 
transform of configuration field $\phi$. The last term in (\ref{densmatrixgcexp}) is the contribution 
of vacuum, that is when $C_k \rightarrow 0~\forall k$. It is a constant and can be included in the 
normalization of wave function, which we fix later in this section. 

Insertion of (\ref{densmatrixgcexp}) in (\ref{genpathint}) gives the generating functional for a system 
initially in state $|\Psi_{GC}\rangle$:
\bea
\mathcal{Z} (J_a,K_{ab};\varrho) & \equiv & e^{i W[J_a,K_{ab}]} = \int \mathcal{D} \Phi^a \mathcal{D} 
\Phi^b \exp \biggl [i S(\Phi^a) + \int d^4x \sqrt {-g} J_a(x) \Phi^a(x) + \nonumber \\ 
&& \frac{1}{2} \int d^4x d^4y \sqrt {-g (x)} \sqrt {-g (y)} \Phi^a (x) K_{ab} (x,y) \Phi^b (y) 
\biggr] \times \nonumber \\
&& \int d^3k' d^3k \biggl [\Phi^{a*}_0\Phi^b_0 \exp \biggl (\int d^3 y F^*_{k'} e^{-ik'y} \Phi^a (y) + 
\int d^3 x F_k e^{ikx} \Phi^b (x) \biggr ) - A^*_{k'}A_k \biggr] \label{genpathintcoher}
\eea
where branch indices $a,b \in \{+,-\}$. $\Phi_0$ and terms in the last line of (\ref{genpathintcoher}) 
are evaluated at the initial time $t_0$. Comparing the contribution of the initial condition with 
the definition of $F[\Phi]$ in (\ref{rhomatrix}) and (\ref{fexpan}), it is clear that only $\alpha_0$ 
and $\alpha_1$ are non-zero. They can be included in the normalization factor and $J$ current, and 
do not induce new diagrams to the effective Lagrangian. Nonetheless, (\ref{genpathintcoher}) 
explicitly shows that as the system is initially in a superposition state, the classical effective 
Lagrangian is a {\it quantum expectation} obtained by summing over all possible states weighed by 
their amplitude. Extension of these results to $|\Psi_{mGC}\rangle$ is straightforward.

\subsection {Initial conditions for solutions of evolution equations} \label {sec:initevol}
In the study of inflation and dark energy, specially through numerical simulations, it is more 
convenient to fix initial conditions, that is the value and variation rate of condensates and 
propagators on the initial equal-time 3-surface rather than boundary conditions at initial and 
final times. Initial conditions for inflation are extensively discussed in the literature, 
see e.g.~\cite{bunchdavis,infinitcond,infinitcond0} and~\cite{infinitcond1} (for review). As 
in this toy model there is not essential difference between (pre)-inflation and dark energy era, 
the same type of initial conditions can be used for both.

We use a Dirichlet-Neumann boundary condition~\cite{infinitcond0,infrenorm,houricond}:
\be
n^\mu \partial_\mu {\mathcal U} = {\mathcal K} {\mathcal U}, \quad \quad g_{\mu\nu} n^\mu n^\nu = 1 
\label{dirichletneumann}
\ee
where $n^\mu$ is a unit vector normal to the initial spacelike 3-surface and $\mathcal{U}$ is a 
general solution of the evolution equation. Assuming a homogeneous, isotropic and spacelike initial 
surface, $n^\mu = (a^{-1}, 0, 0, 0)$ in conformal coordinates. We use boundary conditions similar to 
(\ref{dirichletneumann}) for both condensate and propagators. 

Although $\mathcal {K}$ is arbitrary, it must be consistent with the geometry near initial 
boundary to provide a smooth transition from initial 3-surface~\cite{infinitcond0,infinitcond1}. 
For instance, if we want that for $t \rightarrow t_0$ modes approach to those of a free scalar 
field in flat Minkowski, $\mathcal {K}$ should have a form similar to modes in a static flat space:
\be
\mathcal {K} = i \sqrt{\frac{k^2}{a^2(t_0)} + M^2} = i \omega_k\label{neumanncoeff}
\ee
where $M$ is the effective mass. In this choice (\ref{dirichletneumann}) is a condition on the flow of 
energy from initial surface in Minkowski and de Sitter geometry and is called Bunch-Davis initial 
condition.

The renormalized anti-symmetric propagator must satisfy the condition imposed by field 
quantization~\cite{2picurved}:
\be
\partial_0 G_R^\rho (\vec{x},,t,\vec{y},t) = \frac{i\delta^{(3)}(x-y)}{g^{00} \sqrt{-g}} 
\label{quantumcond}
\ee
At initial time this constraint can be written for mode functions in synchronous gauge as:
\be
\biggl [\um_k^{\rho'} (\eta_0) \um^{\rho*}_k (\eta_0) - \um^{\rho}_k (\eta_0) \um^{\rho*'}_k (\eta_0) 
\biggr ]_R = \frac {-i}{a^2(\eta_0)} \label{derivcond}
\ee
where $\um_k^{\rho'}$ is the derivative of solution $\um^\rho_k$ of the free field equation 
(\ref{homoevolk}) with respect to conformal time $\eta$ at $\eta = \eta_0$\footnote{Here $\um_k$ is 
assumed to be a solution of $\Xi$ rather than its scaled version $\Xi_\chi$}. The bracket and 
index $R$ means that this constraint is applied after subtraction of adiabatic expansion of vacuum, 
which makes the propagator finite. The contribution of fields in the energy-momentum tensor 
imposes a constraint on $G^F_R$, see Sec. \ref{sec:geowave}. It can be used as the second condition for 
fixing integration constants for these propagators.

\subsubsection{Initial conditions for propagators} \label {sec:initpropag}
In what concerns the fields of the toy model, the initial conditions should reflect the absence 
of $A$ and $\Phi$ particles and $\varphi$ condensate at time $t_{0-}$ and their production by decay 
of $X$ at $t_{0+}$. Due to this interaction an initial condition of type (\ref{dirichletneumann}) 
must depend on the solutions of field equations for all the constituent and the constant 
$\mathcal {K}$ includes production/decay rate of one species from/to another. Therefore, 
a boundary condition for the derivative of propagators similar to (\ref{dirichletneumann}) 
which reflects these properties can be defined as the following:
\be
n^\mu \partial_\mu G^F_i = \sum_{j \in \{X,A,\Phi\}} {\mathcal K}_{ij} G^F_j \label{dirichletneumannij}
\ee
In general ${\mathcal K}_{ij}$ depends on $\vec {x}$ and $\vec {y}$, but if we assume that 
interactions are switched on at time $\eta_{0+}$, initially propagators are free and both $G_i$'s 
and ${\mathcal K}_{ij}$ depend only on $\vec {x} - \vec {y}$. In addition, interpretation of 
propagators as expectation value of particle number means that for the model discussed here 
there is a relation between $G_i$'s and ${\mathcal K}_{ij}$ modes in the Fourier space. Notably, in 
interaction model (a) in (\ref{lagrangint}) momentums of decay remnants are determined uniquely
from momentum of decaying particle. In this case, when (\ref{dirichletneumannij}) is written 
in momentum space, convolutions (in momentum space) in the r.h.s. become simple multiplications:
\bea
{G'}^F_i (k) & = & \sum_{j \in \{X,A,\Phi\}} \int d^3p ~ {\mathcal K}_{ij} (\vec {k} - \vec {p}) 
G^F_j (\vec{p}) = \sum_{j \in \{X,A,\Phi\}} {\mathcal K}_{ij} (\vec {k}) G^F_j (\vec{p} (k)) 
\label{dirichletneumannijmom}\\
{\mathcal K}_{ij} (k) & = & {\mathcal K}_i^{vac} (k) \delta_{ij} + \Gamma_{ij} (p (k)) \label {propbound}
\eea
where we have assumed $n^\mu = (a^{-1}, 0, 0, 0)$ in homogeneous conformal coordinates. The 
coefficient ${K}_i^{vac}$ presents the choice of boundary condition for the vacuum. Here we only 
consider Bunch-Davis vacuum defined in (\ref{neumanncoeff}). The constant $\Gamma_{ij}$ is the decay 
width of $j$ to $i$ if $\Gamma_{ij} < 0$, and production rate of $i$ from 
$j$ if $\Gamma_{ij} > 0$~\cite{boltzlepto}. The function $p (k)$ is determined from kinematic of 
decay/production of $i$ to/from $j$. Under the assumption of initial vacuum state for $\Phi$ and $A$, 
only $\Gamma_{X\Phi}$ and $\Gamma_{XA}$ contribute to initial conditions. For model (a) in 
(\ref{lagrangint}) $\Gamma_{X\Phi} = \Gamma_{XA} = \Gamma_X$, where $\Gamma_X$ is the total decay width 
of $X$ particles. We can use perturbative in-out formalism to determine decay rates at initial time - 
even in presence of a condensate - because in the infinitesimal time interval of $[t_0,~t_0 ÷ \epsilon$ 
where these rates are needed the system can be considered as quasi-static. This setup and its purpose 
is very different from effective dissipation rates calculated e.g. in~\cite{scalereffpot}, which 
are time dependent and their purpose is to present 2PI quantum corrections in an effective evolution 
equations for condensates and cosmological matter fluctuations.

Alternatively we can use the following equation as an initial condition:
\be
{G'}^F_i (k) = \sum_{j \in \{X,A,\Phi\}} \int d^3p ~ {\mathcal K}_{ij} (\vec {k} - \vec {p}) 
G^F_j (\vec{p}) = {\mathcal K}_i^{vac} (k) G^F_i (k) + \Upsilon_i (k) \quad i \in \{X,A,\Phi\}
\label{dirichletneumannijsrc}
\ee
where $\Upsilon (k)$ is an external source which must be decided from properties of the model. 
For instance, in the model (a) if the self-coupling of the light scalar field $\Phi$ is much 
larger than its coupling to $X$, we can assume that $\Phi$ particles produced from decay of $X$ in 
the interval $(t_{-0}, t_{+0})$ interact with each other and at $t_{+0})$ all memory about their 
production is lost and particles are distributed according to distribution $\Upsilon_i (k)$, which 
its normalization is determined such that the total energy density of $\Phi$ is equal to the energy 
transferred to this field from decay of $X$ (we neglect the backreaction). This choice of boundary 
condition is specially interesting for numerical simulations because it allows to study all the 
fields in the model in the same range of momentum space. By contrast, 
in (\ref{dirichletneumannijmom}) the range of $k$ and $p$ for modes with largest amplitudes can 
be very different if there are large mass gaps between particles. The disadvantage of 
(\ref{dirichletneumannijsrc}) is that it adds a new arbitrary distribution, namely $\Upsilon_i (k)$ 
to the model. Nonetheless, the assumption of the loss of memory due to many scattering means that 
$\Upsilon_i (k)$ can be well approximated by a Gaussian distribution with zero mean value in the 
frame where initial distribution of $X$ particles has a zero mean value. Its standard deviation, 
however, remains arbitrary, and apriori can be larger than the standard deviation of momentum 
distribution of $X$ particles.

A general solution of field equations can be written as:
\be
\um_k = a^{-1} (c_k U_k + d_k V_k) \label{soldecomp}
\ee
where $U_k$ and $V_k$ are two independent solutions for mode $k$. We have divided the r.h.s. 
of (\ref{soldecomp}) by $a (\eta)$ because solutions $U_k$ and $V_k$ for free fields are usually 
obtained for scaled function $\Xi_\chi \equiv a \Xi$ where $\Xi$ is any of scalar fields of the 
model. Solutions of field equation for some spacial geometries and WKB approximation for general 
case are given in Appendix \ref{app:solution}. If there is initial correlation/entanglement 
between fields, it is implicit in the matrix elements of the state (or equivalently density matrix) 
defined in Appendix \ref{app:initparticle}.

From explicit expression of free propagators with respect to independent solutions given in 
Appendix \ref{app:propag} it is clear that only the difference between arguments of complex 
constants $c_k$ and $d_k$ is observable. In coordinate space this means that free propagators 
depend on $\vec{x}-\vec{y}$ rather than each coordinate separately, and only 3 initial conditions 
(for real rather than complex quantities) are enough to fix integration constants. Therefore 
equations (\ref{derivcond})\footnote{Equation (\ref{derivcond}) is counted as one constraint 
because both sides of the equation are pure imaginary, see Appendix \ref{app:gensolution}.} and 
(\ref{dirichletneumannij}) can fully fix all the propagators and no additional constraint for 
defining $c_k$ and $d_k$ is necessary. However, propagators depend on the normalization of initial 
quantum state $N$ in (\ref{1pdist}), or equivalently the initial momentum distribution discussed 
in the next section. It will be fixed by initial conditions imposed on $T^{\mu\nu}_{eff}$ in 
Sec. \ref{sec:geowave}.
 
Finally, a question must be addressed here: how to calculate decay and scattering rates 
$\mathcal {K}_{ij}$ consistently ? To determine $\mathcal {K}_{ij}$ with respect to renormalized 
masses and couplings we need renormalized propagators and condensate, which in turn need the 
solutions of evolution equations. Thus, the problem seems circular. This issue is not very 
important for the toy model studied here and its simulations, because there is no observational 
constraint for parameters and they can be chosen more or less arbitrarily. They only have to be 
in the physically motivated range and lead to a reasonable cosmological outcome. However, for 
academic interest it is important to know how one would have to proceed, if observed information 
about decay width, scattering cross-section, and masses were available. The interdependence of 
$\mathcal {K}_{ij}$, couplings and masses can be broken if we determine decay width and scattering 
cross sections at perturbative tree order and assume that initial conditions of renormalization 
(\ref{renormphi}-\ref{renormax}) are defined such that $\mathcal {K}_{ij}$ corresponds to observed 
values at renormalization scale. For model (a) in (\ref{lagrangint}) $\Gamma_X$ is calculated 
in~\cite{houridmquin} and we do not repeat it here. 

\subsubsection{Initial distribution} \label{sec:initstate}
In addition to the contribution of density matrix in the generating functional (\ref{genpathint}) 
the density matrix elements $|\Psi_{k_1 k_2 \ldots k_n}|^2$ (for pure states) are needed for determination 
of propagators, see (\ref{propst}-\ref{propless}). As we assume that for both inflation and dark 
energy, no $\Phi$ or $A$ particle exists at initial time, their contribution in the initial state 
$|\Psi\rangle$ is simply vacuum. Thus, only the initial state of $X$ particles is 
non-trivial\footnote{For intermediate states we use numerical value of propagators from previous 
simulation and an analytical expression is not needed.}. In absence of self-interaction for $X$ 
field in the model (\ref{lagrange}) a free initial state without entanglement is justified and 
the many-particle wave-function $|\Psi_{k_1 k_2 \ldots k_n}|^2$ can be factorized to 1-particle functions. 
Moreover, after taking a Wigner transformation, $|\Psi_{k_1 k_2 \ldots k_n}|^2$ can be replaced by a 
momentum distribution $f_X (k, \bar{x},t_0)$ evaluated at the average coordinate $\bar{x}$ of 
$X$ particles~\cite{oneparticledens}, 

1-particle distribution functions of free thermal and single or double Gaussian states discussed in 
Sec. \ref{sec:densitymatrix} are:
\be
f_X (k, \bar{x},t_0) = 
\begin{cases}
\frac{N}{e^{\beta_\mu k^\mu} - 1} & \text{thermal} \\
N e^{- \frac{|\vec{k} - \vec{k}_0|^2}{2\sigma^2}} & \text{Gaussian}
\end{cases}
\label{oneparticledist}
\ee
where $\sigma$ is the standard deviation of the Gaussian; $\beta_\mu$ is proportional to Killing 
vector and can be interpreted as covariant extension of inverse temperature~\cite{covarthermal}
\footnote{More precisely, this a covariant extension of Bose-Einstein distribution. At high 
temperatures $[\beta_\mu\beta^\mu]^{1/2} \rightarrow 0$, and the distribution approaches a 
Maxwell-J\"uttner distribution, see e.g.~\cite{covarthermal0} for a review. Note that this 
distribution is written in the local Minkowski coordinate. As we use it only at the initial time, 
the value of $a(t_0)$ is not an observable and without loss generality we consider $a(t_0) = 1$.} 
In the Gaussian distribution $\vec{k}_0$ is a constant 3-momentum presenting the momentum of the 
center of mass of $X$ particles with respect to an arbitrary reference frame. The factor $N$ is a 
normalization constant. If at $t_0$ the Universe is homogeneous, the distribution $f$ will not 
depend on $\bar{x}$. If simulations present the era after inflation and $m_X \gtrsim 300$~TeV, the 
distribution of $X$ particles could not be in thermal equilibrium with other 
species~\cite{hdmmassunitarity}. This is not an issue for our toy model because at the initial time 
there is no other species. Nonetheless, we preferred to use a Gaussian distribution in our simulations.

Another physically motivated state is a totally entangled state with all particles in one or a few 
momentum states. This is reminiscent to a Bose-Einstein condensate, but is not a Glauber condensate. 
If in addition $X$ has internal quantum numbers (symmetries), other type of entanglement would be 
possible. For instance, in~\cite{infinitsuperpos} an entanglement between different fields of 
a multi-field inflation model is considered. It generates a coherent oscillation between scalar 
fields of the model, which may leave an observable signature on matter fluctuations.

An issue which must be clarified here is the relation between comoving reference frame 
today - defined as the rest frame of far quasars - and the reference frame in which 
$f (k, \bar{x}, t_0)$ and other quantities of the model are defined. Although Lorentz invariance 
assures that final results do not depend on the selection of reference frame, in a multi-component 
system there can be frames in which the formulation of the model is easier, specially when 
approximations are involved. Moreover, when theoretical predictions are compared with observations 
the issue of using the same reference frame for both becomes crucial. If we assume that 
$X$ particles decay significantly or totally before epochs accessible to observations, 
today's comoving frame cannot be directly associated to their rest frame. In this case, it would be 
more convenient to consider the rest frame of $\varphi$, the condensate of $\Phi$, as the reference 
frame. When $\varphi$ is identified with classical inflaton field, reheating at the end of 
inflation is homogeneous in this frame and presumably $\varphi$ frame coincides with the comoving 
frame today. In addition, if the model studied here is supposed to be a prototype for 
formation of a quintessence field during or after reheating, the observed homogeneity of dark 
energy with respect to matter and radiation, which fluctuate, encourages the use of its rest frame 
as reference.

\subsection{Initial condition for geometry} \label{sec:geowave}
The simplest choice for initial geometry is a homogeneous FLRW metric, that is 
$\upphi = {\boldsymbol \psi} = h_{ij} = 0$ in metric (\ref{perturbmetric}). Thus, the 
metric depends only on the expansion factor $a(\eta)$, which its value at initial time is irrelevant 
and without loss of generality can be considered to be $a (\eta_0) = 1$. The value of Hubble 
constant $H \equiv \dot{a}/a$ (or equivalently $\mathcal{H} \equiv a H = a'/a$) must be chosen 
based on the physics of inflation or reheating after inflation, respectively for studying 
condensation of inflaton or dark energy from decay of the heavy particle $X$. 


In a homogeneous FLRW metric only diagonal components of Einstein tensor are nonzero
\footnote{
We assume that equal-time surfaces are defined such that $T^{0i} = T^{ij}|_{i \neq j} = 0$}: 
\bea
&& G_\alpha^{\beta ''} - \delta^{ij} \partial_i \partial_j G_\alpha^\beta + 2 \mathcal {H} G_\alpha^{\beta'} + 
M_\alpha^2(x) a^2 G_\alpha^\beta = \text{\color{blue} [2PI corrections]} \quad \quad 
\alpha \in X,A,\Phi \quad \beta \in {F, \rho} \label{evolinithomo} \\
T^{\eta\eta} &=& 
T_{cl}^{\eta\eta} + \sum_{\alpha \in X,A,\Phi} \frac{1}{2 a^4} \biggl [-G_\alpha^{F''} - 
\delta^{ij} \partial_i \partial_j G_\alpha^F + 4 \mathcal {H} G_\alpha^{F'} + 
M_\alpha^2(x) a^2 G_\alpha^F\biggr ] + \frac{2i}{\sqrt{-g}}\frac{\partial 
\Gamma_2}{\partial g_{\eta\eta}} \nonumber \\
\label {tetaeta} \\
T^{\eta\eta}_{homo} &=&  \frac{3\mathcal{H}^2}{8\pi \mathcal{G}a^4}, \quad \quad \hm \equiv 
\frac{a'}{a} \label{t00inithomo}
\eea
where $T^{\eta\eta}$ is the $00$ component of energy-momentum tensor in homogeneous conformal 
metric and propagators are evaluated at $(\vec{x},\eta,\vec{y}=\vec{x},\eta)$. According to our 
assumptions described earlier, at initial time $T_{cl}^{\eta\eta}$, contributions of $A$ and $\Phi$ 
in the second term of (\ref{tetaeta}), and the last term are all zero.

Spatial components of energy-momentum tensor $T^{ij}$ for homogeneous background metric are:
\bea
\sum_{\alpha \in X,A,\Phi} T_\alpha^{ij} (\eta) &=& \sum_{\alpha \in X,A,\Phi} \frac{1}{2 a^4} 
\biggl [-\delta^{ij} (G_\alpha^{F''} - \mathcal {H} G_\alpha^{F'} + M_\alpha^2(x) a^2 G_\alpha^F) + 
(\delta^{ij} \delta^{kl} - 2 \delta^{ik}\delta^{jl})(\partial_k \partial_l 
G_\alpha^F - \delta_{kl} \mathcal {H} G_\alpha^{F'}) \biggr ] + \nonumber \\ 
&& \frac{2i}{\sqrt{-g}} \frac{\partial \Gamma_2}{\partial g_{ij}} \label{tijinithomo}
\eea 
They do not impose further constraints on the model, but are needed for determination of the 
equation of state defined as $w \equiv p/\rho$, where $\rho = a^2 T^{\eta\eta}$ and 
$P = a^2 \delta^{ij} T^{ij} / 3$. Using Einstein equations $a''$, which is necessary for solving field 
equations, is obtained as:
\be
\frac{a''}{a} = \frac{4\pi \mathcal{G}}{3} \biggl \{a^2 T_{cl} (\varphi) + 
\sum_{\alpha \in X,A,\Phi} \biggl [G_\alpha^{F''} - 
\partial_i \partial^i G_\alpha^F + 2 \mathcal{H}G_\alpha^{F'} + 2 a^2 M_\alpha^2(x) G_\alpha^F + 
\frac{i a^2}{\sqrt{-g}} (\frac{\partial \Gamma_2}{\partial g_{\eta\eta}} - \delta_{ij} 
\frac{\partial \Gamma_2}{\partial g_{ij}}) \biggr ] \biggr \} \label{traceeq}
\ee

Due to coupling between species apriori we cannot define the equation of state separately for each 
species. But, assuming that the coupling is small, a pseudo equation of state can be defined as 
the following:
\bea
w_\alpha = \frac {-G_\alpha^{F''} - \frac{1}{3}\delta^{ij }\partial_i \partial_j G_\alpha^F - 
M^2 a^2 G_\alpha^F + \frac{2ia^4 \delta_{ij}}{3\sqrt{-g}} \frac{\partial \Gamma_2}{\partial g_{ij}}}
{-G_\alpha^{F''} + 4 \mathcal {H} G_\alpha^{F'} - \delta^{ij }\partial_i \partial_j G_\alpha^F + 
M^2 a^2 G_\alpha^F + \frac{2ia^4}{\sqrt{-g}} \frac{\partial \Gamma_2}{\partial g_{\eta\eta}}} 
\quad \quad \alpha \in \Phi, A, X \label{wx}
\eea
where 2PI terms in these expressions are understood to include only terms relevant to field 
$\alpha$. 
Despite its unfamiliar look, eq. (\ref{wx}) has 
expected properties of an equation of state. Specifically, if couplings are small and mass term 
dominates over spatial variation and variation due to the expansion of the Universe, 
$w \rightarrow 0$ and species behave as a cold matter. On the other hand, if $M \rightarrow 0$, 
$w \rightarrow 1/3$ as expected for relativistic particles.

\subsection {Wave function and vacuum renormalization} \label{sec:wfrenorm}
In Appendix \ref{app:propag} we show that for free fields $G_i^F(x,y)$ depends on $x-y$ and on the 
average coordinate $\bar{x}$ through possible dependence of particle distribution of the state on 
which the propagator is defined. Therefore, if the initial distribution of $X$ particles 
$f(k,\bar{x},\eta_0)$ defined in (\ref{1pdist}) is homogeneous and independent of $\bar{x}$, initial 
$G_i^F(x,x)$ and its time derivatives do not depend on $x$. Nonetheless, position derivatives are 
not zero because they are taken with respect to $x$ and then $x=y$ is applied. Using these properties 
and field equations (\ref{evolinithomo}), the term proportional to $G''_k$ can be eliminated from 
(\ref{tetaeta}) and under the assumption that initially the effective mass $M$ does not depend on 
space coordinates, which is consistent with renormalization conditions 
(\ref{renormphi}-\ref{renormxacond}), constraints (\ref{evolinithomo}-\ref{t00inithomo}) can be 
written as:
\bea
\frac{1}{(2\pi)^3} \int d^3k \biggl [3 \mathcal {H} G_i^{F'}(k) + (k^2 + M_i^2 (\eta_0) a^2) 
G_i^F(k) \biggr ] & = & 0 \quad \quad i \in \Phi,A \label{phiprogconstrint} \\
\frac{1}{(2\pi)^3} \int d^3k \biggl [3 \mathcal {H} G_X^{F'}(k) + (k^2 + M_i^2 (\eta_0) a^2) 
G_X^F(k) \biggr ] & = & \frac{3 \mathcal{H}^2}{8\pi \mathcal {G}} - a^2 \rho_{cl} (\varphi (t_0)) 
\label{xprogconstrint}
\eea
where $\rho_{cl} (\varphi (t_0))$ is the energy density of initial condensate field. Here we have 
used the momentum space because we want to show that (\ref{phiprogconstrint}) and 
(\ref{xprogconstrint}) constraints on the contribution of fields in the initial energy-momentum 
tensor and Friedmann equation (\ref{t00inithomo}) determine the remaining arbitrary constants in 
the renormalized model, namely the constant term in $T_{eff}^{\mu\nu}$ calculated in (\ref{enermom1}) 
and the wave function normalization. Indeed, if we had not dropped (\ref{enermom1}) from 
$T_{eff}^{\mu\nu}$, we had to to add $-\frac{i}{2} tr \ln G_i^{-1},~ {i =A,\Phi,X}$ to l.h.s. of 
equations (\ref{phiprogconstrint}) or (\ref{xprogconstrint}) according to the relevant species. 
This term and integrals in the l.h.s. of these equations depend on the wave function 
normalization of species. We assume that normalization factors are chosen such that the equality of 
l.h.s. with observables on the r.h.s. is satisfied. This procedure finalizes the renormalization of 
the model. We notice that in our fully quantum field theoretical approach to a cosmological model, wherever 
a contribution to vacuum arises, it can be included in the wave function normalization and does not 
affect observable quantities. From this observation we conclude that if dark energy is the Cosmological 
Constant, its origin cannot be anything else than the quantization of gravity, which is not considered 
here.\footnote{It is also intriguing that QFT models need gravity for being fully renormalized and 
meaningful. Notably, in cosmology Friedmann equation replaces Born rule in quantum mechanics that 
determines normalization of wave function using its interpretation as a probability distribution.
See~\cite{houriqmgr} for more discussion about inherent relation between quantum mechanics and 
gravity.}.

For a gas of free particles the constraint (\ref{xprogconstrint}) can be expanded to mode functions 
by using (\ref{propfu1part}). Then, the normalization factor $N$ in (\ref{1pdist}) can be determined 
as:
\bea
N & = & \biggl |\biggl (\frac{3 \pi^2\mathcal{H}^2}{2 \mathcal {G}} - (2\pi)^3 a^2 \rho_{cl} 
(\varphi (t_0))\biggr ) \biggl \{ \int d^3k |\psi_k|^2 \biggl [3 \mathcal {H} 
\biggl (\um'_{kX} (\eta_0) \um_{kX}^* (\eta_0) + \um^{'*}_{kX} (\eta_0) \um_{kX} (\eta_0) \biggr ) + 
\nonumber \\
&& k^2 \biggl (\um_{kX} (\eta_0) \um_{kX}^* (\eta_0) + \um^*_{kX} (\eta_0) \um_{kX} (\eta_0) 
\biggr ) \biggr ] \biggr \}^{-1} \biggr | \label{distnormaliz}
\eea
where terms corresponding to (\ref{enermom1}) are included in N. 

\subsection{Initial conditions for condensate} \label{sec:initcondcond}
Similar to propagators, the evolution equation of condensate (\ref{dyneffa}) is of second order 
and needs two initial or boundary conditions. However, due to the setup of the model discussed here, 
they are not independent of initial conditions for propagators, which were discussed in previous 
sections. 

Consider the state of $\Phi$ particles produced through decay of $X$ in the infinitesimal time 
$\Delta \eta = \eta_{0+} -\eta_0$ in an initially homogeneous Universe. If the self-interaction 
between $\Phi$ particles in the decay remnant is neglected, their quantum state can be expanded as:
\bea
|\Psi_\Phi (\eta_0 + \Delta \eta)\rangle & = & |\Psi_\Phi (\eta_0)\rangle \otimes \sum_{i=0}^{N_X} \int d^3p_1 
\cdots d^3p_i \frac {C_{p_1} \cdots C_{p_i} (\Delta \eta)^{i/2}}{(2\pi)^{3i} i!} f_\Phi(p_1) \cdots 
f_\Phi(p_i) a^\dagger_{p_1} \cdots a^\dagger_{p_i} |0\rangle \label{initphistate} \\
N_X & = & \frac{V}{(2\pi)^3} \int d^3 k f_X (k) \rightarrow \infty \label{xnum}
\eea
where $|\Psi_\Phi (\eta_0)\rangle$ is the initial state of $\Phi$ particles before switching on $X$ 
decay and $V$ is the volume of the Universe. Here in what concerns $\Phi$ field, 
$|\Psi_\Phi (\eta_0)\rangle = |0\rangle$ (or a condensate states for intermediate simulations). 
Coefficients $C_{p_i}$ are amplitudes of modes $p_i$ of $\Phi$ 
particles produced from decay of $X$ particles. The distribution $f_\Phi (p)$ can be related to 
initial momentum distribution of $X$ particles $f_X$ defined in (\ref{1pdist}) and is calculated 
in Appendix \ref{app:distphia}. Because our aim from expanding the state of $\Phi$ particles is to 
calculate initial conditions for evolution of condensate, it is more convenient to write the state 
in Schr\"odinger picture. In this case, creation operator $a^\dagger_p$ in (\ref{initphistate}) is 
time-independent; amplitude $C_p$ is time-dependent; and $|C_p|^2$ is the probability of production 
of a $\Phi$ particle with momentum $p$ from decay of a $X$ particle with momentum $k$:
\be
|C_p|^2 \Delta \eta \approx (1 - e^{-\frac{\Gamma_X a \Delta \eta}{\gamma_X}}) \approx 
\frac{\Gamma_X a \Delta \eta}{\gamma_X} \label{cpapprox}
\ee
where the invariant width $\Gamma_X$ for model (a) is 
$\Gamma_X = 8\pi^2 \mg^2 P / m^2_X$~\cite{houridmquin} and 
$P = ((m_X^2 - m_\Phi^2 - m_A^2)^2 - 4 m_\Phi^2 m_A^2)^{1/2} / 2m_X$. The boost Lorentz factor 
$\gamma_X = k^0 (p) / M_X$ where $k^0$ is the energy of decaying $X$ particles and can be related to 
momentum $p$ of the remnant $\Phi$, see Appendix \ref{app:distphia} for details. 

In presence of self-interaction scattering of $\Phi$ particles rapidly uniformizes their 
distribution and the second term in (\ref{initphistate}) approaches fully or partially to a 
condensate state, and condensed fraction would depend on self-coupling $|\lambda / n|$. Moreover, 
if momentum distribution of $X$ particles has a relatively small standard deviation, momentums 
in (\ref{initphistate}) will be very close to each others and state of newly produced particles 
in (\ref{initphistate}) approaches to a condensate. More generally, the state $|\Psi_\Phi\rangle$ can 
be decomposed to condensate and noncondensate states:
\bea
|\Psi_\Phi\rangle & = & \mathcal{N}_\varphi |\Psi_C\rangle + \mathcal{N}_\phi|\Psi_{NC}\rangle, 
\quad \quad |\mathcal{N}_\varphi|^2 + |\mathcal{N}_\phi|^2 = 1 \label{phistatedecomp} \\
|\Psi_C\rangle & \equiv & \sum_{i=0}^\infty \int d^3p_1 \cdots d^3p_i \frac{C_{p_1} 
\cdots C_{p_i}(\Delta \eta)^{i/2}}{(2\pi)^{3i} i!} f_\Phi(p_1) \cdots f_\Phi(p_i) \delta^{(3)} (p_i - p_{i-1}) 
\cdots \delta^{(3)} (p_2 - p_1) a^\dagger_{p_1} \cdots a^\dagger_{p_i} |0\rangle \nonumber \\
& = & \frac {1}{(2\pi)^3} \sum_{i=0}^\infty \int d^3p\frac{C_p^i(\Delta \eta)^{i/2}}{i!} 
f_\Phi^i(p) a^\dagger_{p_1=p} \cdots a^\dagger_{p_i=p} |0\rangle = \frac{1}{(2\pi)^3} 
\int d^3p~e^{C_p f_\Phi (p) a_p^\dagger} |0\rangle \label{coherphistate}
\eea
where $|\Psi_{NC}\rangle$ is the remaining non-condensate and $\mathcal{N}_\varphi$ is a normalization 
factor. The coherent component $|\Psi_C\rangle$ is a generalized Glauber coherent state with amplitude 
$C_p f_\Phi (p)$. By definition $\langle \Psi_{NC}|\hat{\Phi}|\Psi_{NC}\rangle = 0$ ant it does not 
contribute in $\varphi (t_{0+}) = \langle\Psi_\Phi|\hat{\Phi}|\Psi_\Phi\rangle$. Thus, using the definition of a condensate, the initial time 
derivative of condensate field $\varphi'$ is determined:
\be
\varphi' (\mathbf{x}, \eta_0) = \frac{|\mathcal{N}_\varphi|^2}{(2\pi)^3} \int d^3p ~ f_\Phi (p) \biggl 
(C_p \mathcal {U}_p (\eta_0)~ e^{-ip.\mathbf{x}} + C^{*}_p \mathcal {U}^*_p (\eta_0) ~ e^{ip.\mathbf{x}} \biggr )
\label {initcondvalderiv}
\ee
As $\Phi$ is a real field $\mathcal {U}_{-p} (\eta_0) = \mathcal {U}^*_p (\eta_0)$. Thus, if 
$f_\Phi (\vec{p}) = f_\Phi (-\vec{p})$, (\ref {initcondvalderiv}) takes the familiar form of an inverse 
Fourier transform:
\be
\varphi' (\mathbf{x}, \eta_0) = \frac{2|\mathcal{N}_\varphi|^2}{(2\pi)^3} \int d^3p ~ f_\Phi (p) 
C_p \mathcal {U}_p (\eta_0)~ e^{-ip.\mathbf{x}}
\label {initcondvalderiv}
\ee
The normalization factor $\mathcal{N}_\varphi$ determines the initial rate of condensation, but its 
determination from first principles is not straightforward. To get an insight to its amplitude, we 
use scattering rate of high energy $\Phi$ particles. Their dissipation rate which leads to cascade 
formation of $\Phi$ particles and their condensation can be estimated as: 
\be
\frac{\Gamma_\varphi}{H} \sim (2\pi)^{10} \lambda^2 \biggl (\frac{M_X}{2H}\biggr )^4 \label{dissiprate}
\ee
where we assume $M_\Phi, M_A \ll M_X$ and neglect annihilation of $\Phi$ by interaction with $X$. For 
the value of parameters used in simulations described in the next section the initial formation rate of 
condensate $\Gamma_\varphi/H < 1$. However, considering the small standard deviation of $X$ particles 
energy distribution, even without scattering the momentum of $\Phi$'s are close enough to each 
other\footnote{In the next section we also discuss an example simulation for which 
$\frac{\Gamma_\varphi}{H} \gg 1$.}. Therefore, in the simulations we assume that the state of $\Phi$ at 
$\eta_{0+}$ is a condensate, $|\mathcal{N}_\varphi| \approx 1$, and $|\mathcal{N}_\phi| \approx 0$.

\section {Simulations} \label {sec:simulation}
Evolution equations (\ref{dyneffa}-\ref{evolgrho}) cannot be solved analytically. Moreover, 
due to nonlinear and nonlocal interaction terms in 2PI formalism, evolution equations of 
propagators and condensates are integro-differential and their numerical simulation is more 
difficult and CPU intense than classical multi-field inflation models~\cite{infmfieldsimul} and 
reheating~\cite{reheatrev}. Besides, the model developed here includes multiple fields with very 
different masses running over some 39 orders of magnitude. Consequently, 
the numerical model is stiff and it is not possible to rend quantities close to unity by scaling 
them. For these reasons we were obliged to perform separate simulations with different time 
(or equivalently expansion factor) steps, because despite using an adaptive time step, a single rule 
cannot be used for the totality of the simulated interval and at some point numerical errors make the 
simulation unreliable. 

To reduce CPU time we used smaller time steps at early times. High densities of species at this epoch 
cause high rate of interactions and more precise evolution of dynamics is crucial for the correctness 
of simulations at later epochs. Inversely, the expansion of the Universe at later epochs decreases 
the effective coupling between particles. Division of simulation to multiple steps explained in the 
previous paragraph and gradual increase of time steps inevitably induce discontinuities and numerical 
uncertainties. Nonetheless, repetition of simulations at different breaking points and with different 
adaptive time intervals has convinced us that essential properties of the model obtained from these 
simulations and their interpretations are reliable.

In numerical simulations on a lattice in momentum space, the size of simulation box $|k_{max}|$ plays 
the role of a UV cutoff and is identified with the scale $\mu_0$ in (\ref{renormphi}-\ref{renormax}). 
In an expanding universe, in which the physical size of the coordinate lattice increases with time, 
the initial value of masses and couplings can be considered as their renormalized - physical - 
value at UV limit. On the other hand, the size of simulation box in real space imposes an IR cutoff. 
It must be enough large such that it contains the physically interesting IR limit, namely the horizon 
at each epoch. The dependence of simulation results on the lattice volume can be estimated by varying 
the initial volume while the size of cells are kept constant. Unfortunately, we were not able to 
investigate the dependence of effective masses and couplings on UV and IR cutoffs for fixed lattice 
size because the procedure quickly increases the amount of necessary memory and execution time. 
Nonetheless, as physical size of the box is determined by the inverse of initial Hubble constant 
$H^{-1 (t_0)}$, simulations at high $H$ - presumably for inflation - and low $H$ - presumably for a 
lately produced dark energy - demonstrate the variation of effective mass and couplings of $\Phi$ and 
its condensate with scale.

\subsection{Parameters} \label{sec:param}
We consider a $9^3$ dimensional cubic lattice on which the three quantum fields $X$, $A$, and $\phi$, 
and the condensate field $\varphi$ live. For calculation of closed time path integrals in the 
evolution equations we sum over the past $10$ time steps. We also tested the simulation of early epochs 
with summation over $30$ past steps, and found little difference between the two cases. Therefore, we 
continued with smaller number of summations, which made execution time more affordable. To decrease 
memory request for these operations we work in momentum space and neglect the dependence on average 
coordinate in the integrals. This is an approximation which should be added to other uncertainties 
and imprecisions of these simulations.

We performed two series of simulations, one presenting inflation era and the other condensation of a 
light scalar field from end of reheating to present time. The main difference between these simulations 
is the value of initial Hubble constant, which in addition to fixing the initial expansion rate, its 
inverse is used as distant scale to determine the physical size of the simulation box, cell size, and 
momentum modes.

Only simulations for a $\Phi^4$ self-interaction potential are reported here. In most simulations 
initial masses and couplings are: $m_X = 10^{-3} M_P,~m_A=10^{-15} M_P,~m_\Phi = 10^{-36} M_P$, and 
$\lambda = 10^{-14},~\mg/M_P = 10^{-17}$. They correspond to renormalized values at IR scale for 
$\varphi_R = 0$ defined in (\ref{renormphi}-\ref{renormax}). In dark energy simulations we also tried  
smaller $m_X$ and other values for couplings. According to these choices $X$ presents a heavy field - 
presumably from Planck or GUT scale physics; $A$ is a prototype for fields at electroweak symmetry 
breaking scale; and $\Phi$ is a light field, which may be considered as inflaton, quintessence or both. 
However, an important result of these simulations, explained in more detail later in this section, 
is the crucial role of all the fields and their interactions in triggering inflation and late 
accelerating expansion.

In both series of simulations we assume a vacuum initial state for $A$ and $\Phi$ and null initial 
value for the classical condensate field $\varphi$. We remind that one of the main objectives of 
this work is to understand formation and evolution of a condensate in an expanding universe, 
whether and how it preserves its quantum coherence at cosmological scales, and whether and how an 
effective potential which supports an accelerating expansion during inflation and at late times may 
emerge.

For the field $X$, which is initially the only contributor in the effective classical energy-momentum 
density, we assume a Gaussian distribution similar to (\ref{oneparticledist}) with mean value at 
$k=0$ and standard deviation $\sigma_X = m_X/ 10$. This choice has both practical and physical 
reasons. As we explained in Sec. \ref{sec:density}, for a Gaussian initial condition we can use 2PI 
formulation of a vacuum state. Moreover, it is well known that a particle more massive than a 
few hundreds GeV leads to an overdense Universe if it were ever in thermal equilibrium with the 
Standard Model species. Therefore, a random Gaussian initial distribution for $X$ seems a more 
realistic assumption than a thermal initial condition. We use the same distribution for both 
inflation simulations and those beginning after reheating. 

Momentum modes of the lattice are determined such that:
\be
|k^i_{max}| \sim \pi H(t_0) \quad i=1,~2,~3 \label{knorm}
\ee
where $H(t_0)$ is the Hubble constant at initial time. Therefore, simulation box in momentum space 
initially includes both subhorizon and superhorizon modes. Moreover, they are all inside $1\sigma$ 
deviation from mean value of $X$ particles distribution.

As we discussed in the introduction section, one of the main objectives of this study is the 
investigation of the contribution of quantum and condensate components in the effective 
energy-momentum tensor and their fluctuations, which are the principle cosmological observables. 
The tensor $T^{\mu\nu}_{eff}$ in (\ref{effenermomdet}) can be divided into 3 components: the 
condensate, which despite its quantum origin can be treated as a classical field; the 1PI 
contribution, that is the second bracket in (\ref{effenermomdet}) and includes the contribution 
of perturbatively free particles; and finally 2PI non-equilibrium interactions. In the following 
subsections we discuss evolution of these components and their effects on the cosmological 
expansion. All dimensionful quantities in the plots are in $M_p$ units.
 
\subsection{Inflation} \label{simulinf}
For these series of simulations the chosen initial value of Hubble function is $H(t_0) = 10^{-6} M_P$. 
There is not a generally accepted consensus about the energy scale of inflation~\cite{infscale}. 
An upper limit of $\sim 10^{16}$~GeV~$ \lesssim M_{GUT}$ can be estimated from upper limit of tensor 
to scalar perturbation ratio $r$ from Planck observations, based on comparison with predictions of 
monomial or hybrid inflation models~\cite{infplanck}. In this case, the choice of a mass larger 
than inflation scale for $X$ particles means that they are produced by physics at Planck or GUT 
scale and can be considered as cold matter. Therefore, the cosmology of this model is initially matter 
dominated.

\subsubsection{Evolution of expansion factor}
\begin{figure}
\begin{center}
\begin{tabular}{p{9cm}p{9cm}p{9cm}}
a) & \hspace{-4cm} b) & \hspace{-8cm} c) \\
\vspace {-1cm}\includegraphics[width=9cm]{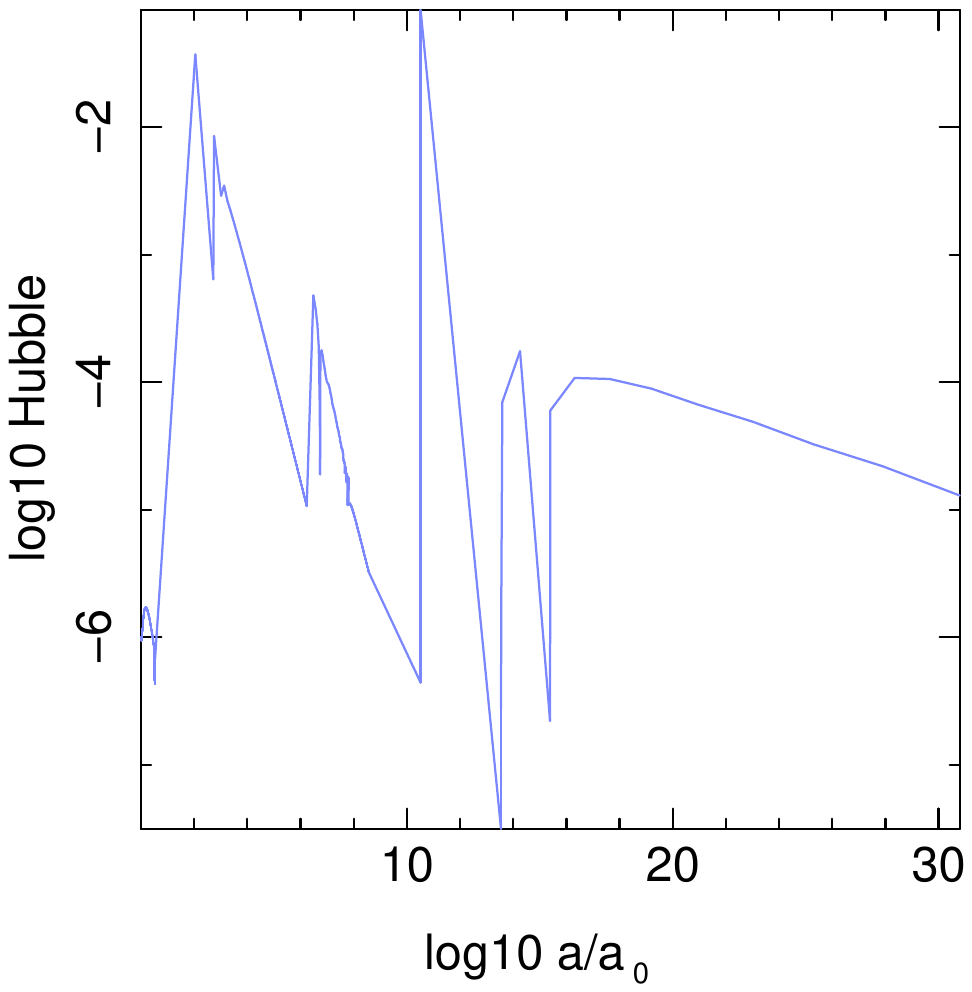} & \vspace {-1cm}\hspace{-4cm}\includegraphics[width=9cm]{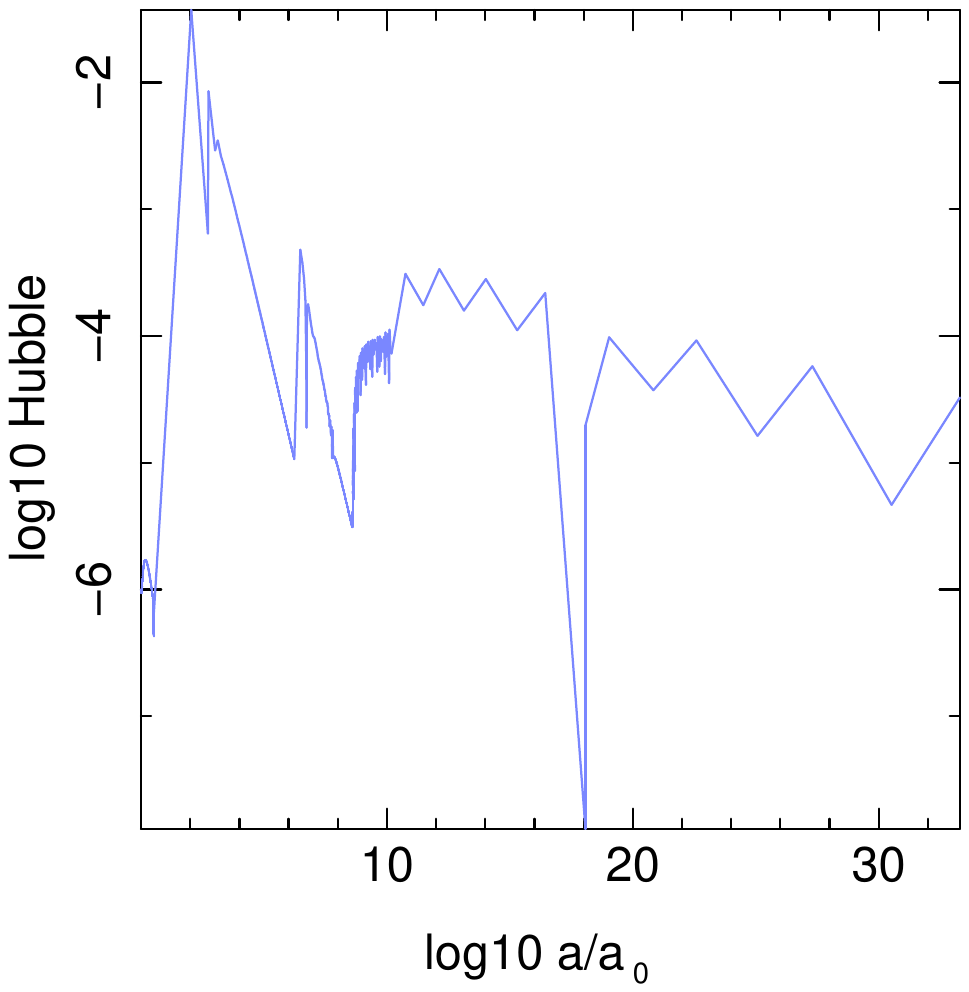} & \vspace {-1cm}\hspace{-8cm}\includegraphics[width=9cm]{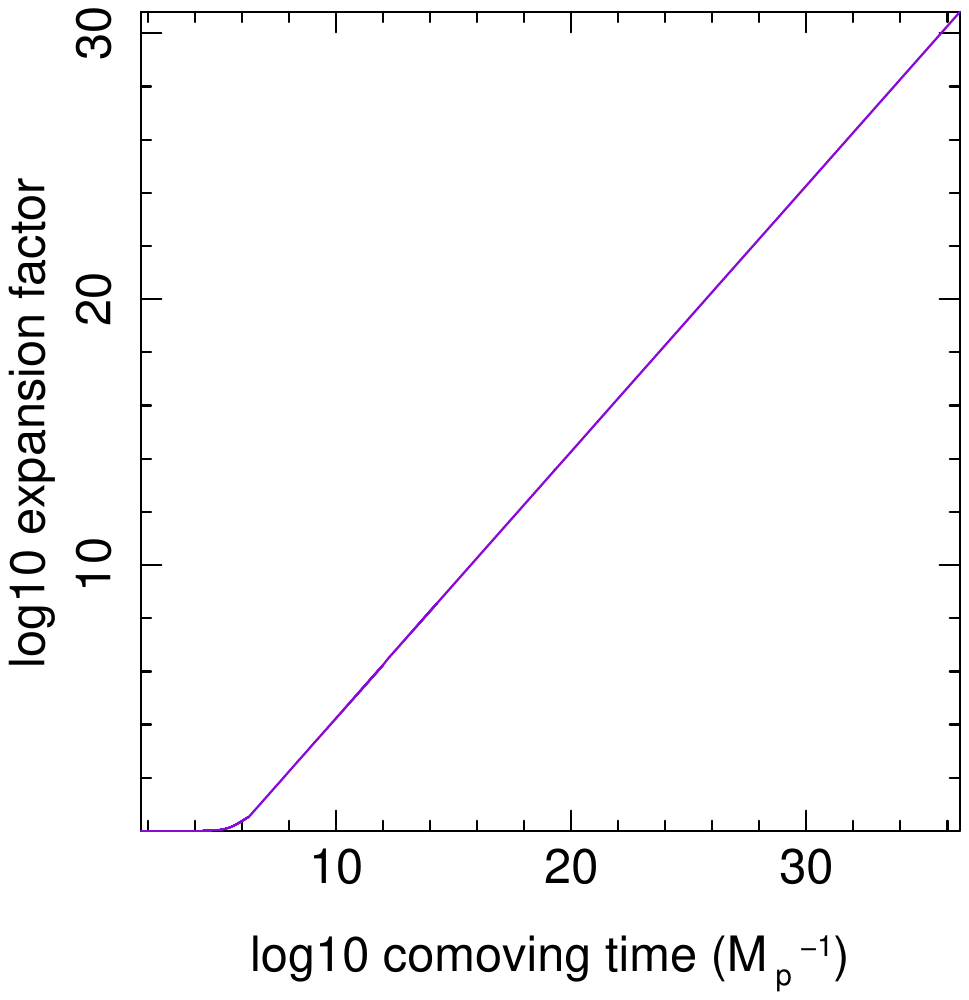}
\end {tabular}
\end {center}
\caption{a) and b): Evolution of Hubble function with expansion factor $a(t)$. 
They are obtained from series of 5 separate successive simulations with different rules for time 
incrementation to reduce accumulation of numerical errors. Rapid variations may be in some 
extend numerical artefacts. The difference between plots a) and b) is the initial $a(t)/a_0$ in one 
of the simulations at intermediate $\log_{10} a(t)/a_0 \sim 8.5$, see the text for details. 
c) Evolution of expansion factor with time for simulations shown in a). \label{fig:inf-hubble}}
\end{figure}

Fig. \ref{fig:inf-hubble}-a shows the evolution of Hubble function $H$ with respect to the expansion 
factor $a(t)$. The evolution of expansion factor with time is shown in Fig. \ref{fig:inf-hubble}-c. 
At late times $a / a_0 \sim (t/t_0)^\alpha,~\alpha \gtrsim 1$. Thus, the inflation generated in this 
model has a power-law profile. We remind that in the classical models, power law inflations are 
usually generated with an exponential potential~\cite{infpowerlaw}, which does not have a 
renormalizable quantum counterpart and must be considered as an effective potential.

The initial increase and oscillation of $H$ is due to the rapid evolution of energy-momentum density 
from being dominated by cold $X$ particles to a binding energy dominated {\it plasma} through 
non-equilibrium interaction between the three constituents of the model. Large oscillations in the 
Hubble function before the onset of inflation are mainly due to the chaotic behaviour of nonlinear 
evolution equations. Indeed, approximate analytical solution of evolution equations of the model 
in~\cite{houricond} shows the presence of a parametric resonance, see following sections for discussion 
of processes causing such behaviour. Indeed what is happening here is analogous to preheating and 
exponential particle production at the end of inflation~\cite{reheatrev}. Moreover, because in these 
simulations the metric is evolved consistently, the effect of particle production and interaction 
between various components induces instabilities in the expansion rate and the Hubble function, which 
backreact on the evolution of densities and may stimulate further instabilities.

\subsubsection{Artefact issue} \label{sec:artefact}
 Numerical simulations in general include glitches and artefacts and we cannot rule out that some of 
the features in our results are artefacts induced by approximations used to simplify computations and 
by low resolution of these simulations. 

To qualify numerical uncertainties we truncated simulations shown in Fig. \ref{fig:inf-hubble}-a 
at $\log (a/a_0) \sim 8.5$ and continued with slightly different time steps. 
Fig. \ref{fig:inf-hubble}-b shows the Hubble function obtained from this second series of simulations. 
Although details of plots and numerical values of physical quantities in this series of simulations 
are somehow different from the first one, their overall behaviour is very similar. For instance, 
in the case of $H(a)$ in Fig. \ref{fig:inf-hubble}-b, despite oscillations at late stages of the  
simulation, the average slope, i.e. the average $d\log H / d\log a \equiv -\epsilon_1$~\cite{infparam} 
is similar to that of smooth evolution of the first simulation series shown in 
Fig. \ref{fig:inf-hubble}-a. Therefore, in the following sections we only discuss the results of the 
first series of simulations and restrict our conclusions to overall aspects rather than details, which 
may not be reliable. 

Better simulations are necessary for verifying to which extend results and conclusions of these 
simulations are correct. Comparison with more or less similar simulations is another way of 
cross-checking the results. For instance, large oscillations of energy-momentum tensor and expansion 
rate before the onset of inflation are reported by other authors~\cite{desitterinstab1} and 
compared to the instability of QED vacuum. Therefore, despite inevitable numerical effects, initial 
oscillations in these simulations seem real, and as explained above, a consequence of dynamical 
instabilities. 

\subsubsection{Inflation parameters} \label{sec:simulinfparam}
For determining characteristics of the inflationary epoch in this model - defined as when the Hubble 
function varies slowly with increasing expansion factor - we fit $\log H(a)$ using parameters 
$\epsilon_i,~i=1,2$ defined in~\cite{infparam}. We obtain $\epsilon_1 \sim 0.01 - 0.04$ and 
$\epsilon_2 \sim -0.14~ - ~0.35$, depending on the choice of time steps used for the fitting. 
In classical treatment of inflation models 
$\epsilon_i$ parameters can be analytically related to the spectral index of scalar fluctuations 
$n_s -1= -\epsilon_2 - 4\epsilon_1$ and tensor to scalar ratio $r =16\epsilon_1$. Comparison of 
values obtained for $\epsilon$ parameters from our simulations and corresponding values for $n_s$ 
and $r$ shows that according to these relations the model is not consistent with the CMB 
observations~\cite{infplanck}. Even when the value of $n_s$ is consistent with observations, $r$ is 
too large. However, in Sec. \ref{sec:simulspect} we show that the value of both these parameters 
obtained directly from simulations are indeed consistent with observations. Therefore, the relation 
between $\epsilon$'s and properties of the spectrum of fluctuations obtained from classically treated 
scalar field models cannot be applied to fully quantum non-local approach. We should also remind that 
the choice of parameters for the simulations were motivated by the results of classical interacting 
quintessence models studied in~\cite{houridmquin} and no adjustment was performed to reproduce CMB 
observations.

\subsubsection{Evolution of densities} \label{sec:simuldense}
Effective potential of inflation is a very important quantity because apriori it can be extracted 
from angular spectrum of CMB and LSS fluctuations~\cite{cmbinf}. As for the time being cosmological 
observations are the only accessible conveyor of the physics of early Universe and high energy 
scales, it is crucial to understand the relation between effective classical quantities 
extracted from cosmological data and the underlying fundamental model. Observations of the Planck 
satellite shows no significant non-Gaussianity and is consistent with a small tensor to scalar ratio 
of $r \lesssim 0.05$. These results indicates a flat effective potential and a small field 
inflation~\cite{infplanck}.

\begin{figure}
\begin{center}
\begin{tabular}{p{6cm}p{6cm}p{6cm}}
a) & b) & c) \\
\hspace{0.8cm}\includegraphics[width=4.75cm]{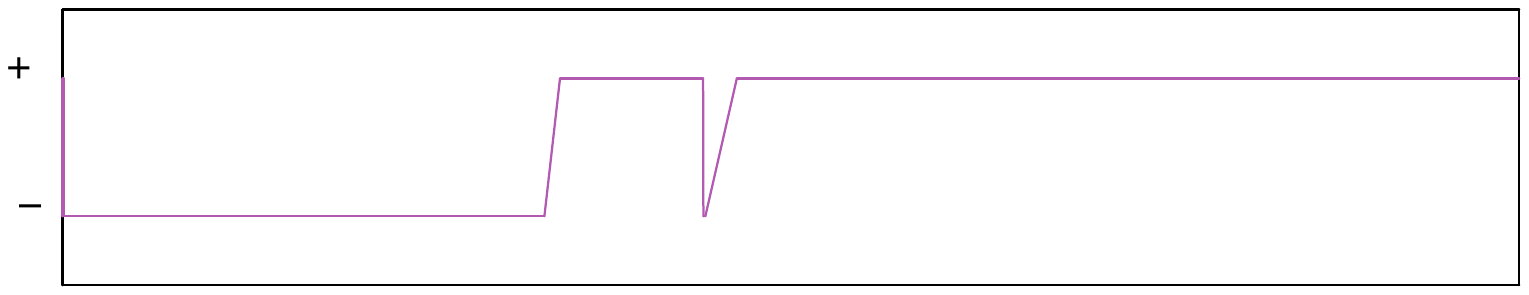} & \hspace{0.55cm}\includegraphics[width=4.75cm]{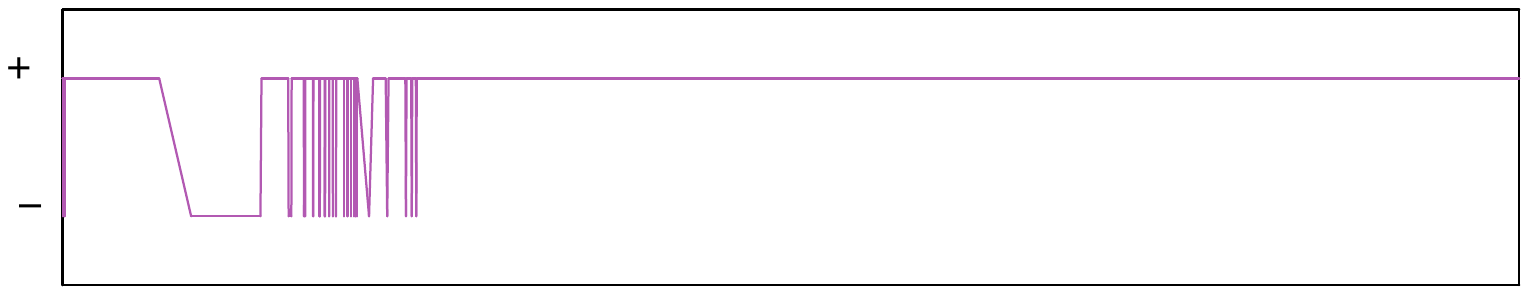} & \hspace{0.8cm}\includegraphics[width=4.75cm]{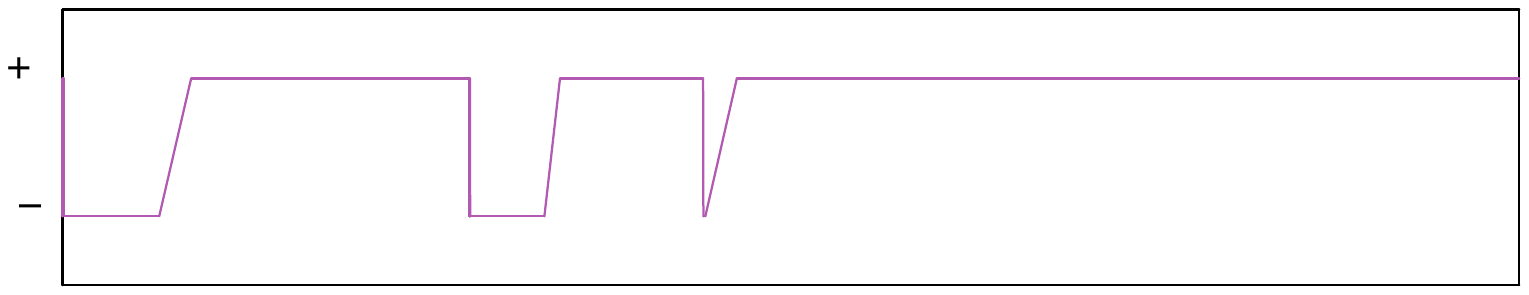} \\
\vspace{-3.9cm}\includegraphics[width=6cm,angle=90]{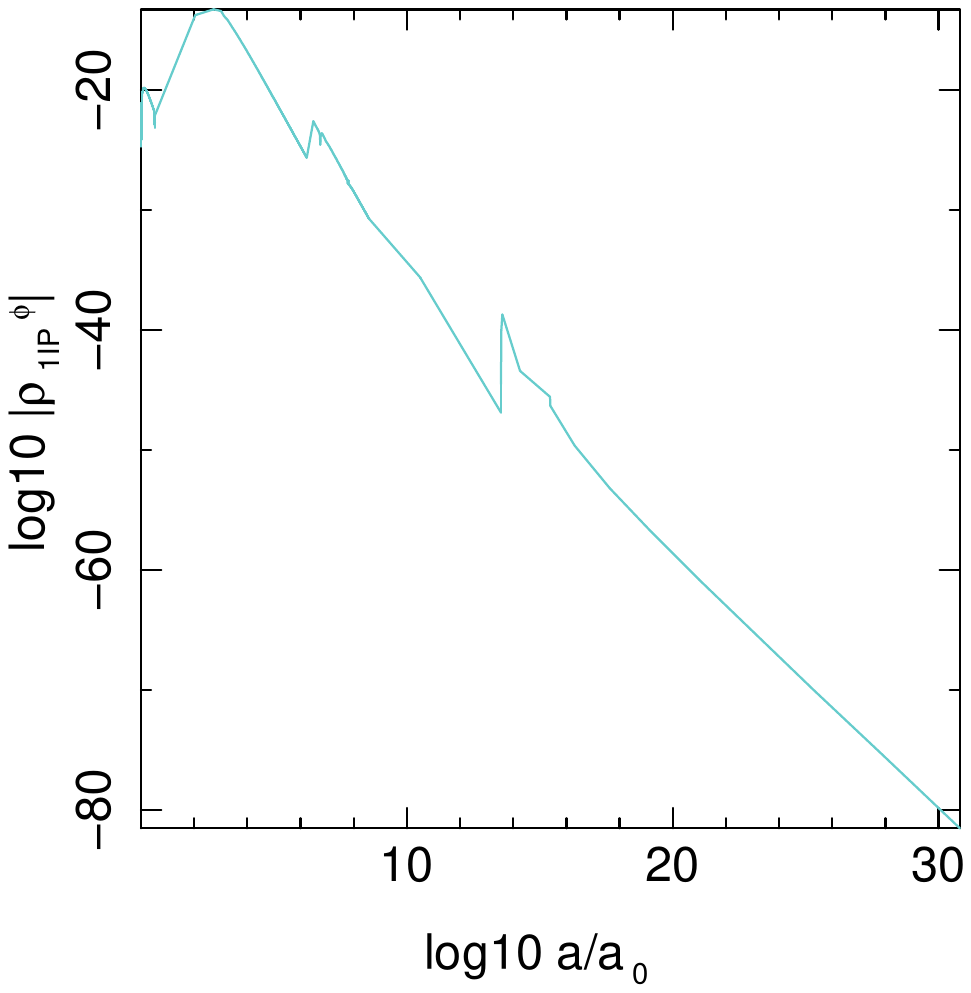} & \vspace{-2.9cm}\includegraphics[width=5cm]{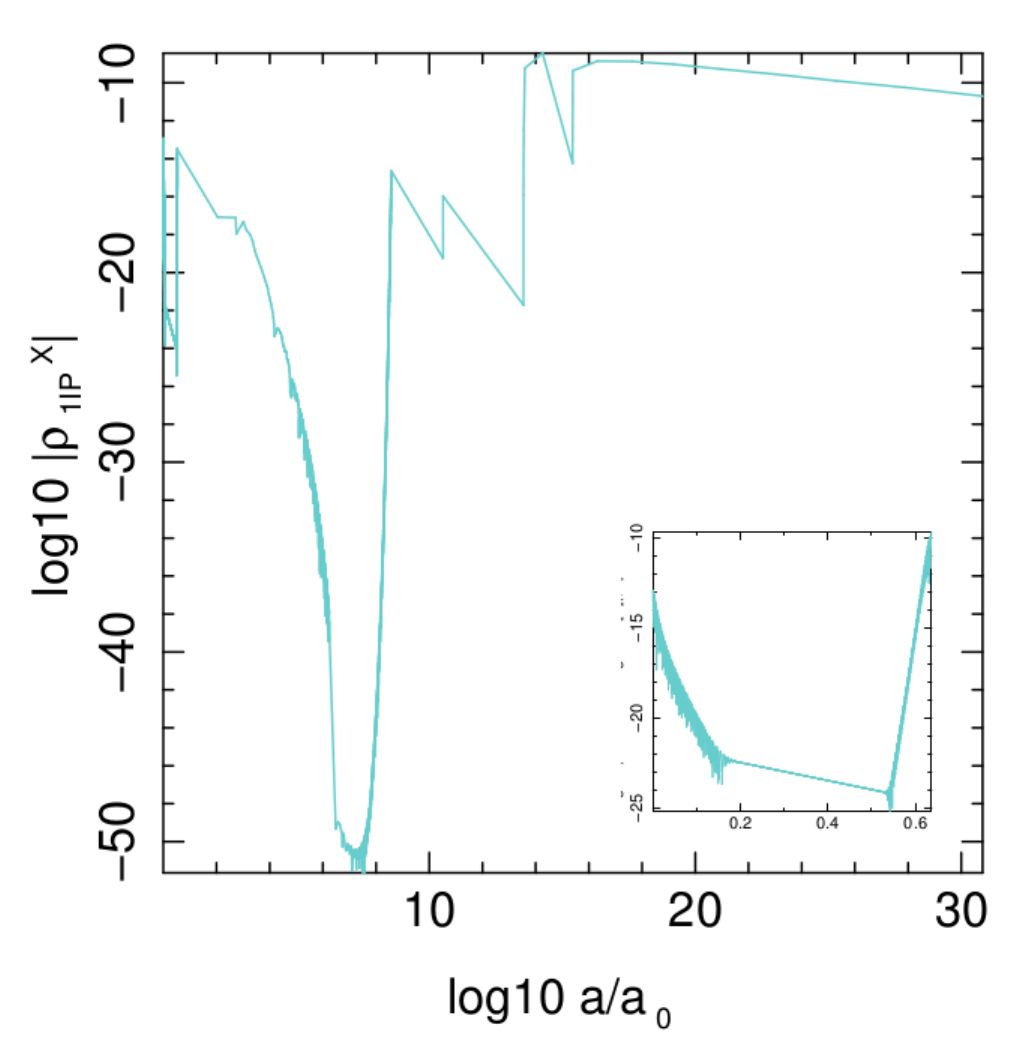} & \vspace{-3.9cm}\includegraphics[width=6cm,angle=90]{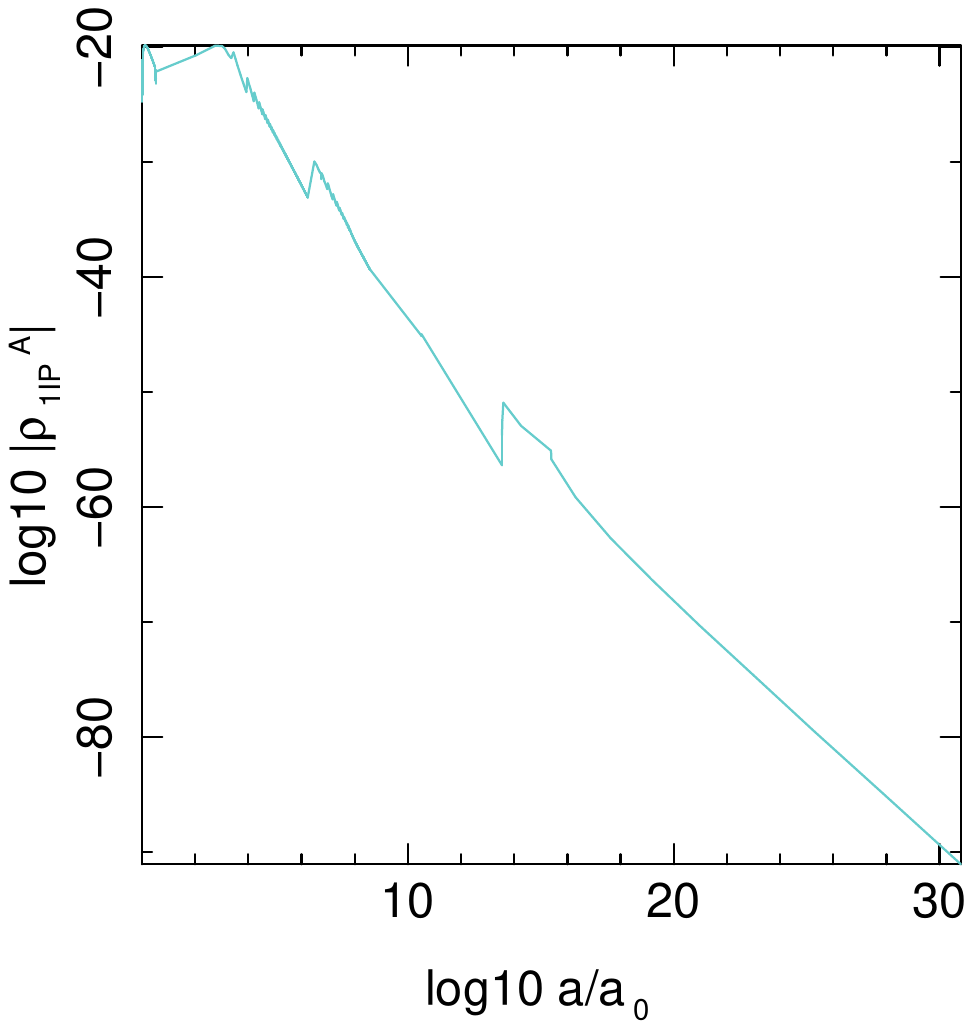}
\end {tabular}
\end {center}
\caption{a), b), c): Contribution of 1PI terms $\rho^i_{1PI},~i=X,A,\Phi$ in the total energy density. 
The upper plots show the sign of these terms and lower plots their amplitude. The inset in b) is a 
zoom on the early evolution of $\rho^X_{1PI}$. \label{fig:pi1}}
\end{figure}

\begin{figure}
\begin{center}
\begin{tabular}{p{4cm}p{4cm}p{4cm}p{4cm}}
a) & b) & c) & d)\\
\vspace {-0.5cm}\includegraphics[width=4.5cm,angle=90]{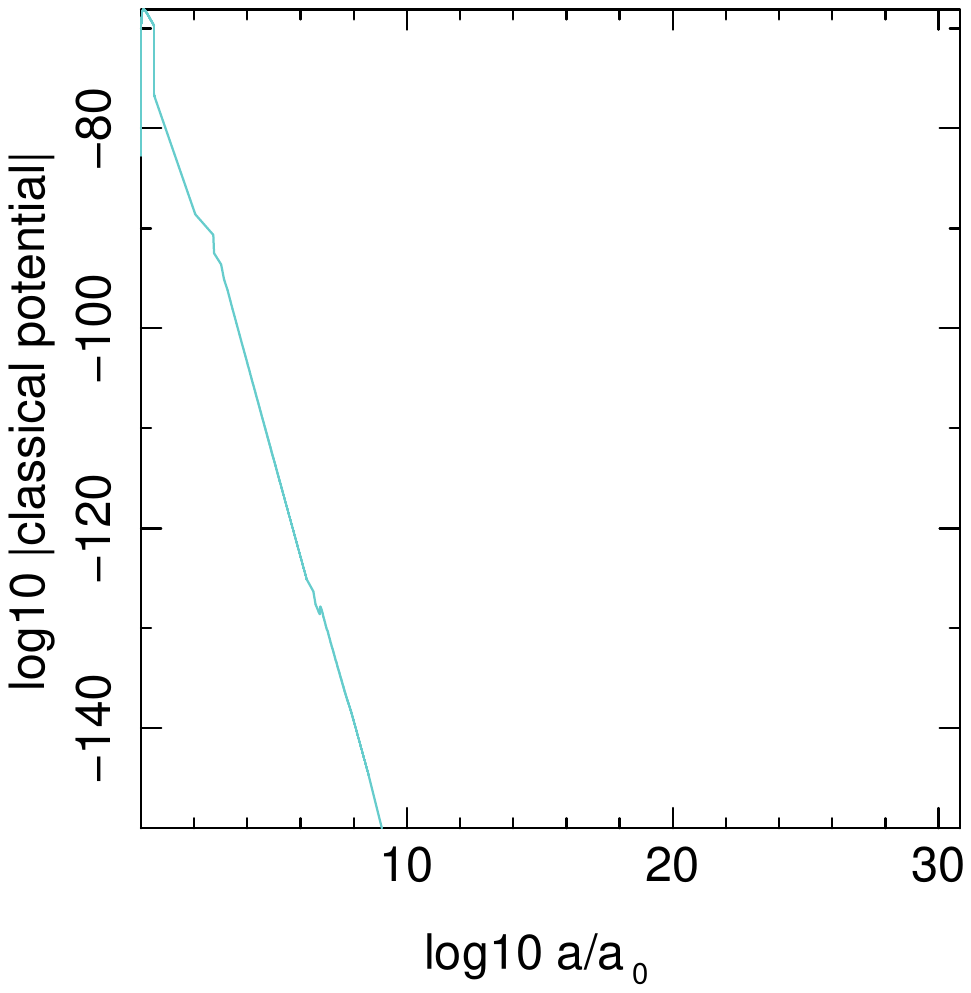} & 
\vspace {-0.8cm}\includegraphics[width=5.2cm,angle=90]{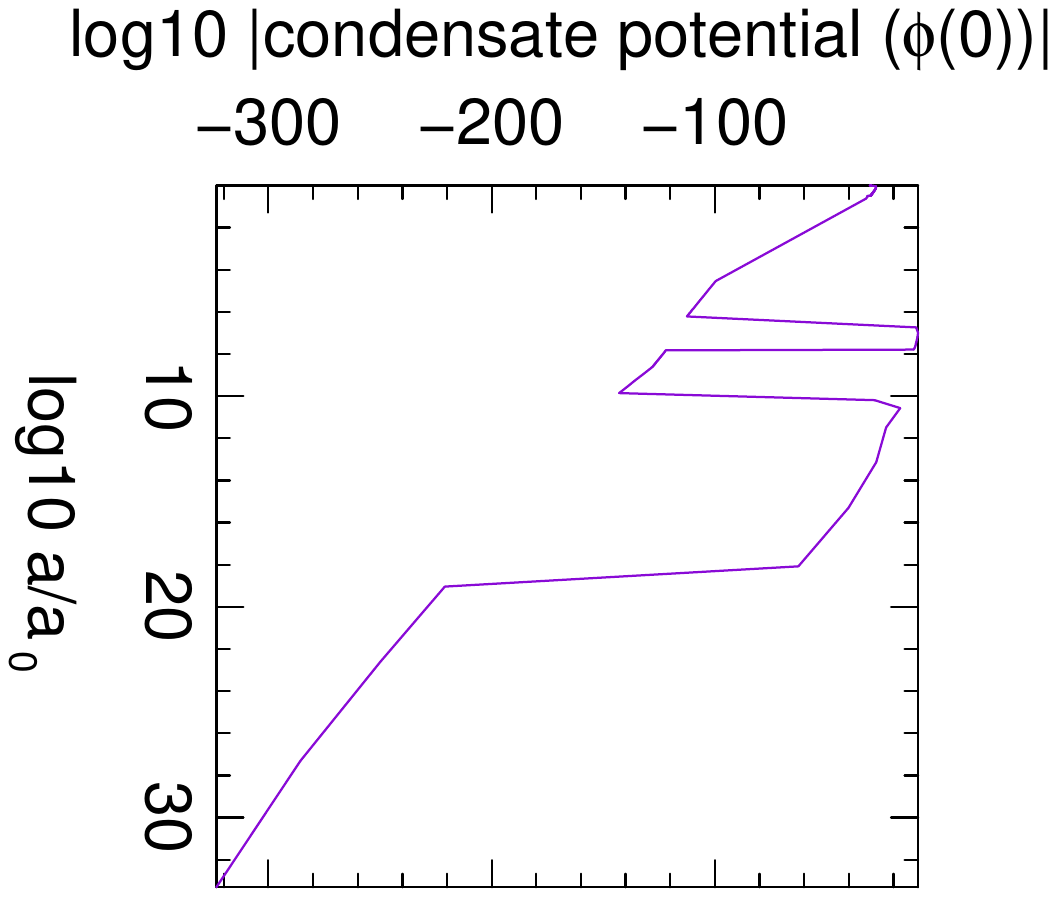} & 
\vspace {-0.8cm}\includegraphics[width=5.2cm,angle=90]{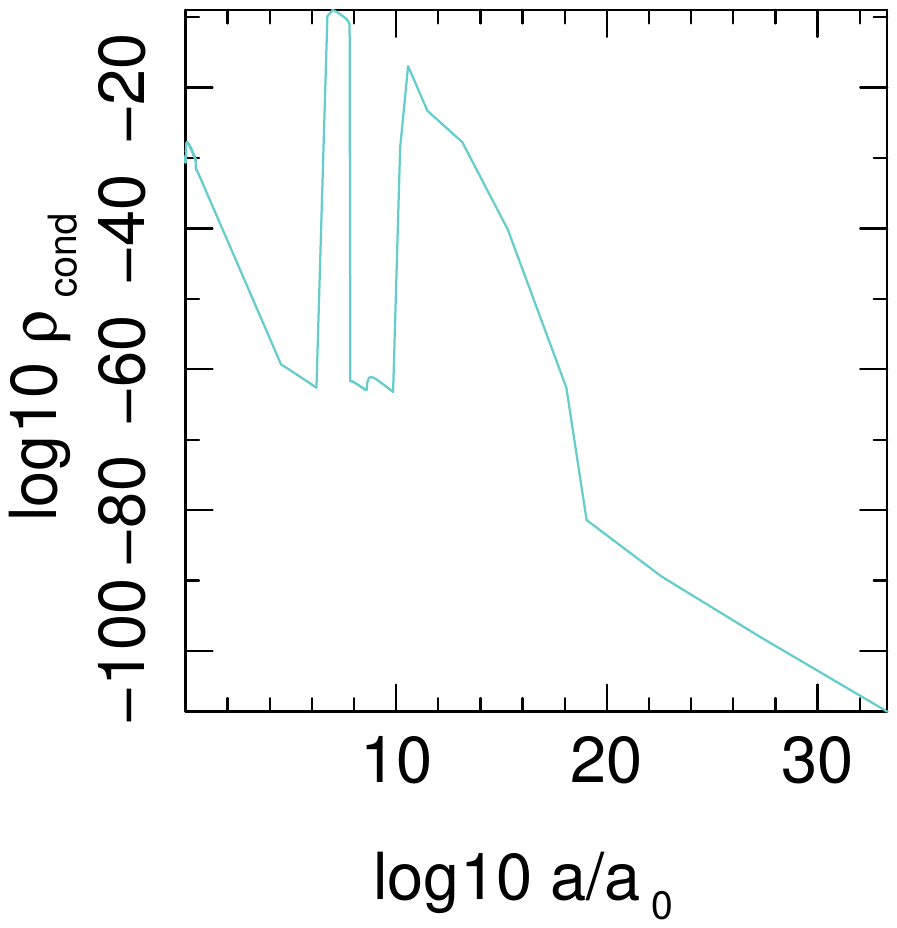} & 
\vspace {-0.5cm}\includegraphics[width=4.5cm,angle=90]{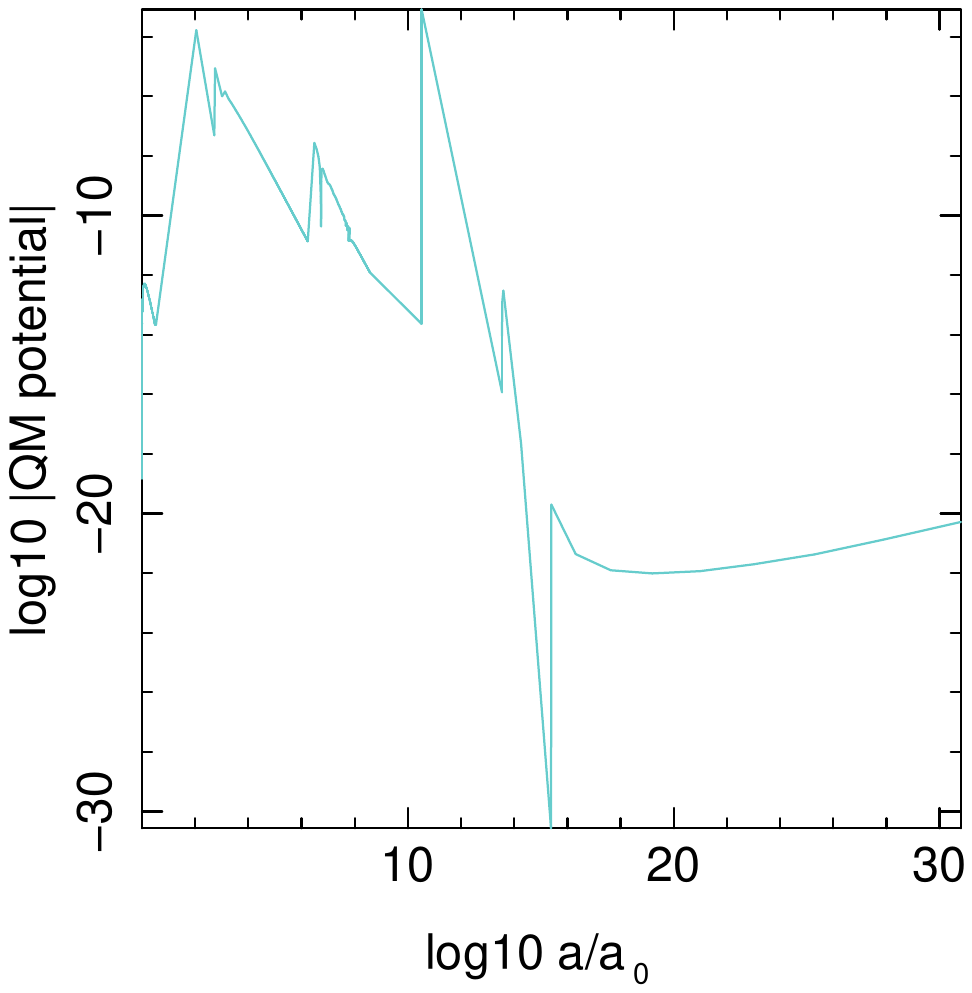}
\end {tabular}
\end {center}
\caption{a): Classical potential of condensate $\varphi$; b) Effective potential of condensate, 
including quantum corrections; c) Effective energy density of condensate; d) Contribution of 2PI 
terms in the total energy density. \label{fig:rhos}}
\end{figure}

Fig. \ref{fig:pi1} shows the evolution of the 1PI terms in the density $\rho$, that is the second 
bracket in (\ref{rhotot}), for the three constituents of the model, which from now on we call them  
$\rho^i_{1PI},~i=X,A,\Phi$. 
It is easy to verify that the initial rapid decay of $\rho^X_{1PI}$ is not due to what we may call 
{\it semi-classical decay}, i.e. the lowest order tree diagram of $X$ particles decay into $A$ and 
$\Phi$. With the value of $\gm$ chosen for these simulations the decay width of $X$ through this 
channel is comparable to the present value of Hubble constant, and consequently the lifetime of free $X$ 
particles is comparable to the present age of the Universe. Our tests show that the slope of 
this decay depends on the self-coupling $\lambda$ of $\Phi$ and is induced by the sudden increase in 
the number of these particles and their interaction with $X$, see plots in Fig. \ref{fig:rhos} which 
show the evolution of classical potential of the condensate $\varphi$, its effective energy density, 
and contribution of 2PI terms in the total energy density. 

\begin{figure}
\begin{center}
\begin{tabular}{p{6cm}p{6cm}p{6cm}}
a) & b) & c) \\
\vspace {-1cm}\includegraphics[width=6cm,angle=90]{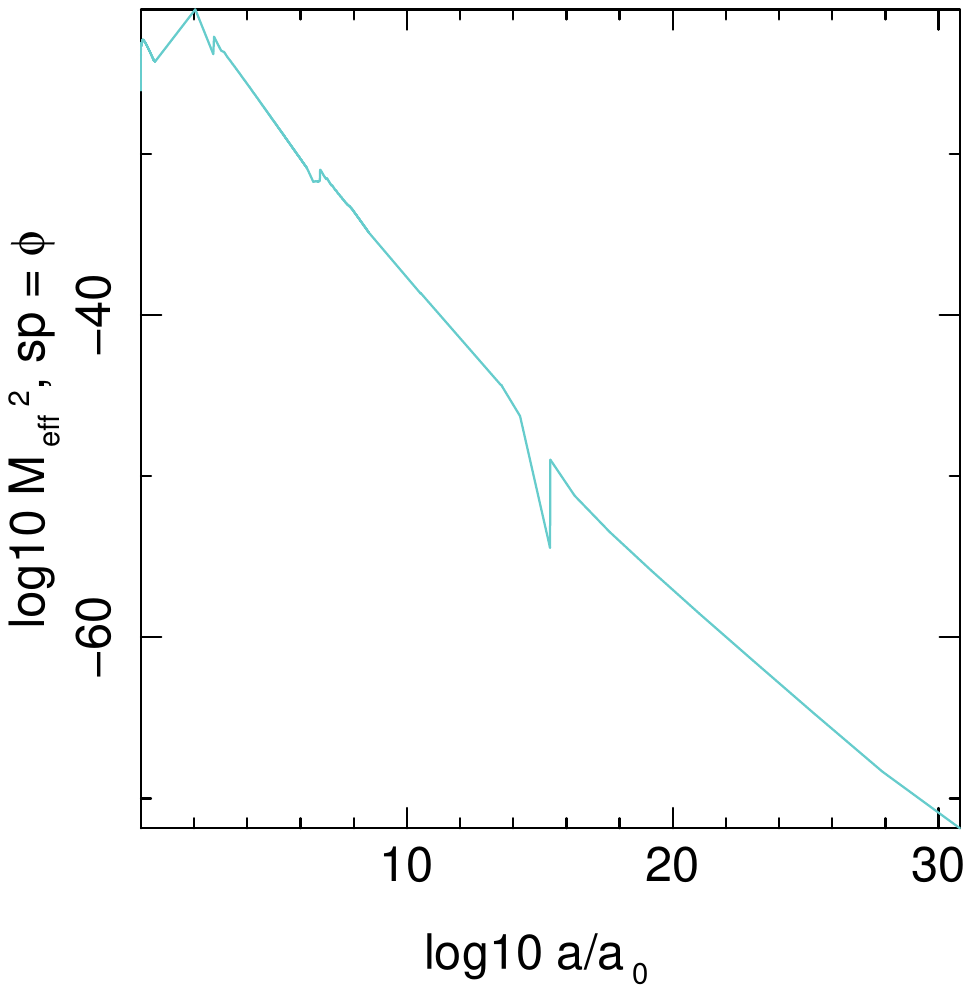} & \vspace {-1cm}\includegraphics[width=6cm,angle=90]{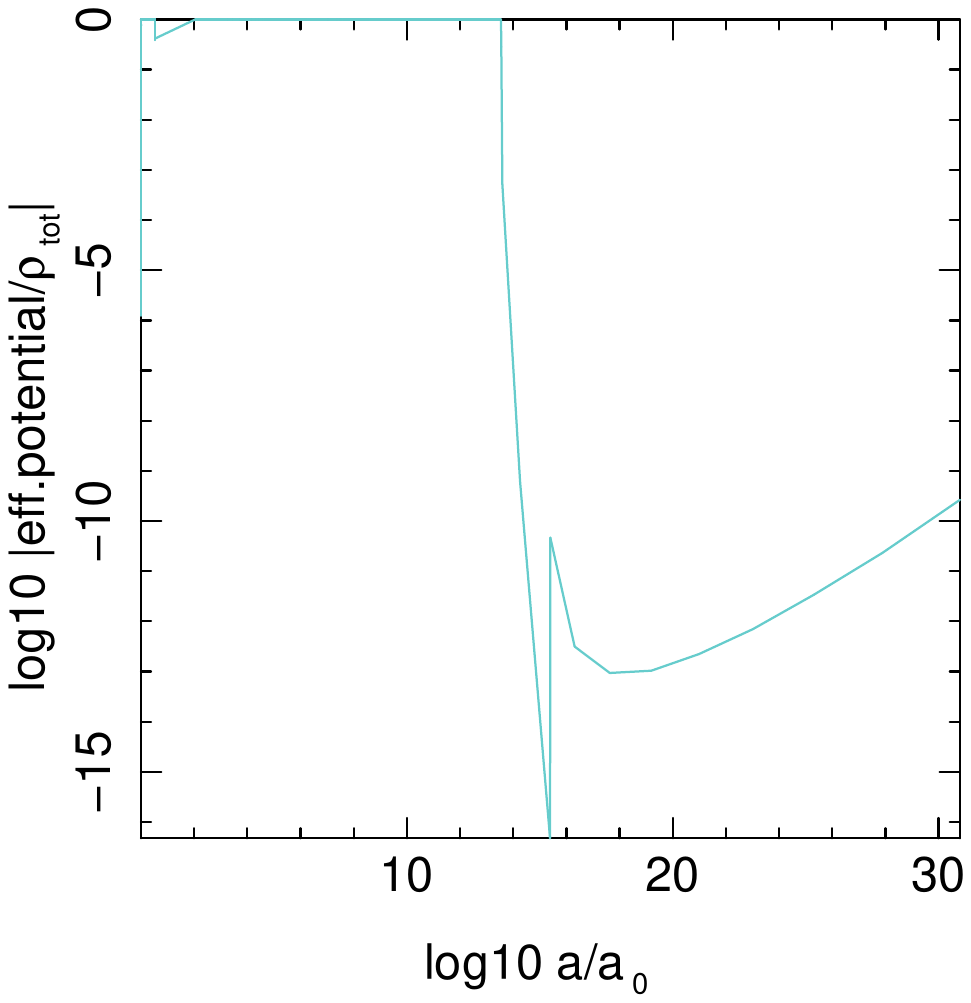} & \vspace {-1cm}\includegraphics[width=6cm,angle=90]{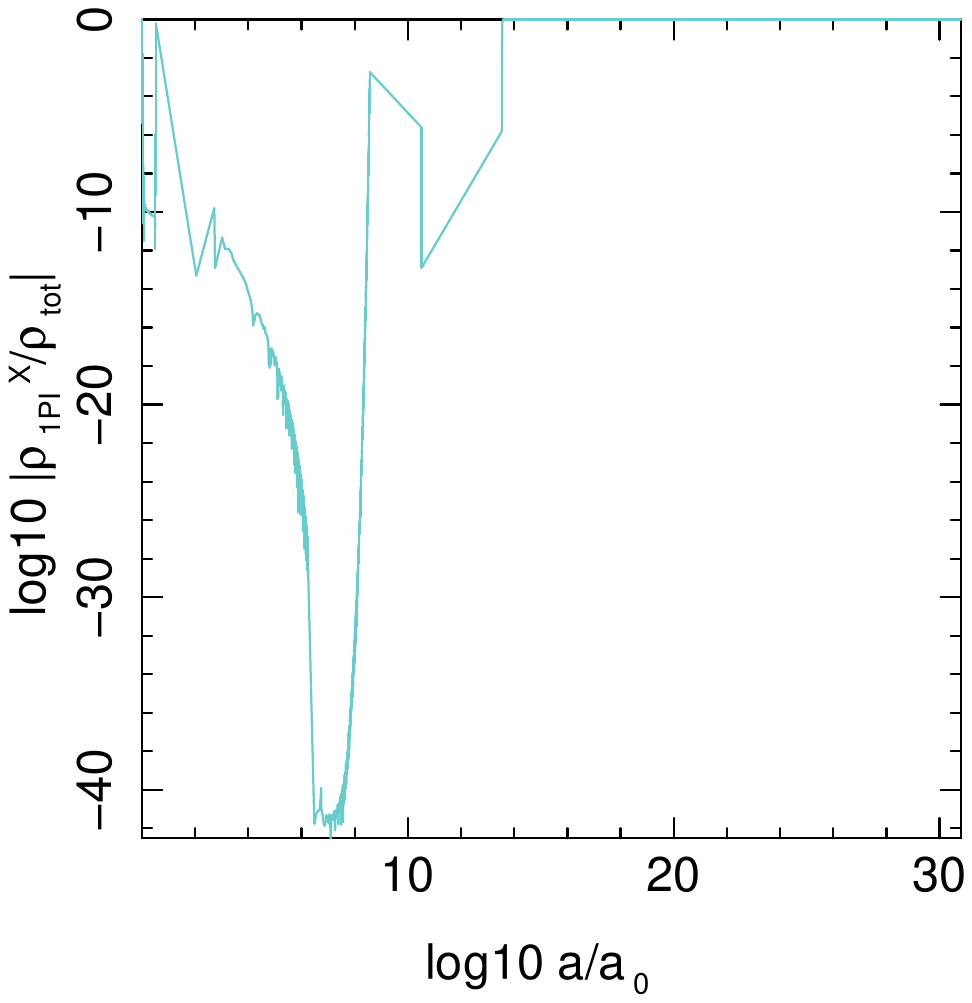}
\end {tabular}
\end {center}
\caption{a): Effective $M^2_\Phi (x=0)$; b) Fraction of zero mode, i.e homogeneous component of 
effective quantum binding energy density to total average density; c) Ratio of average $\rho^X_{1PI}$ 
to total density.\label{fig:effmass2}}
\end{figure}

We argue that the large mass difference between $X$ and $\Phi$ and self-interaction of the latter is 
enough to quickly initiate a cascade production of $\Phi$ particles. In turn, their interaction with 
energetically dominant but numerically rare $X$ particles transfers their energy to a non-equilibrium 
quantum binding energy corresponding to 2PI terms in energy-momentum tensor (\ref{effenermomdet}) or 
equivalently (\ref{rhotot}). Therefore, despite small couplings the state of matter during this 
era is non-perturbative and comparable with a strongly coupled {\it plasma}, which its instabilities 
lead to large oscillations - parametric resonance - of densities and Hubble function\footnote{This 
process is analogous to small-x regime of deep inelastic scattering at high energy, where large number 
of soft QCD gluons make the model effectively non-perturbative. Here light $\Phi$ field behaves 
similar to soft gluons.}. The large effective masses of $\Phi$ and its condensate shown in 
Fig. \ref{fig:effmass2}-a are manifestation of their effectively strong interaction during this 
non-perturbative regime.

It is useful to compare these results with an $O(N)$ model studied in strong coupling regime and 
Minkowski spacetime~\cite{qftfixedpoint}. It finds that in the non-perturbative regime, after initial 
parametric resonance, low momentum modes dominates and the system exhibits an effective weak coupling. 
Despite apparent contradiction, this finding is consistent with the above results, because due to the 
limited resolution of these simulations, initially they are concentrated on high momentum modes and 
parametric resonance of these modes~\cite{houricond} manifests itself in large oscillation of average 
density and Hubble function. However, with the expansion of the Universe, momentums are redshifted to 
the domain where coupling between fields are effectively weak and densities and Hubble function 
behave smoothly, as predicted by~\cite{qftfixedpoint}.

The strong effective coupling of fields does not last for long because at the same time 
the effective mass of $\Phi$ increases, see Fig. \ref{fig:effmass2}-a, and its backreaction 
decreases the rate of decay of $\rho^X_{1PI}$, see the inset in Fig. \ref{fig:pi1}-b and the amplitude 
of condensate field $\varphi$ in Fig. \ref{fig:rhos}-a. A slow decay rate of $\rho^X_{1PI}$ continues 
for some time before the latter and the density of $\varphi$ increase again, see 
Fig. \ref{fig:rhos}-b. Repetition of the same processes leads to oscillation of the total density 
$\rho_{tot}$ reflected in the oscillation of the Hubble function. However, due to the expansion 
of the Universe, gradually the amplitude of quantum binding energy, shown in Figs. \ref{fig:rhos}-c 
and its contribution to the total energy density, shown in Fig. \ref{fig:effmass2}-b, decreases and 
a slow evolution of total density leads to a power-law inflation. It is driven by the transfer of 
quantum binding energy to $X$ field, shown in Fig. \ref{fig:effmass2}-c.

As expected, the effective masses of $\Phi$ and its condensate $\varphi$ initially include a 
significant contribution from self-interaction and coupling with heavy field $X$. However, this 
effect is restricted to high energy modes, see the description of the spectrum of fluctuations in the 
next Sec. \ref{sec:simulspect}. We remind that accelerating expansion reduces the physical size 
of simulated modes $k/a(t)$ and the effect of local quantum corrections diminishes. 
Consequently, the effective mass of quantum fluctuations of $\Phi$ and its condensate $\varphi$ 
approaches its renormalized value at IR scale and $\varphi \rightarrow 0$. This may be an evidence 
that a shift symmetry for neutralizing the effect of quantum corrections on the mass of light field 
$\Phi$ would not be necessary, because rapid expansion automatically suppresses the effect of 
quantum corrections. This observation also shows the shortcomings of numerical simulations of 
cosmological models, which are unable - without increasing resolutions - to follow the evolution of 
growing distance scales with the same precision.

Apriori the effective mass of quantum and condensate components of $\Phi$ are not equal, see diagrams 
in Figs. \ref {fig:condensatediag} and \ref {fig:propagdiag}. However, in this series of simulations 
their difference is much smaller than numerical precision and Fig. \ref{fig:effmass2}-a presents 
the mass of both components. We should also remind that $A$ and $X$ fields have no self-interaction, 
and thereby no local quantum correction to their mass.

\subsubsection{Evolution of pseudo-free particles} \label{sec:simul1pi}
The behaviour of 1PI contributions of $A$ and $\Phi$ are very different from that of $X$, see 
Fig. \ref {fig:pi1}. Their densities $\rho^A_{1PI}$ and $\rho^\Phi_{1PI}$ vary much slower than 
$\rho^X_{1PI}$, which its variation with both time and expansion factor is very sharp and steep, and 
reminiscent to multiple first order phase transitions. In comparison, the densities of lighter fields 
behave similar to a slow and continuous second order phase transitions. Moreover, their variation is 
asynchronous with respect to $\rho^X_{1PI}$. The reason behind these differences is not only the large 
difference in  their mass, but also their interactions. Notably, due to its self-interaction $\Phi$ 
field which has the smallest mass, attains much higher densities than $A$. The systematically 
asynchronous onset of features and sign changes for different fields and components are the evidence 
that despite low quality of these simulation complex behaviour of fields and their properties must 
be grossly genuine and cannot be completely numerical effects. These features coarsely demonstrate 
the fully non-equilibrium nature of underlying processes. However, as we discussed at the beginning 
of this section and demonstrated with Figs. \ref {fig:inf-hubble}-a and \ref{fig:inf-hubble}-b, we 
do not rely on the details in our conclusions.

\begin{figure}
\begin{center}
\begin{tabular}{p{6cm}p{6cm}p{6cm}}
\hspace{2cm} a) & b) & \hspace{-2cm} c) \\
\hspace{2cm}\vspace {-1cm}\includegraphics[width=6cm]{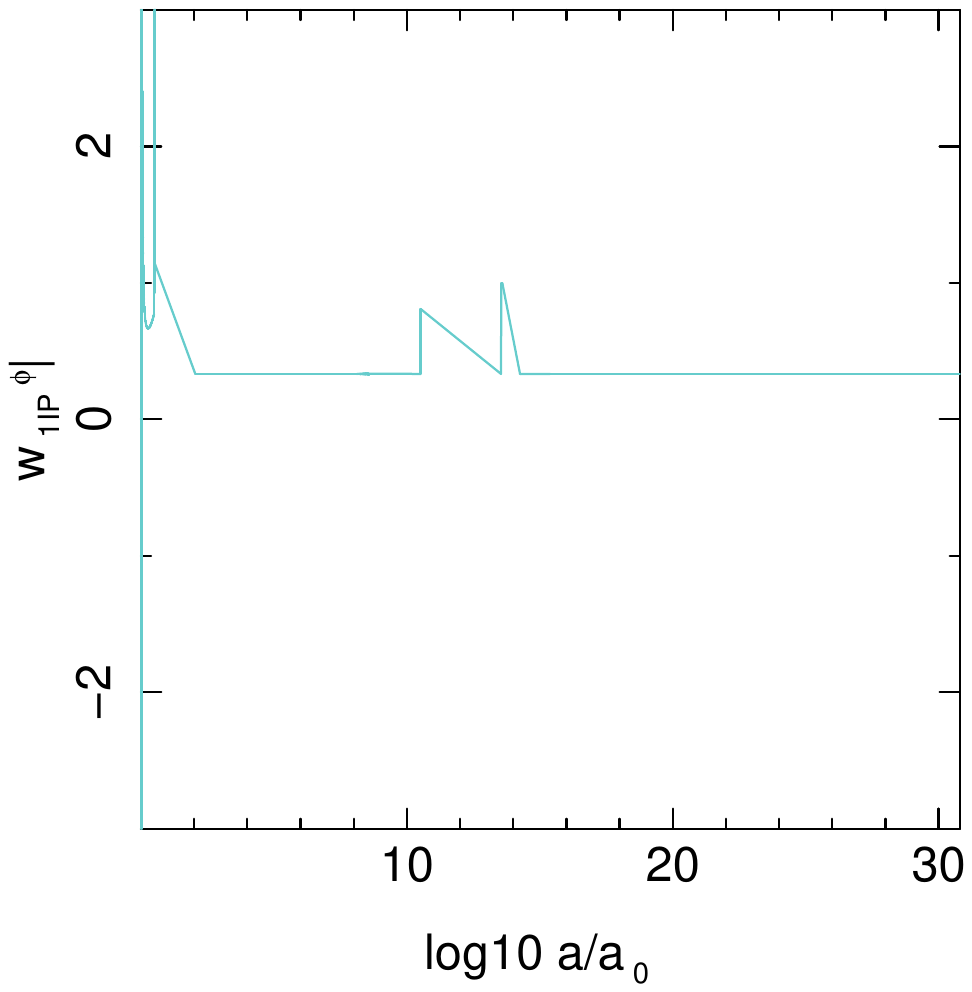} & \vspace {-4.2cm}\includegraphics[width=6cm]{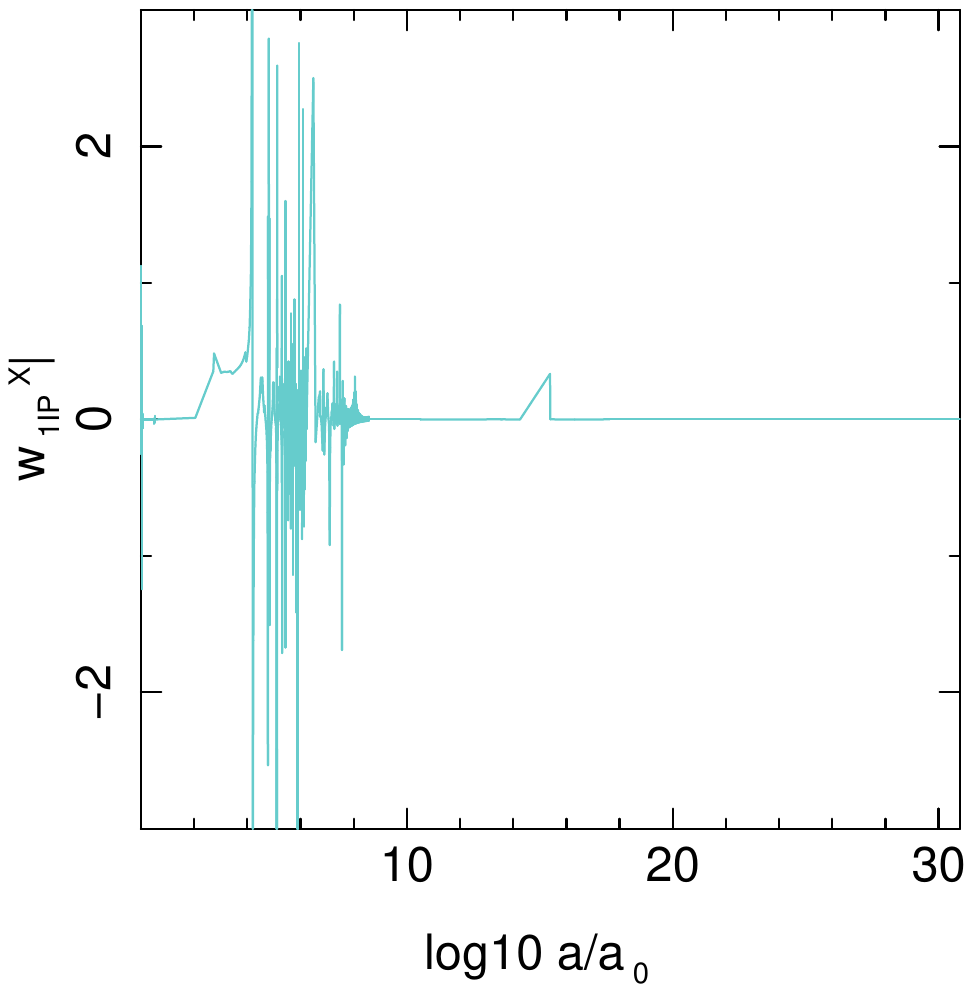} & \hspace{-2cm}\vspace {-1cm}\includegraphics[width=6cm]{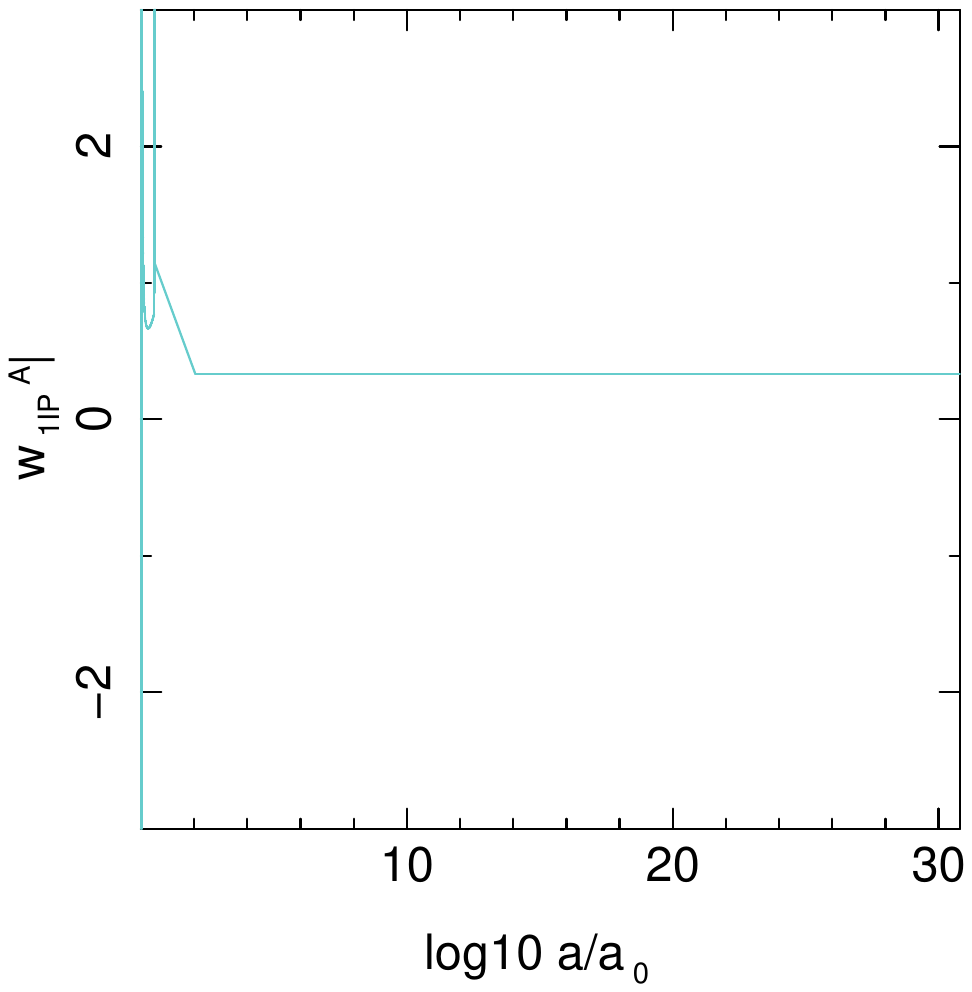}
\end {tabular}
\end {center}
\caption{a), b), c): Equation of state for 1PI components of the energy-momentum tensor of $\Phi$, 
$X$ and $A$, respectively.\label{fig:w1pi}}
\end{figure}
Another interesting characteristic of 1PI contributions of the fields is the negative sign of 
$\rho^i_{1PI},~i=X,A,\Phi$ in some era. This means that these components of total density cannot 
be considered as belonging to truly {\it free} particles. Nonetheless, after initial instabilities, 
their equation of state defined as: $w^i_{1PI} \equiv p^i_{1PI} / \rho^i_{1PI}$, approaches to zero for 
$X$ and to $1/3$ for $A$ and $\Phi$, see Fig. \ref{fig:w1pi}. Thus, they behave similar to 
non-relativistic and relativistic free particles, respectively. This observation justifies the 
interpretation of $\rho^i_{1PI},~i=\Phi,X,A$ as {\it pseudo-free} particles and shows that the 
process of inflation and particle production are inseparable. Because in this toy model $X$ is much 
heavier than other fields and its lifetime as free particles is very long, at the last stages of 
inflation it dominates as a cold matter. However, it is conceivable that if its life time is shorter, 
at the end of inflation light fields become dominant and induce a radiation domination era, as 
expected in a hot Big Bang model. We did not study such a case.

\subsubsection{Effective potential and condensate} \label{sec:simuleffpoten}
\begin{figure}
\begin{center}
\begin{tabular}{p{6cm}p{6cm}}
a) & b) \\
\hspace{0.65cm} \includegraphics[width=4.75cm]{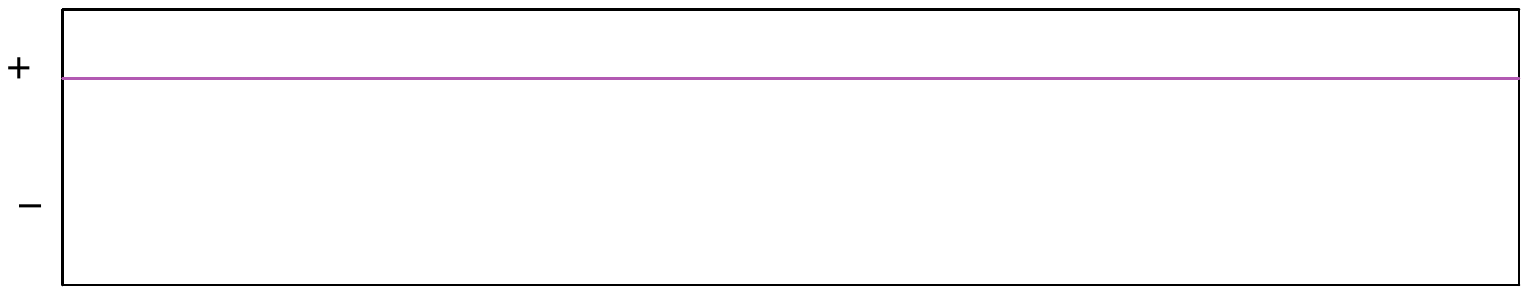} & \hspace{0.65cm} \includegraphics[width=4.75cm]{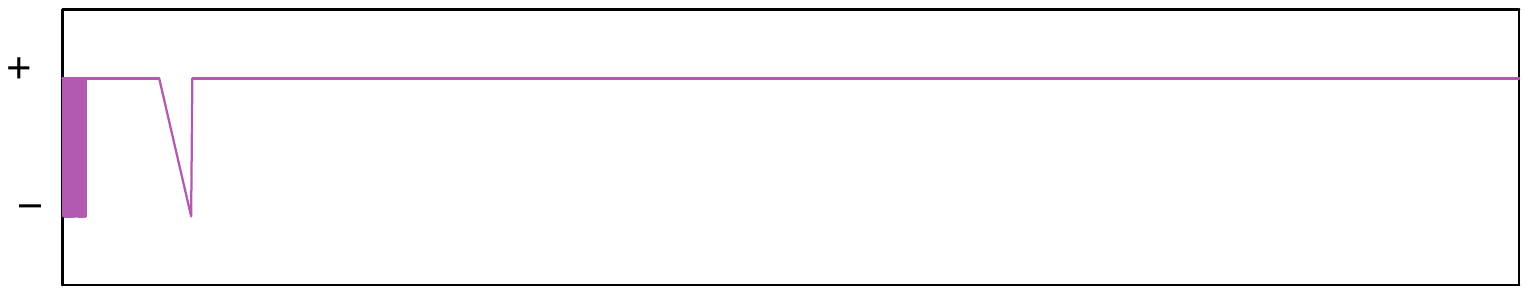} \\
\vspace{-3.9cm}\includegraphics[width=6cm,angle=90]{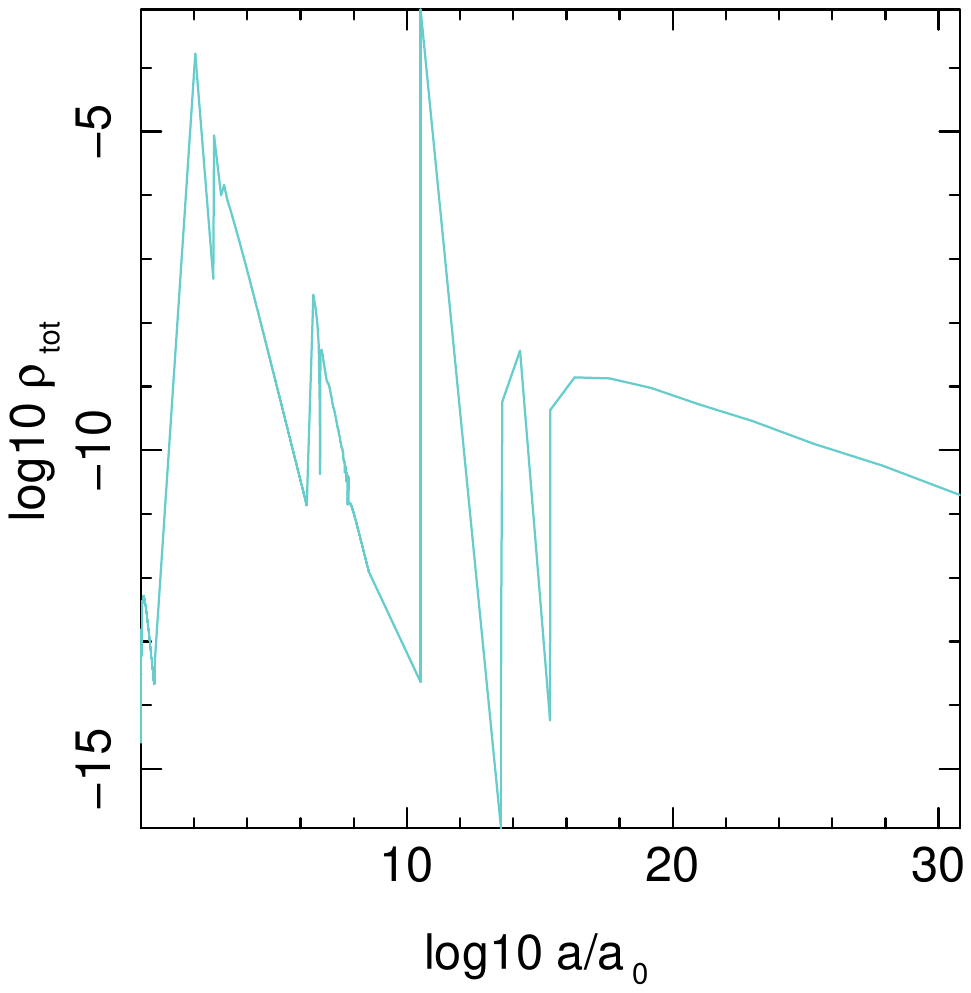} & \vspace{-3.9cm}\includegraphics[width=6cm,angle=90]{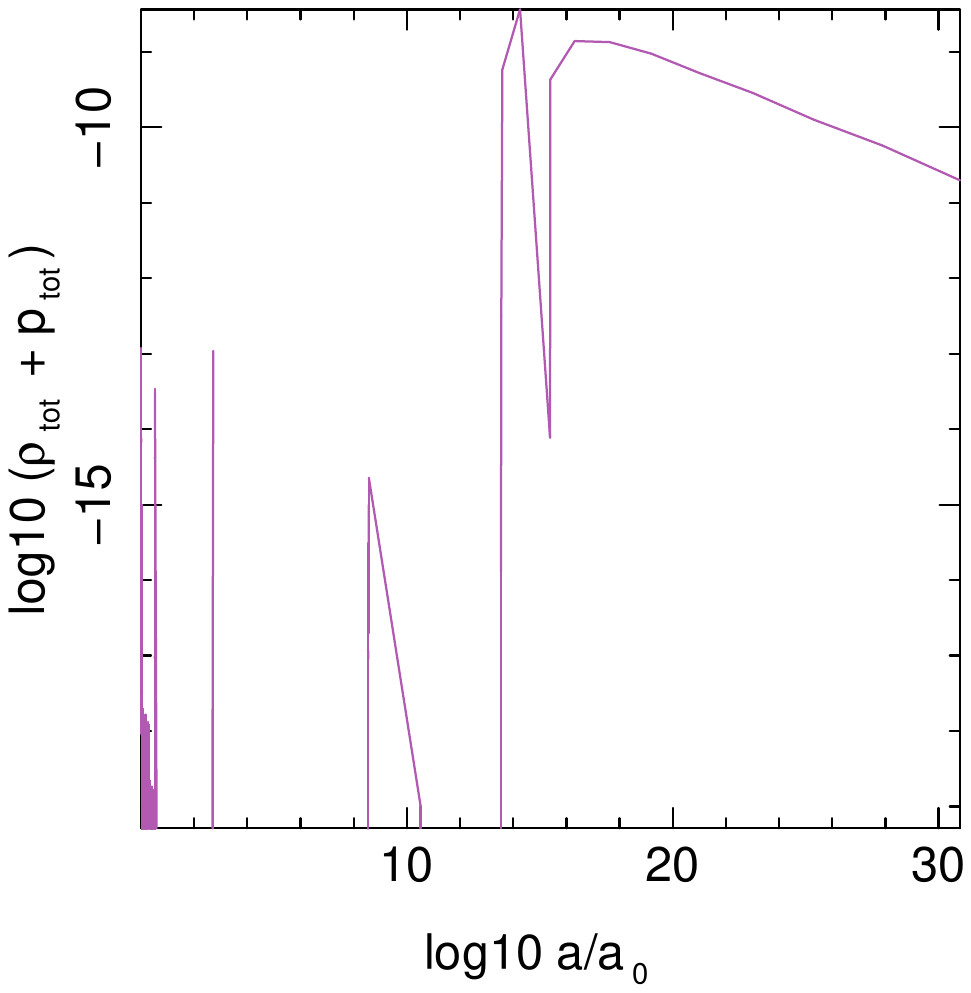}
\end {tabular}
\end {center}
\caption{a): Total energy density; b) Total specific enthalpy $\rho + P$. Upper plots: sign; Lower 
plots: amplitude. \label{fig:totrho}}
\end{figure}

Total density $\rho$ and specific enthalpy, defined as $\rho + p$, are shown in Fig. \ref{fig:totrho}. 
They are both positive (up to numerical errors for the latter) and there is no violation of null energy 
principle in the simulations. 

Comparison of condensate density shown in Fig. \ref{fig:rhos}-b with the total density and other 
components of energy-momentum demonstrates that its contribution is completely negligible. Moreover, 
the comparison of Figs. \ref{fig:rhos}-b and \ref{fig:rhos}-c shows that after the onset of inflation, 
assumed to be at $log (a/a_0) \gtrsim 15$, the energy density of condensate $\rho_\varphi$ is dominated 
by its kinetic energy (not shown here) rather than its effective potential. These observations are 
consistent with approximate analytical results reported in~\cite{houricond}, which show that the 
condensate can grow during radiation domination era, when expanding is relatively slow. But it decays 
during matter domination and by extension during inflation eras, which have faster expansion rate.
 
Figs. \ref{fig:effmass2}-b and \ref{fig:effmass2}-c show that in these simulations inflation is 
supported by the decay of quantum binding energy to particles. They also indicate that during inflation 
a significant fraction of quantum binding energy goes to the formation of $X$ particles. However, this 
may be in part due to the stiffness of the model and imprecision of simulations, which capture more 
easily the heavy $X$ particles rather than lighter fields. Nonetheless, as mentioned earlier, this 
observation is consistent with some analytical calculations for simpler models~\cite{desitterinstab1}. 
Moreover, early works and some recent studies of the evolution of scalar quantum fields in an expanding 
universe show particle production processes and their impact on the expansion~\cite{infparticleprod}. 
Therefore, our results may be a confirmation of previous studies. In any case, this aspect of the model 
needs confirmations by better simulation. on the other hand, in contrast to some 
studies~\cite{preinfnongauss}, the initial large oscillations of densities do not leave observable 
oscillations in the dominant matter component $X$ - presumably dark matter - at late times. Further 
arguments in favour of claim will be given in Sec. \ref{sec:simulspect}.

The above results and observations indicate that the relation between properties of inflation 
parameters $\epsilon_i$, extracted from observations, and characteristics of the underlying 
model, e.g. self-interaction of inflaton field, is not straightforward. For instance, although the 
light field $\Phi$ and its condensation have very important role in the control of quantum processes 
which lead to inflation, its contribution in the classical effective energy-momentum density may be 
insignificant. Moreover, in contrast to single field monomial models, the energy density 
of condensate $\varphi$ during inflation may be dominated by its kinetic energy rather than its 
potential. However, this property would be undetectable from $\epsilon_i$. In slow-roll monomial 
models of inflation by definition the potential energy must dominate the energy density. However, 
as the inflation in the model studied here is conducted by other components, this is not a 
necessary condition for making the model consistent with observations. Nonetheless, the 
dominance of kinetic energy of the condensate has an impact on the spectrum of fluctuations of the 
condensate, which we will discuss in Sec. \ref{sec:simulspect}.

\subsubsection {Stronger self-coupling} \label{sec:simulstrongcoupl}
\begin{figure}
\begin{center}
\begin{tabular}{p{6.cm}p{6.cm}p{6.cm}}
\hspace{-1cm} a)\includegraphics[width=6cm,angle=90]{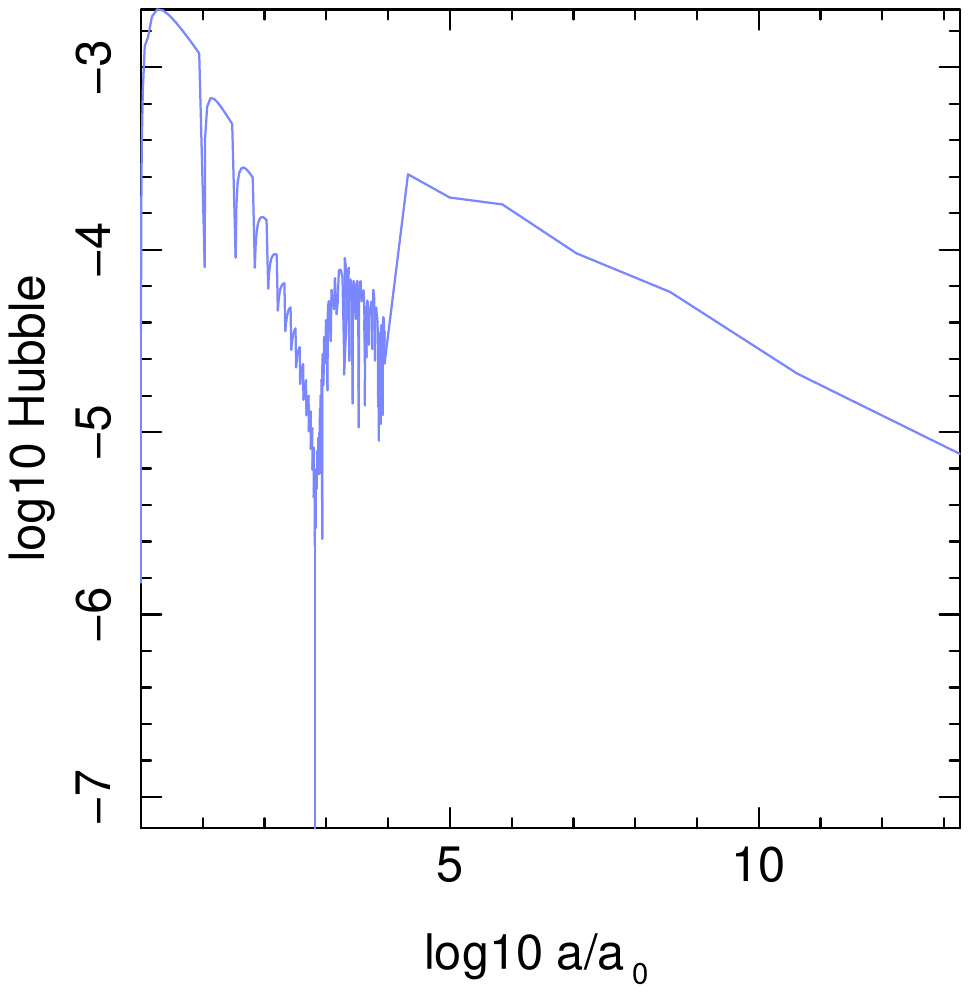} & 
\hspace{-1cm} b)\includegraphics[width=6cm,angle=90]{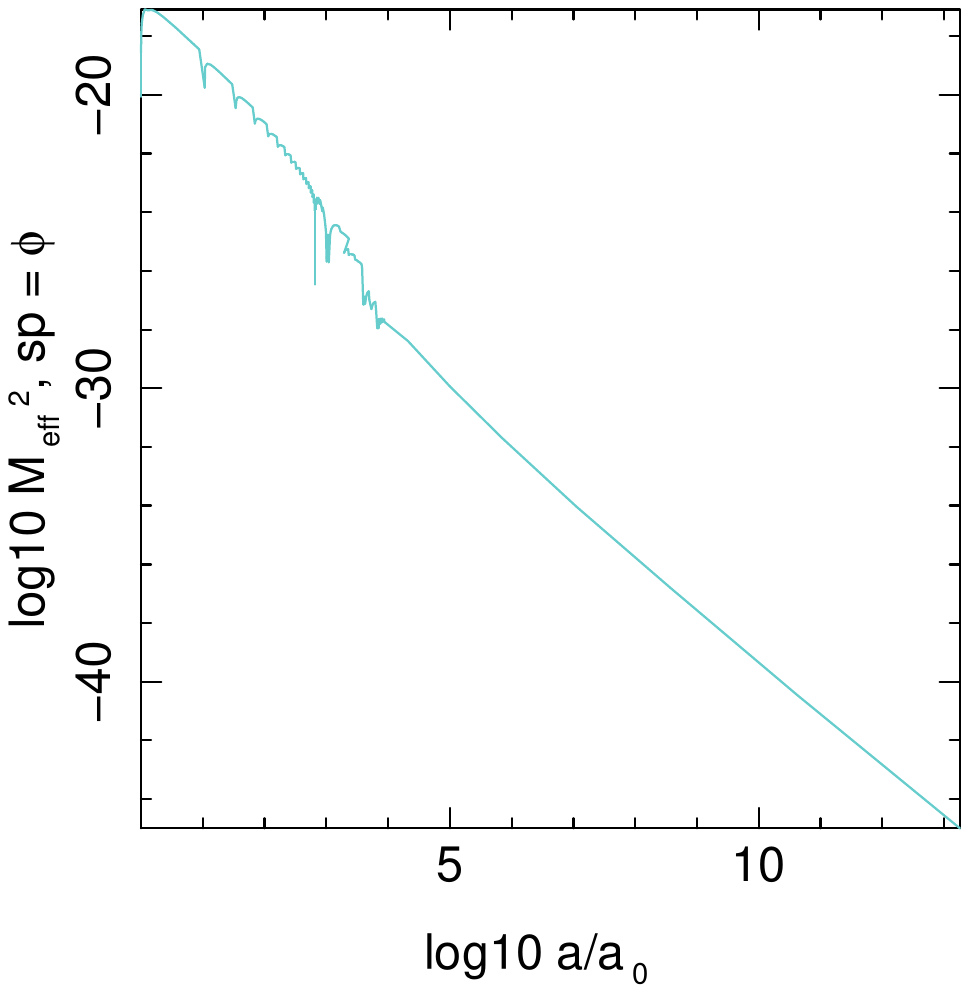} & 
\hspace{-1cm} c)\includegraphics[width=6cm,angle=90]{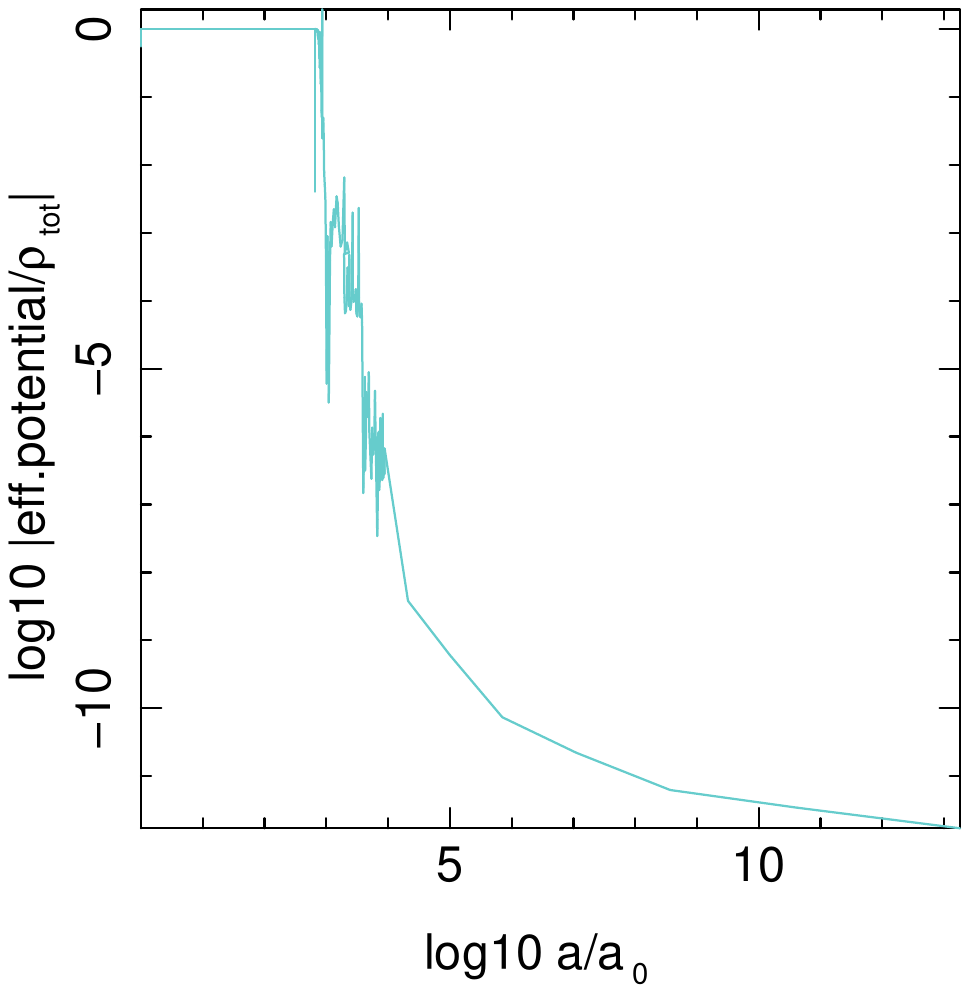} \\
\hspace{-1cm} d)\includegraphics[width=6cm,angle=90]{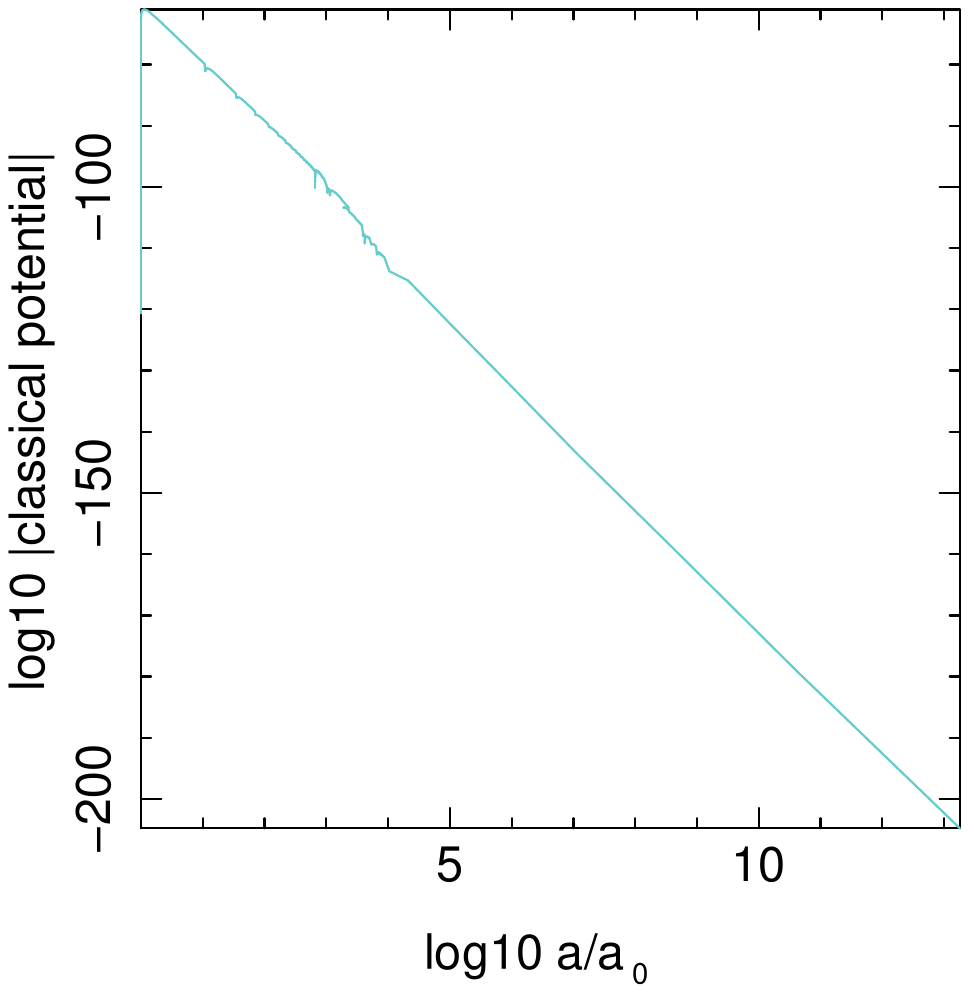} & 
\hspace{-1cm} e)\includegraphics[width=6cm,angle=90]{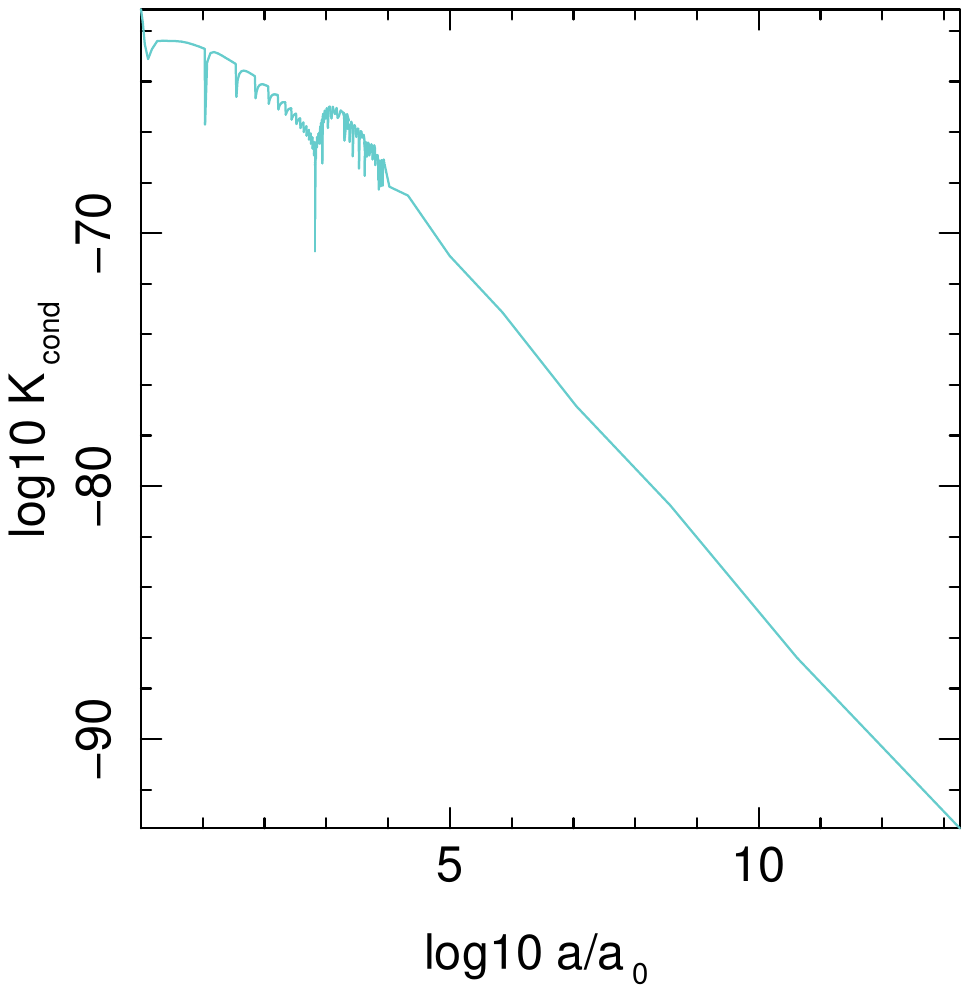} & 
\hspace{-1cm} f)\includegraphics[width=6cm,angle=90]{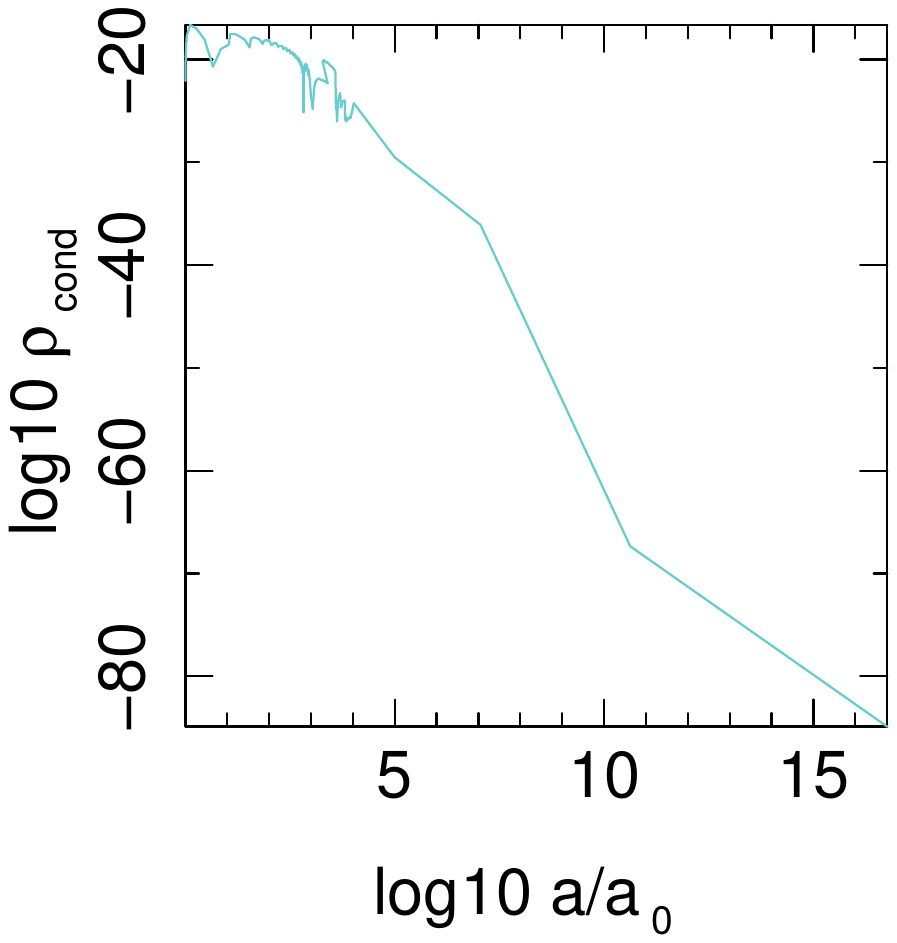} 
\end {tabular}
\end {center}
\caption{Properties of the model with $\lambda = 10^{-8}$. a) Hubble function; b) Effective mass 
$M^2_\Phi$\; c) Ratio of quantum binding energy to total energy density; d) Classical potential of 
condensate $\varphi$; e) Effective potential of condensate; f) Effective energy density of 
condensate $\varphi$;  \label{fig:lambda8}}
\end{figure}

For the sake of comparison Fig. \ref{fig:lambda8} shows the 
evolution of Hubble function, effective mass of $\Phi$, ratio of quantum 2PI binding energy to total 
energy density, and properties of the condensate $\varphi$ for a model with $\lambda = 10^{-8}$ and 
other parameters the same as the simulations discussed above. In this case the estimation of initial 
rate of condensate formation (\ref{dissiprate}) gives $\Gamma_\varphi / H \gg 1$ at initial time. 
Therefore, it is expected that condensate has a more significant contribution in the total energy 
density of the Universe. Indeed, we observe significant differences between properties of condensate 
in this model and simulations with $\lambda = 10^{-14}$. Notably, the heavy particle production and 
inflation begin much earlier. This is due to the higher effective mass of $\Phi$ particles at a 
given epoch, that is a fix $a$, see Fig. \ref{fig:lambda8}-b. However, although the larger coupling 
constant increases classical potential energy, it remains much smaller than quantum binding energy 
and kinetic energy of the condensate shown in Fig. \ref{fig:lambda8}-e. On the other hand, the 
effective energy density of condensate, shown in Fig. \ref{fig:lambda8}-f, is dominated by quantum 
corrections during inflation, and consequently its equation of state $w_\varphi \sim -1$. Nonetheless, 
during inflation $\rho_\varphi$ is not constant and decreases as $\sim a^{-2}$. Apparently, this 
violates the usual relation between $w$ and evolution of density with expansion factor. However, 
the density of condensate alone is not conserved and its interaction with other components of the 
model must be taken into account. If the condensate continues the same trend after inflation, 
it cannot be a candidate for dark energy. However, at the end of inflation if its decline slows 
down and its density asymptotically approaches to a constant density, as analytical approximations 
has shown~\cite{houricond}, the small leftover may explain the observed accelerating expansion of 
the Universe at present era. Unfortunately, for the time being simulations cannot be extended to 
these late epochs with enough precision to capture these details. 

These observations raise the issue of the end of inflation. In the present simulations we do not 
observe an end to inflation. However, based on earlier behaviour of model we expect that a 
change in the contribution of different components of energy-momentum induces again a 
{\it phase transition}. An evidence for such behaviour is the gradual increase of quantum binding 
energy at the end of simulations in Figs. \ref{fig:rhos}-d and \ref{fig:lambda8}-c. However, we cannot 
be sure that this is not an artefact, specially because it includes only a few time steps in our 
simulations.

\subsubsection{Spectrum of fluctuations} \label{sec:simulspect}
Although horizon flow and its derivatives $\epsilon_i,~i=0,1,2,3,\cdot$ are usually used for 
parametrizing inflation models, only for the simplest among them, in particular a single scalar 
field in slow-roll regime, they can be considered as reliable proxies for spectrum of primordial 
fluctuations. Therefore, for a stiff multi-field model in a non-equilibrium state, as the one 
discussed here, it is better to investigate the spectrum of fluctuations directly.

\begin{figure}
\begin{center}
\begin{tabular}{p{10cm}p{10cm}}
\vspace{0.5cm} a) & \vspace{0.5cm} b) \\
\hspace {-1cm}\includegraphics[width=14cm]{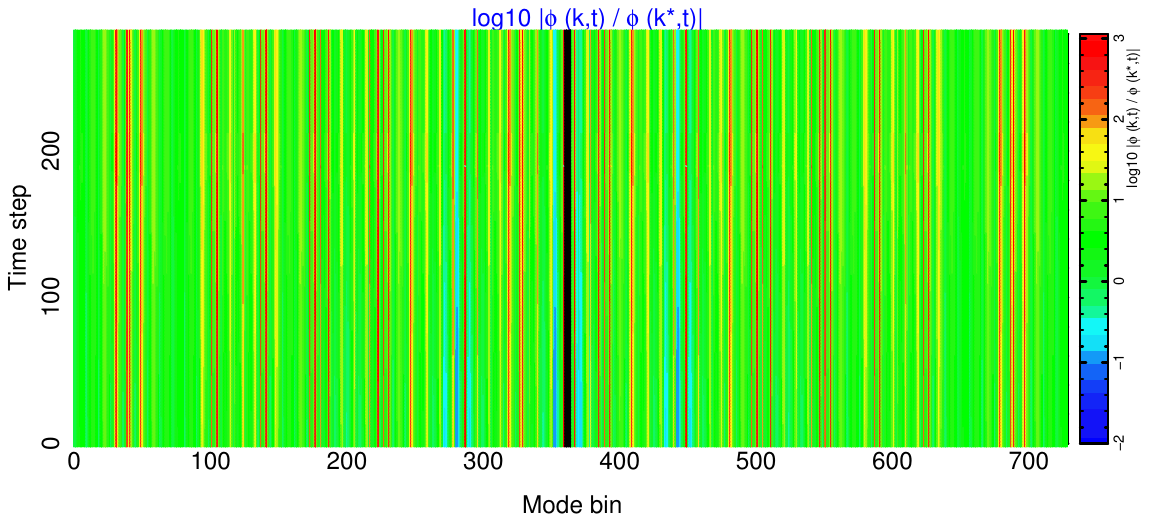} &
\hspace {-2cm}\includegraphics[width=14cm]{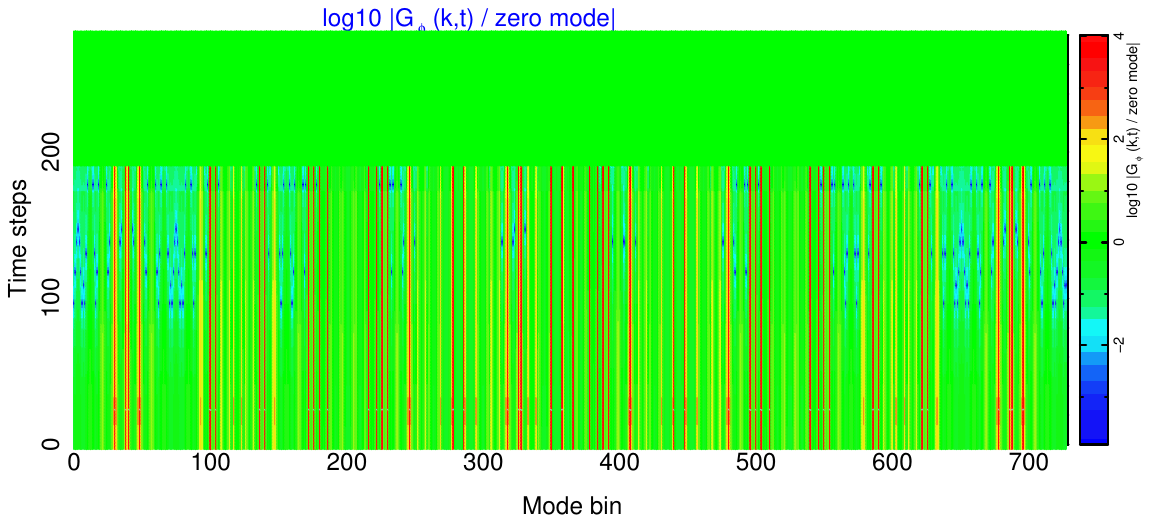} \\
\vspace{-5cm} c) & \vspace{-5cm} d) \\
\vspace{-6cm}\hspace {-1cm}\includegraphics[width=10cm,angle=90]{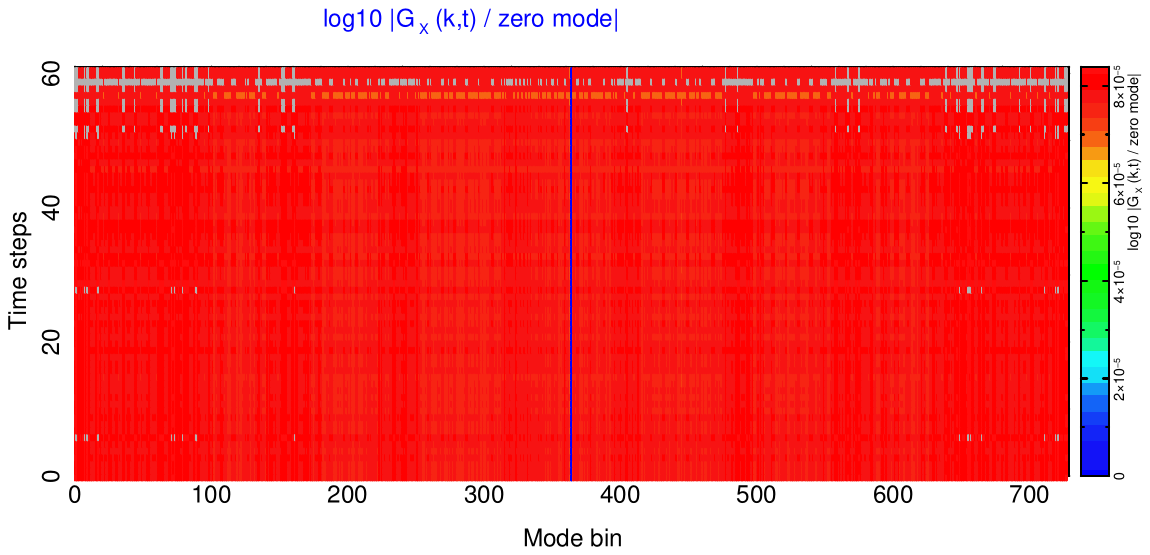} & 
\vspace{-6cm}\hspace {-2cm}\includegraphics[width=14cm]{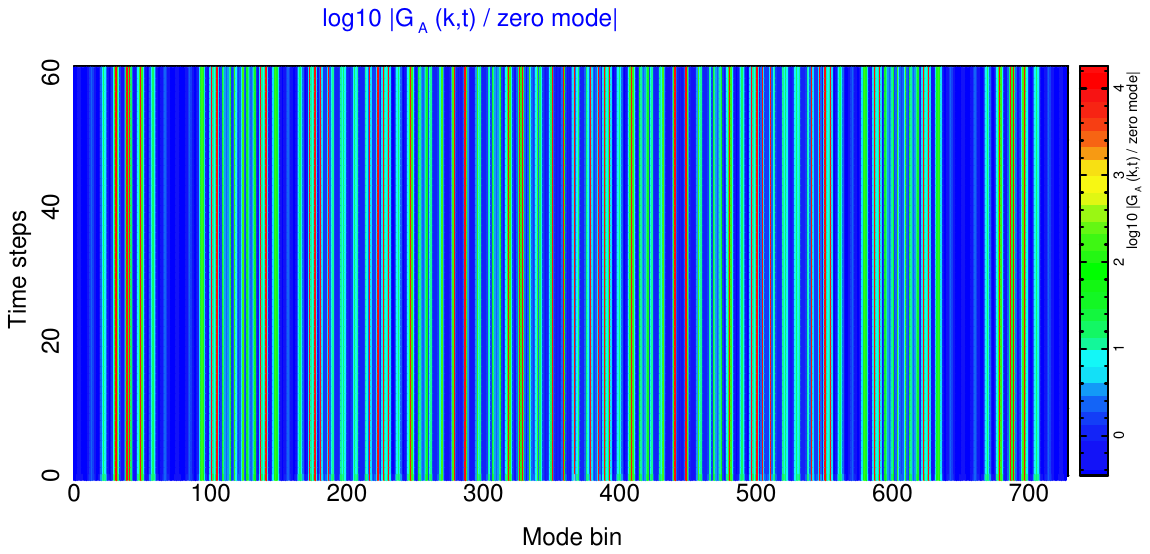}
\end {tabular}
\end {center}
\vspace{-5cm}\caption{a) Color coded normalized spectrum of condensate $\varphi (k,t) / 
\varphi (k^*,t)$ where $k*$ has the largest amplitude in the simulation box. b), c) and d) Color 
coded normalized exact propagators $G^F_i(k, t) / G^F_i(k=0, t),~i=\Phi,~X,~A$, respectively. 
The x-axis presents a cube of $9^3$ channels in the mode space. They are arranged such that $|k| = 0$ 
corresponds to channel $364$. An example of the 3D modes is shown in Fig. \ref{fig:t00spect}-f. 
The value of $k$ in this plot is with respect to conformal coordinate and does not depend on time. 
The corresponding physical (comoving) mode is $k/a$. The y-axis presents simulation time steps as 
explained in footnote \ref{foottimestep}. To better highlight variation in amplitude of modes in a) 
and b) all the registered time intervals are used, but c) and d) show only data during inflation. 
The apparently abrupt change in the spectrum of $G^F_\Phi(k, t) / G^F_\Phi(k=0, t)$ is partly because 
of adaptive time steps which varies in different stages of simulation, and partly due to the 
absence of some intervals from plots, as explained in footnote \ref{foottimestep}. \label{fig:propag}}
\end{figure}

\begin{figure}
\begin{center}
\begin{tabular}{p{8cm}p{8cm}}
a) & b) \\
\hspace{-1.5cm}\includegraphics[width=10cm,angle=90]{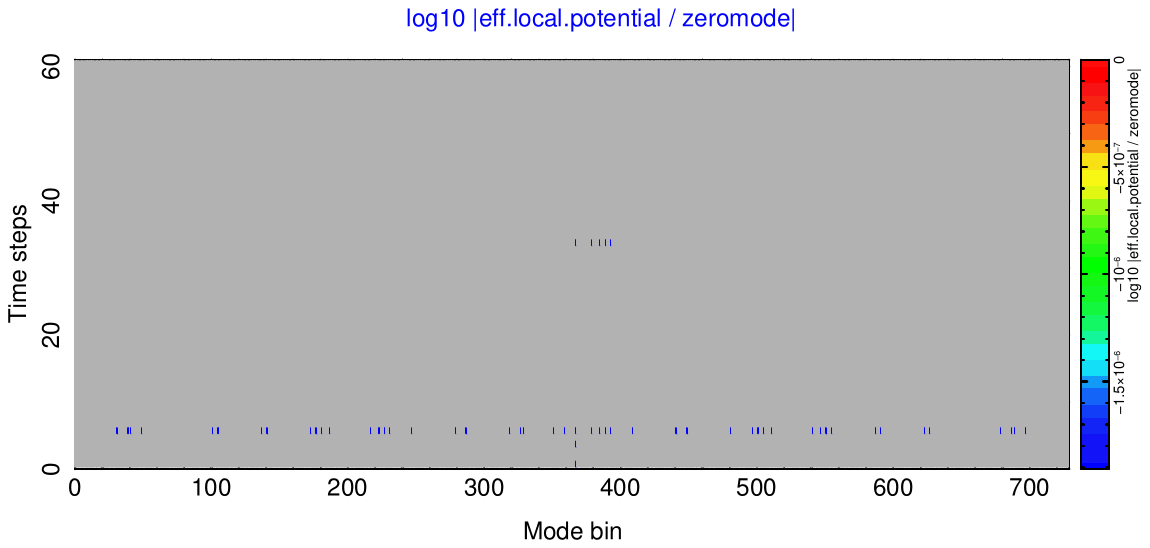} &
\hspace{-0.5cm}\includegraphics[width=10cm,angle=90]{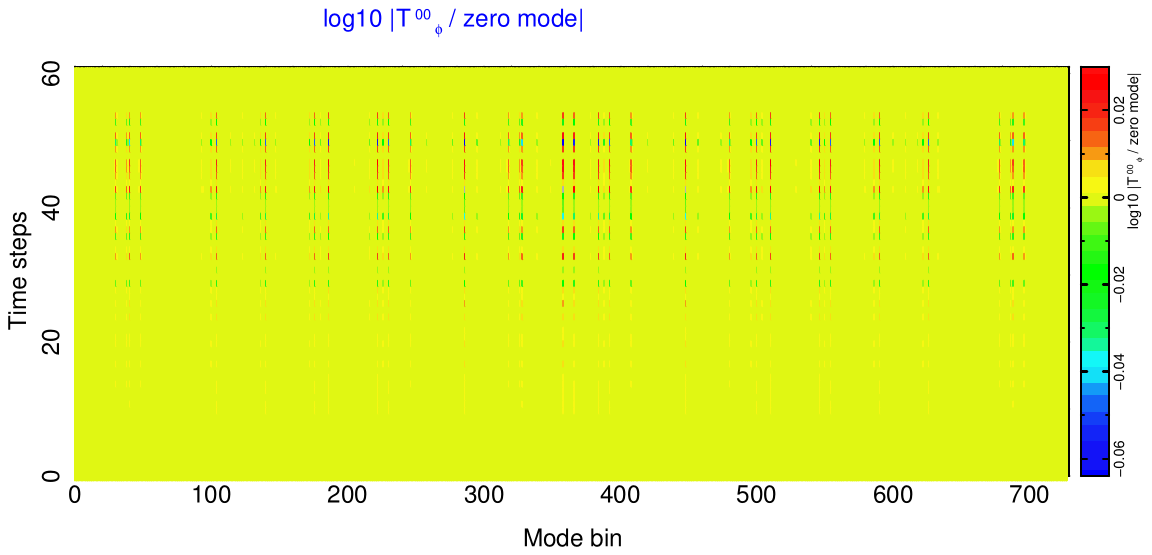} \\
\vspace{-4.5cm} c) & \vspace{-4.5cm} d) \\
\vspace{-5cm}\hspace{-1.5cm}\includegraphics[width=10cm,angle=90]{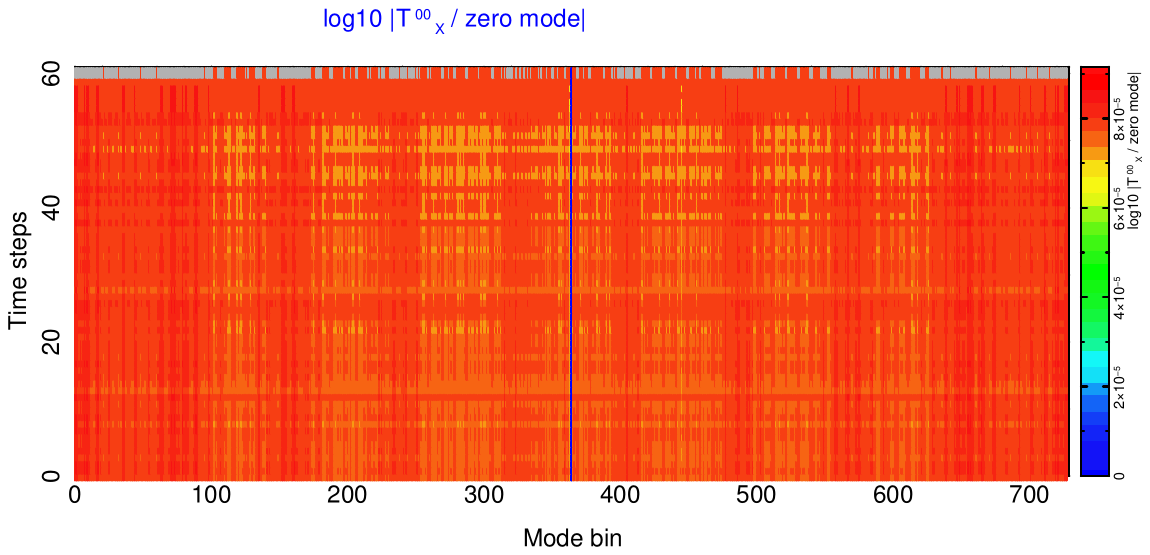} & 
\vspace{-5cm}\hspace{-0.5cm}\includegraphics[width=14cm]{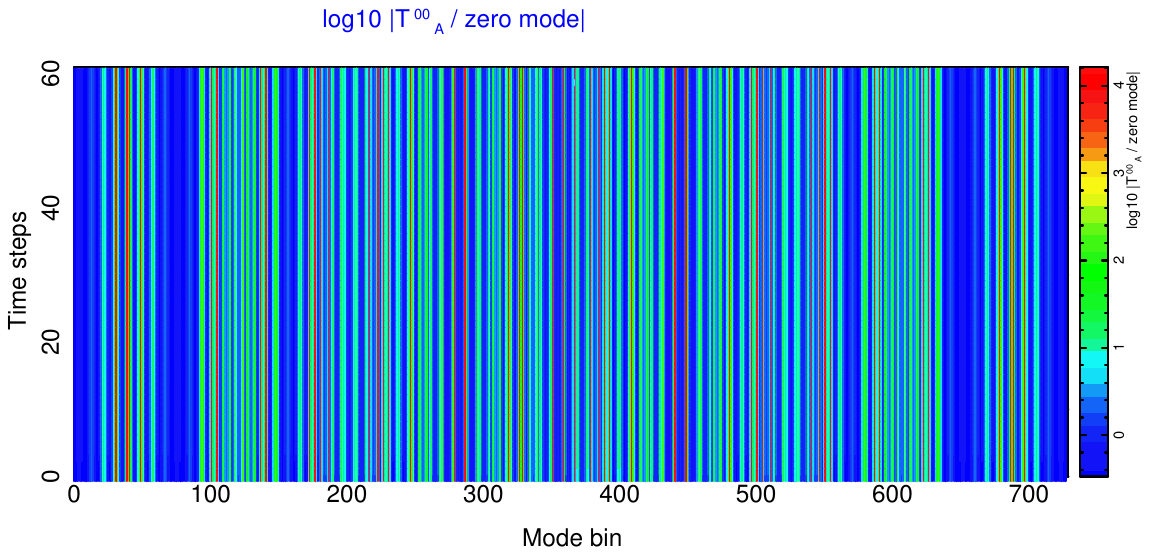} \\
\vspace{-4.5cm}e) & \vspace{-4.5cm}f)\\
\vspace{-5cm}\hspace{-1.5cm}\includegraphics[width=14cm]{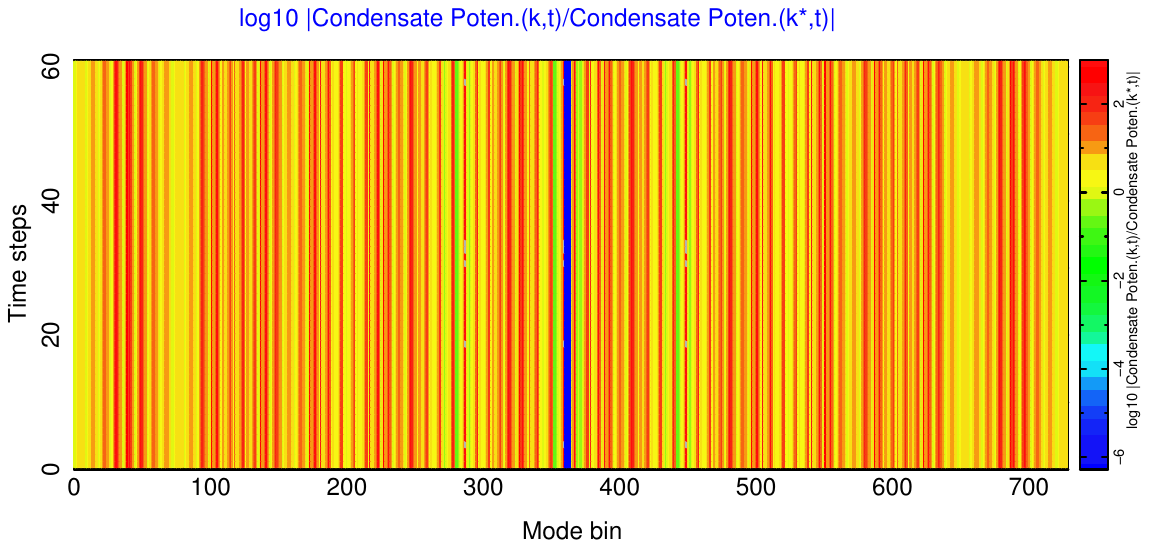} & 
\vspace{-5cm}\hspace{1cm}\includegraphics[width=9cm]{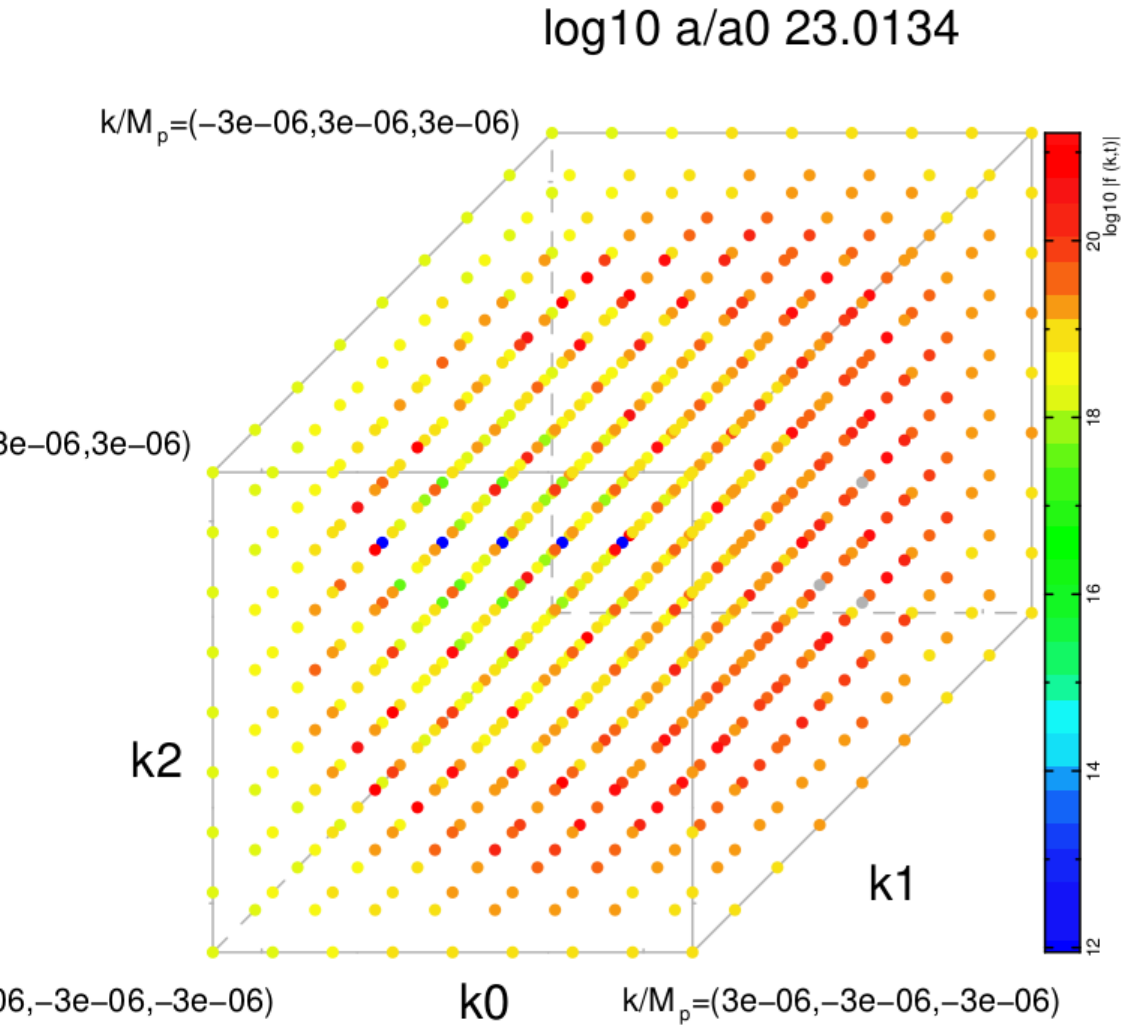}
\end {tabular}
\end {center}
\vspace{-4cm}\caption{Spectrum of energy density components: a) 2PI quantum binding energy density 
normalized to its zero mode. Color gray corresponds to upper limit, which in this plot is zero and 
corresponds to an scale invariant amplitude; 
b), c), d) $T^i_{1PI} (k, t) / T^i_{1PI} (k=0, t),~i=\Phi,~X,~A$, respectively; e) energy density of 
condensate normalized to its zero mode. Description of axis is 
the same as in Fig. \ref{fig:propag}; f) Color coded amplitude of $a\varphi$ modes in the mode cube 
during one simulation time step. Values of modes are in $M_p$ unit.\label{fig:t00spect}}
\end{figure}

Fig. \ref{fig:propag} shows the evolution of normalized exact propagators $G^F_i(k,t) / G^F_i(k=0,t),
~i=\Phi,X,A$ during inflation era, calculated numerically up to second perturbative order and under 
approximations discussed at the beginning of this section\footnote{For reducing the volume of output 
during simulations we registered the data for 1 out of $n$ time steps, which the value of $n$ 
depended on the length of simulations. What is called {\it time step} in the spectrum plots 
corresponds to registered steps rather than real time steps, which was much larger. 
\label{foottimestep}}. We remind that for free fields $G^F_i(k,t) \equiv 2N_k + 1$ where $N_k$ is 
proportional the expectation value of particle number in mode $k$. For interacting fields $N_k$ can 
be considered as an effective number. Fig. \ref{fig:t00spect} shows the spectrum of fluctuations 
of various components of $T^{00}$, which for a homogeneous background metric corresponds to energy 
density. 

As we discussed in Sec. \ref{sec:simul1pi}, at late stages of inflation the density of the Universe 
is dominated by $X$ particles, or more precisely the 1PI component of energy-momentum tensor. 
The first conclusion from these plots is that the amplitude of fluctuations in this model 
is ${\mathcal O}(1) \times 10^{-5}$, thus consistent with observations. Moreover, the evolution of 
fluctuations is very close to adiabatic, defined as $\delta N_k(a(t)) / N_0(a(t)) = const.$, 
see e.g.~\cite{adiabaticdef} for review. The amount of variation with time of this quantity, assumed 
as presenting isocurvature fluctuations, is $\lesssim 10\%$. Considering low resolution of our 
simulations, this value is roughly consistent with the Planck constraint on the fraction of 
isocurvature perturbation of a few percents~\cite{infplanck}.

Fluctuations of 2PI quantum corrections are very small even at early stages of inflation. This 
reflects the non-local nature of this component, which couple different scales together and wash 
out their differences, even when they are superhorizon. 1PI fluctuations of $X$, which is the 
dominant component at the end of our simulations, is of order $\mathcal{O}(1) \times 10^{-5}$ and 
comfortably consistent with observations. However, we observe significant fluctuations in the energy 
density of condensate $\varphi$ and in 1PI contributions of $\Phi$ and $A$ fields. 1PI components 
of $\Phi$ and $A$ include tree Feynman diagrams with $X$ as interaction field. Considering the 
large mass of $X$ particles, these diagrams can be approximated by local interactions. For this 
reason the fast expansion of the Universe during inflation suppresses interaction at superhorizon 
scales and only induces oscillations at shorter scales. This process is analogous to Doppler peaks 
in the present power spectrum, generated by interaction of photons with baryons and free streaming. 
However, in contrast to baryons, the contribution of $\Phi$ and $A$ in the total energy density and 
their fluctuations are highly subdominant and would not be observable. On the other hand, as we 
discussed in Sec. \ref{sec:simuleffpoten} the dominance of $X$ at the end of these simulations may 
be due to the stiffness of the model. Moreover, a larger coupling $\mg$ decreases the lifetime of 
$X$ particles and thereby their final density. Then, the self-interaction of $\Phi$ and its 
interaction with $A$ exchanging virtual heavy $X$ particles should uniformize their fluctuations 
during a regime analogous to reheating. On the other hand, if large fluctuations survive reheating, 
they may have some effect at small distant scales as seeds and contribute in the formation of 
galaxies and/or supermassive black holes\footnote{We did not investigate whether these large 
fluctuations may lead to formation of primordial black holes. Such inquiry needs much better 
spectral resolution.}.

\begin{figure}
\begin{center}
\begin{tabular}{p{6cm}p{6cm}p{6cm}}
a) & b) & c) \\
\hspace{-1cm}\includegraphics[width=10cm]{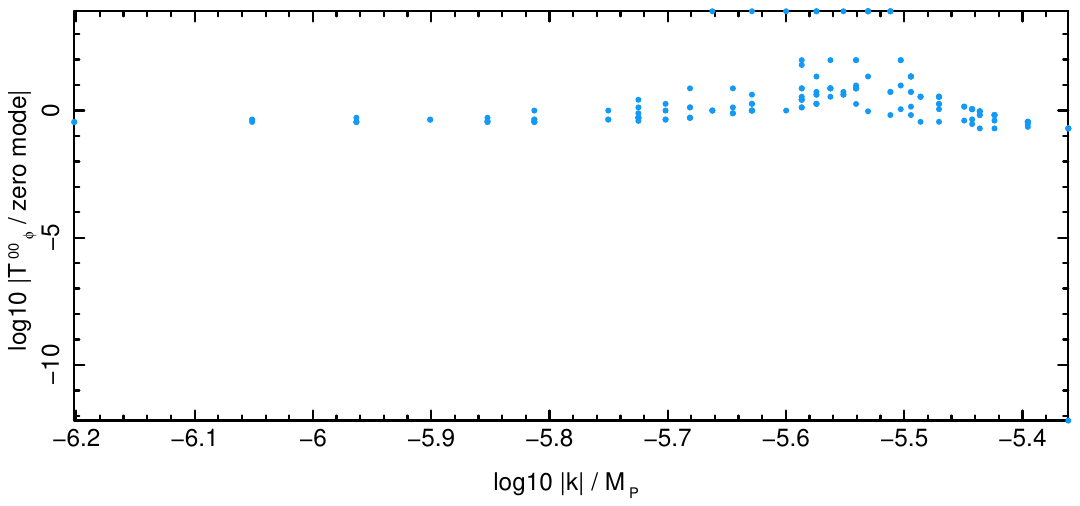} &
\hspace{-1cm}\includegraphics[width=7cm,angle=90]{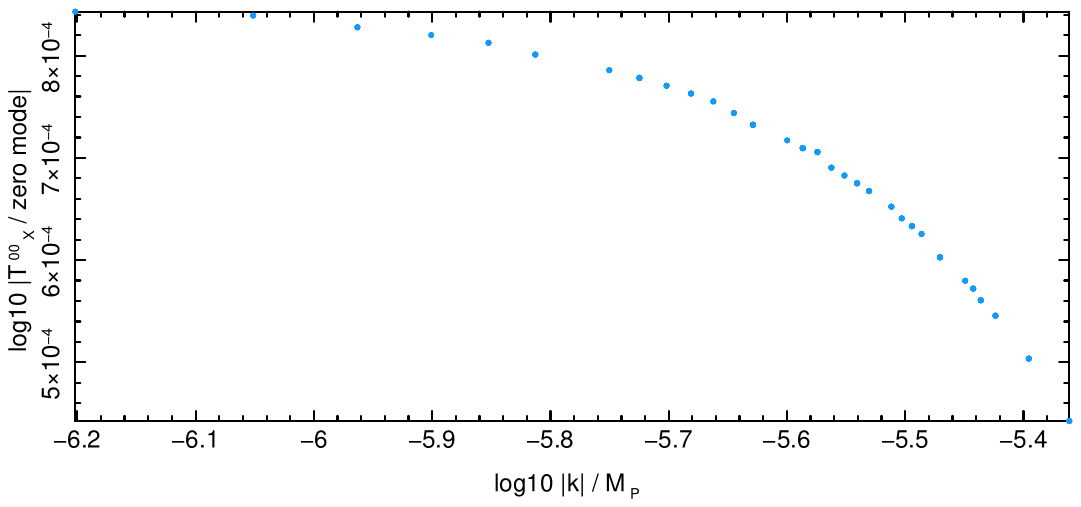} &
\hspace{-1cm}\includegraphics[width=10cm]{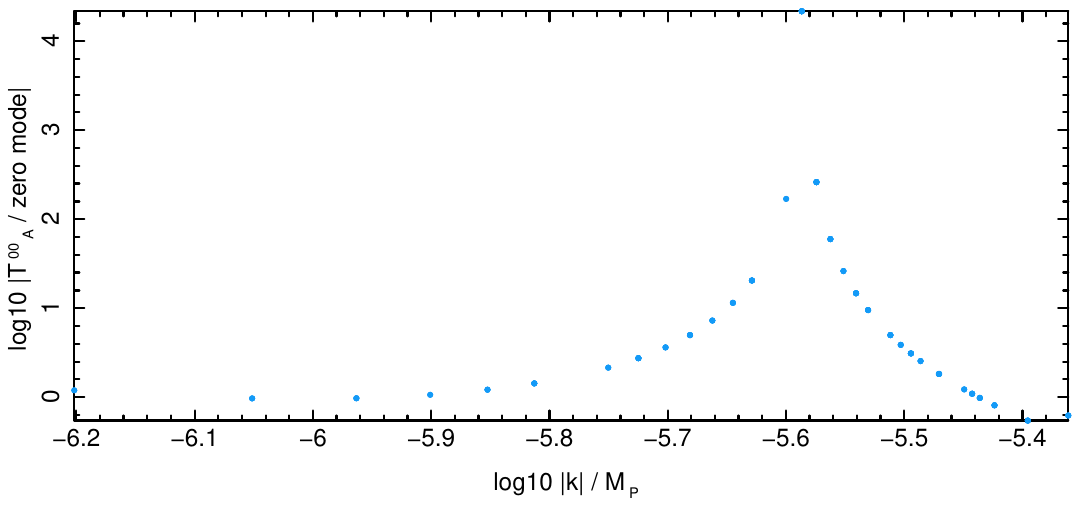}
\end {tabular}
\end {center}
\vspace{-3cm}\caption{Spectrum of 1PI components at the end of simulations with 
$\lambda = 10^{-8}$. Deviation from a line in a) is due to numerical effects, which have slightly 
violated isotropy in 3D mode space. \label{fig:1dpowerspect}}
\end{figure}
The properties of the power spectrum is better discernible in 1D plots. Fig. \ref{fig:1dpowerspect} 
shows 1D power spectrum of $T^i_{1PI} (k, t) / T^i_{1PI} (k=0, t),~i=\Phi,~X,~A$ at the end of 
simulations for the case of $\lambda = 10^{-8}$. It is evident that they are not a power law. 
Nonetheless, the spectrum of $T^X_{1PI}$ which is the dominant component of energy momentum tensor 
at late times is self-similar and close to a power-law with $n_s -1 \lesssim 0$. Thus, it is 
consistent with CMB observations. 
We emphasize that we have not adjusted parameters of the simulations to reproduce observed 
cosmological quantities and the purpose of comparison with observations is to see whether their 
general characteristics are close to observations. For instance, simulations with 
$\lambda = 10^{-14}$ leads to $n_s-1 \approx 0$ or very slightly positive.

\begin{figure}
\begin{center}
\begin{tabular}{p{6cm}p{6cm}p{6cm}}
a) & b) & c) \\
\hspace{-1cm}\includegraphics[width=10cm]{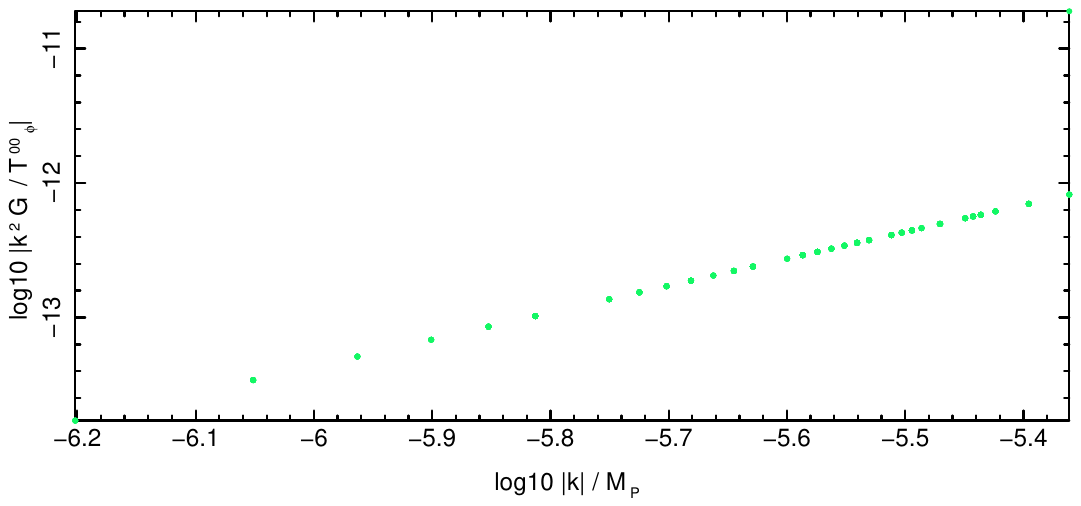} &
\hspace{-1cm}\includegraphics[width=10cm]{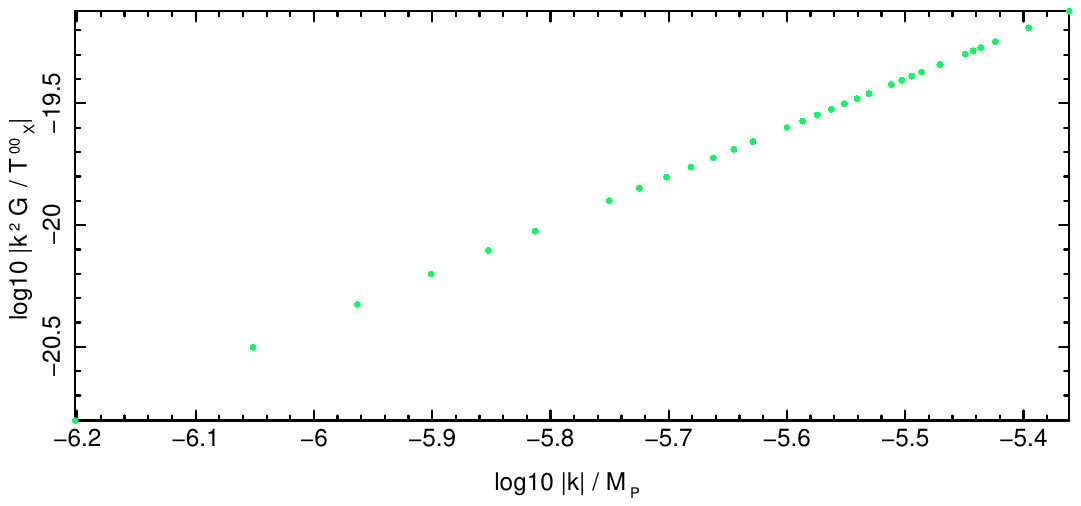} &
\hspace{-1cm}\includegraphics[width=10cm]{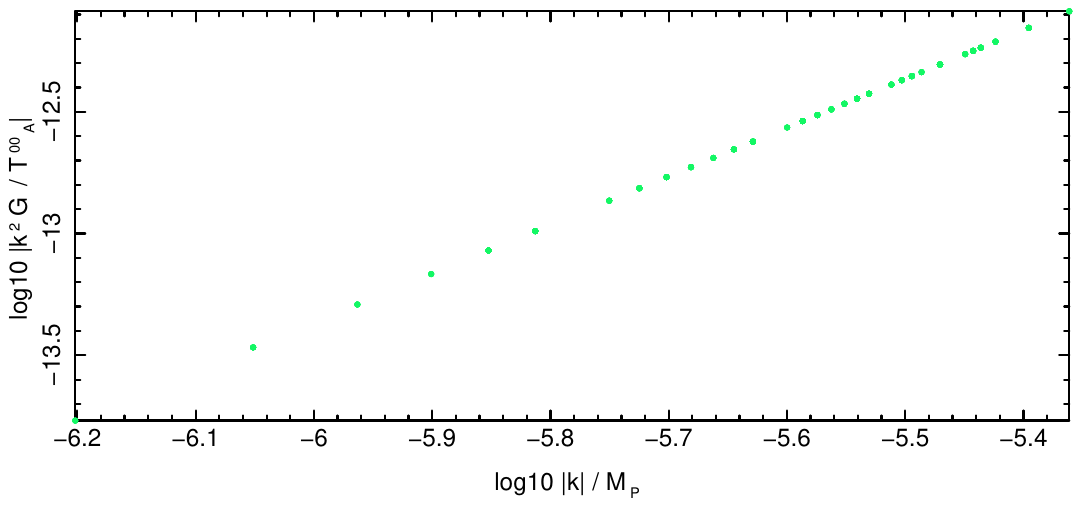}
\end {tabular}
\end {center}
\vspace{-3cm}\caption{Ratio of anisotropic shear to scalar fluctuations for the fields of the 
model as an order of magnitude estimation of tensor to scalar ratio $r$. 
\label{fig:1dpowerspectgw}}
\end{figure}
Although our simulations use a homogeneous metric and cannot determine tensor modes, here we try 
to find an order of magnitude estimation for tensor to scalar ratio $r$. In the effective 
fluid description of energy-momentum tensor the anisotropic shear (\ref{sheartot}) generates 
tensor fluctuations $h_{ij}$ in the metric (\ref{perturbmetric}). If we calculate 
the shear generated in a homogeneous background metric, it is straightforward to see that 
$\Pi^{\mu\nu} (k,t) \propto k^ik^j G (k,t)$. Therefore, the amplitude of gravitational 
waves (without taking into account their backreaction) is $\sim k^2 G (k,t)$. 
Fig. \ref {fig:1dpowerspectgw} shows $k^2 G / T_{1PI}$ as a function of $k$ at the end of 
simulations for the three fields of the model. The ratios are very small for all the fields. 
This result is expected because none of components of the energy-momentum tensor in this 
model becomes at any moment (trans)Planckian. A more precise estimation of $r$ needs simulations 
which include evolution of metric fluctuations.

\subsection{Dark energy} \label{simulde}
As we described in the Introduction, the model studied here was first suggested and investigated 
as a candidate explanation for dark energy. The purpose of the present work was to extend earlier 
studies to a full non-equilibrium quantum field theoretical formulation. According to this model 
dark energy is the condensate of the light scalar field $\Phi$. The condensate might have been 
produced during inflation and evolved in such a way that its present effective equation of state 
$w_\varphi \sim -1$ and its density approximately constant. In this case, it can be considered as 
the remnant of inflation. Alternatively, dark energy condensate may be associated to the decay of 
a heavy particle - presumably dark matter or a constituent of it - produced after inflation. 

Simulations presenting inflation and evolution of various components of the model in the previous 
subsection showed that the fast expansion of the Universe during this epoch significantly suppresses 
the condensate. Consequently, its remnant may become too diluted with the expansion of the Universe 
to be consistent with the observed density and equation of state of dark energy. To see whether 
the second option, that is the decay of a heavy particle after inflation, can produce a dark energy 
condensate, we simulated the same model with an initial value of Hubble function expected for the 
epoch after reheating of the Universe, namely $H_0 = 10^{-15}-10^{-13}~M_P \sim 10^4-10^6$~GeV. In 
addition to the same parameters as the case of inflation, we also performed simulations with 
$\lambda = 10^{-17},~ \mg/ M_P = 10^{-20}$ and $m_X = 10^{-8} M_P$. The reason for reducing the mass of 
main matter source is lower energy scale of physical processes after preheating. Due to limited 
numerical resolution of simulations, we were also obliged to reduce its coupling to other fields, 
otherwise we had to reduce time steps, which made simulations too long. Unfortunately, even in this 
modified model we were only able to have a crud simulation of late time evolution. Here we present 
the results of these simulations and describe features that we judge reliable. However, better 
simulations are necessary to confirm them. 

We call simulations with $\mg/ M_P = 10^{-17}$ and $m_X = 10^{-3} M_P$ {\it Model 1} and 
simulations with $\mg/ M_P = 10^{-20}$ and $m_X = 10^{-8} M_P$ {\it Model 2}. Up to precision of our 
simulations Model 1 behaves very similar to inflation described in the previous section. For 
this reason we do not explain it in detail. Nonetheless, it demonstrates that the model described 
here behaves in a self-similar manner and a shift of initial time, or equivalently initial Hubble 
constant, simply shift in time the accumulation of quantum binding energy, which ultimately leads 
to inflation. Thus, the model is not fine-tuned.

\begin{figure}
\begin{center}
\begin{tabular}{p{6cm}p{6cm}p{6cm}}
a) & b) & c) \\
\hspace{-1cm}\includegraphics[width=9cm]{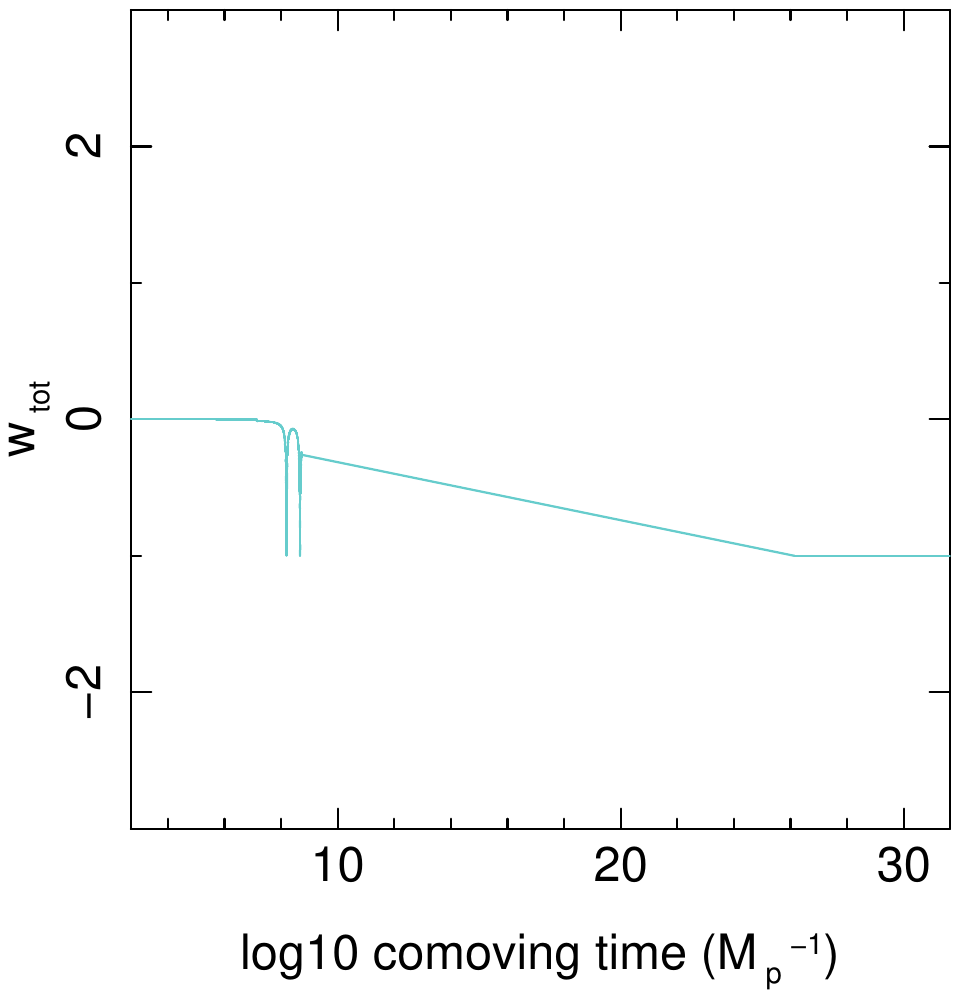} &
\hspace{-1cm}\includegraphics[width=9cm]{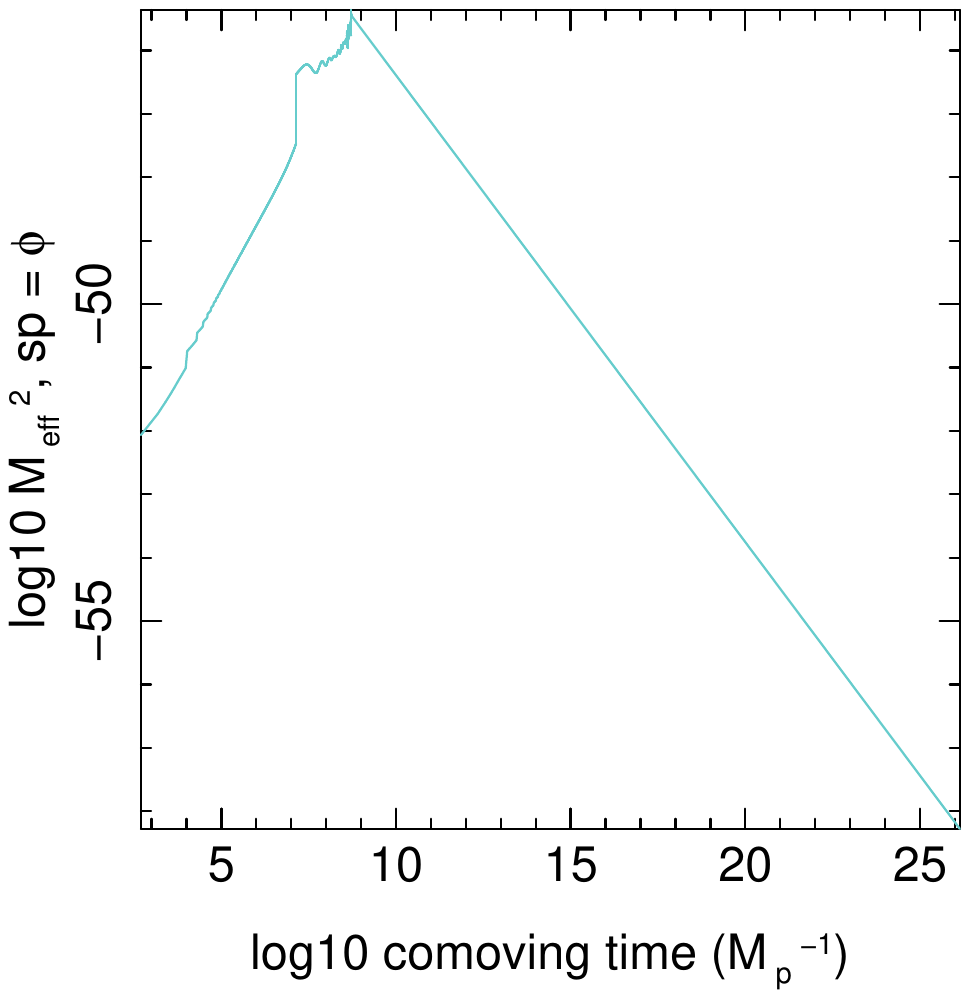} &
\hspace{-1cm}\includegraphics[width=9cm]{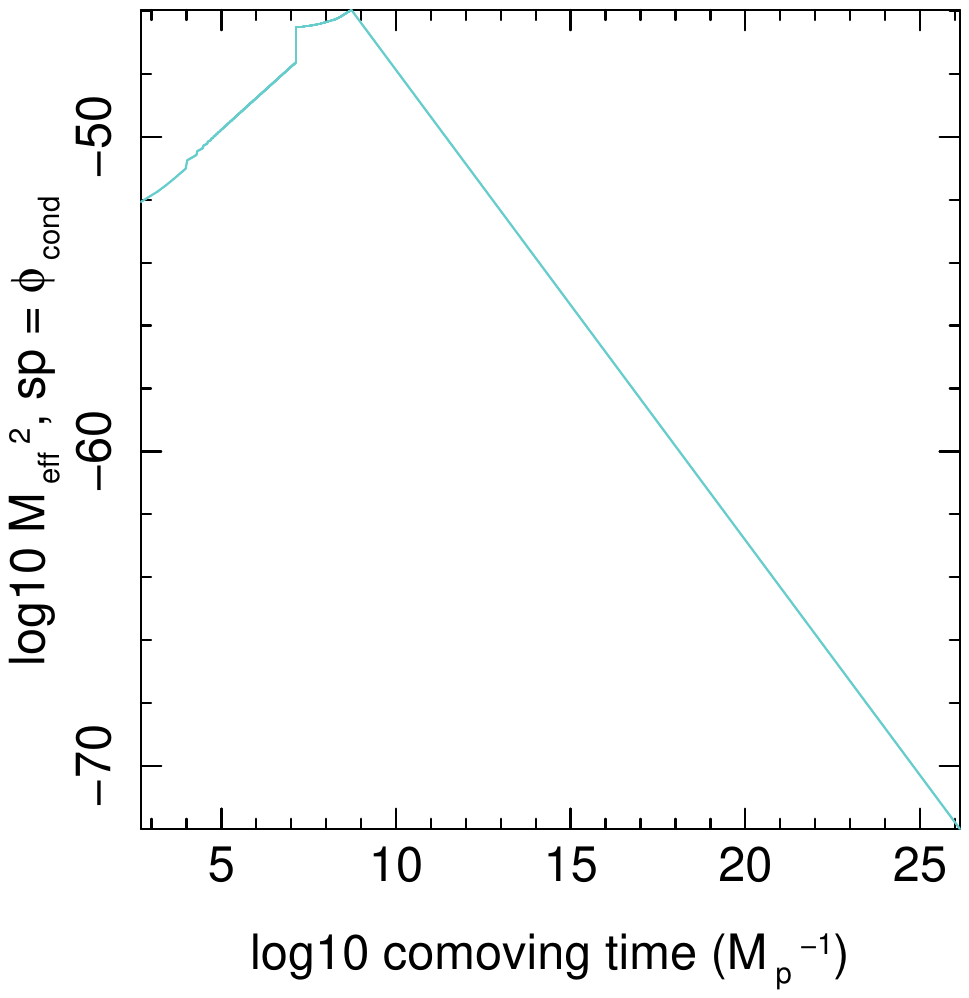}
\end {tabular}
\end {center}
\vspace{-1cm}\caption{Evolution of effective quantities with time in Model 2 simulating late 
accelerating expansion of the Universe: a) Evolution of total equation of state; b) Evolution of 
effective mass of $\Phi$ field; c) Evolution of effective mass of condensate. \label{fig:dewtot}}
\end{figure}

Fig. \ref{fig:dewtot}-a shows the evolution of $w_{eff} \equiv \rho / p$ in Model 2, where $\rho$ 
and $p$ are defined in (\ref{rhop}). As expected, it evolves from matter domination, that is 
$w_{eff} = 0$ to $w_{tot} \approx -1$. We notice that the beginning of transition from matter 
domination is much earlier than what is observed in cosmological data. However, this regime of the 
simulations includes only a few time steps and some deviation from real cosmologies in a toy model 
is expected. We remind that the heating of the Universe occurs in the SM sector, which is not 
present in our simplistic model of early Universe.

Figs. \ref{fig:dewtot}-b and \ref{fig:dewtot}-c show the effective $M^2_\Phi$ and $M^2_\varphi$. Similar 
to the case of inflation, after the initial increase of the effective mass due to accumulation and 
condensation of $\Phi$ field, its value sharply decreases and approaches its IR limit during the 
phase transition from matter domination to an accelerating expansion. We notice a difference between 
the effective mass of $\Phi$ and $\varphi$ such that $M^2_\Phi > M^2_\varphi$ at any time. As discussed 
earlier, this is due to the difference in Feynman diagrams which contribute to these effective 
masses. The comment about too early onset of phase transition discussed for Fig. \ref{fig:dewtot}-a 
applies here too.

\begin{figure}
\begin{center}
\begin{tabular}{p{6cm}p{6cm}p{6cm}}
a) & b) & c)\\
\hspace{-1cm}\includegraphics[width=6.5cm,angle=90]{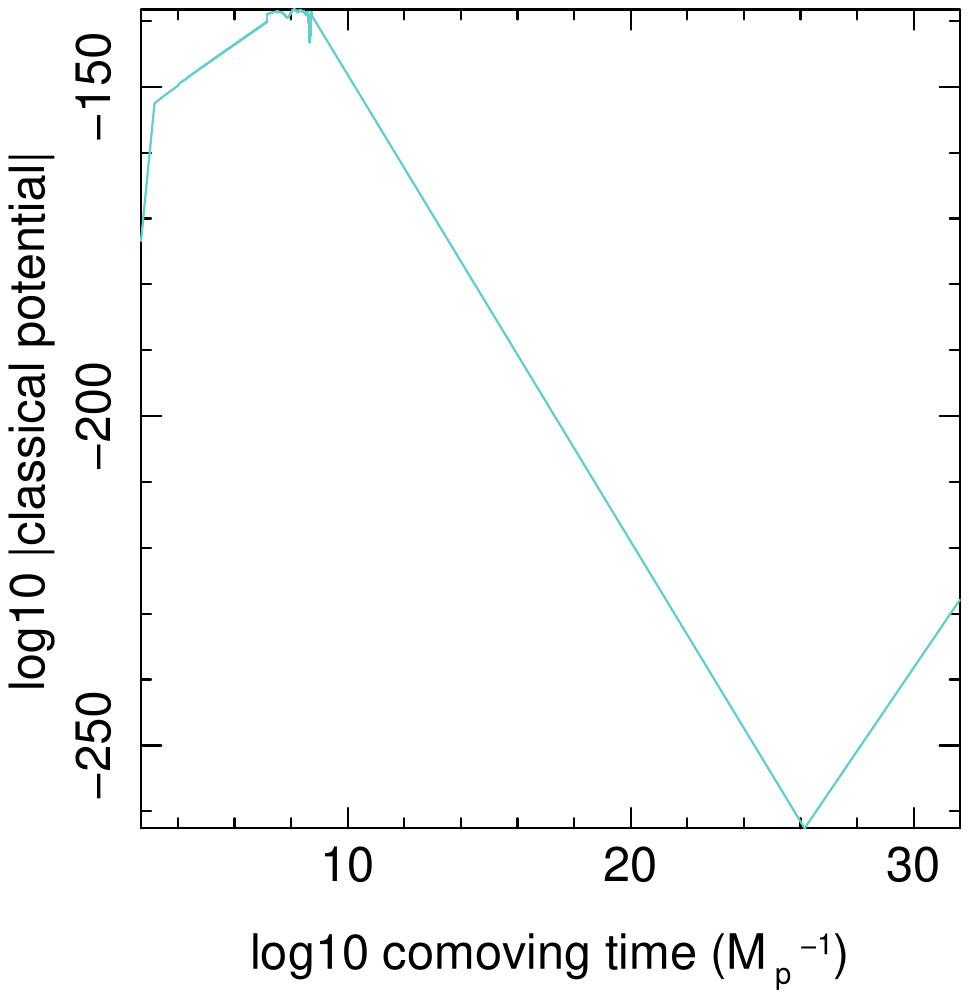} & 
\hspace{-1cm}\includegraphics[width=9cm]{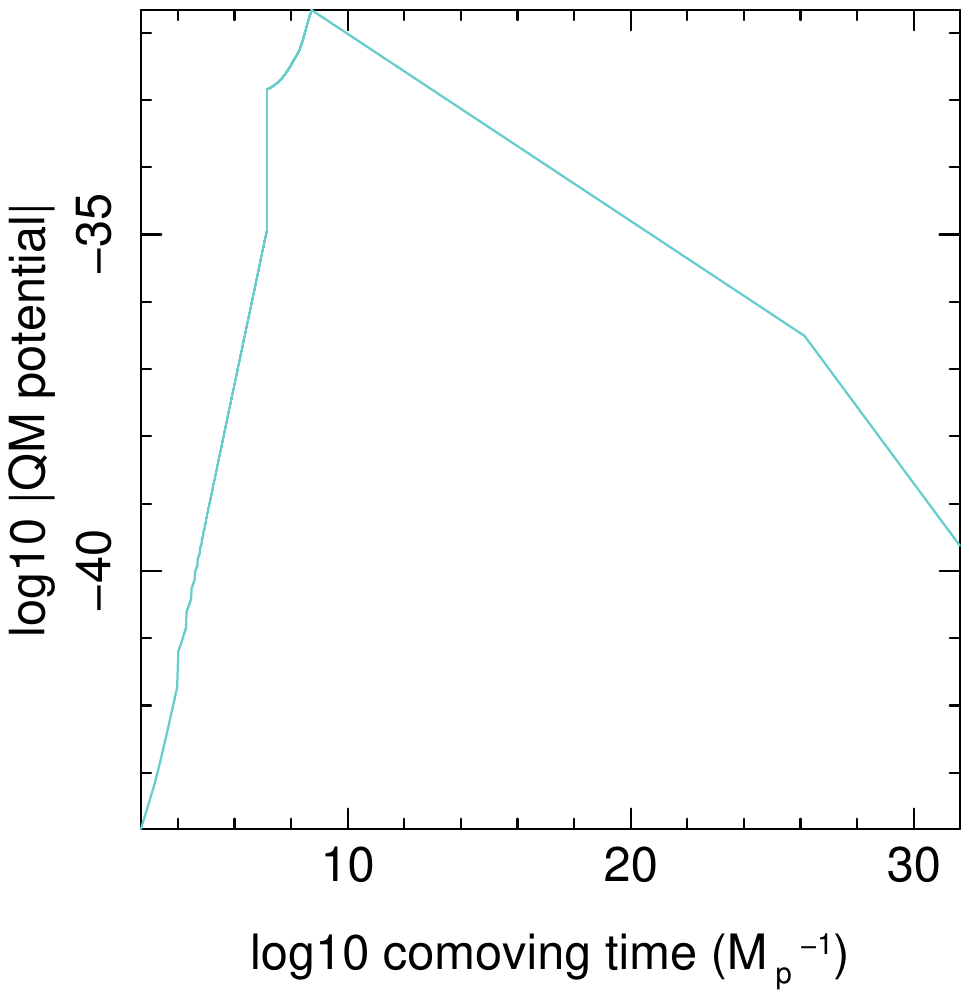} & 
\hspace{-1cm}\includegraphics[width=9cm]{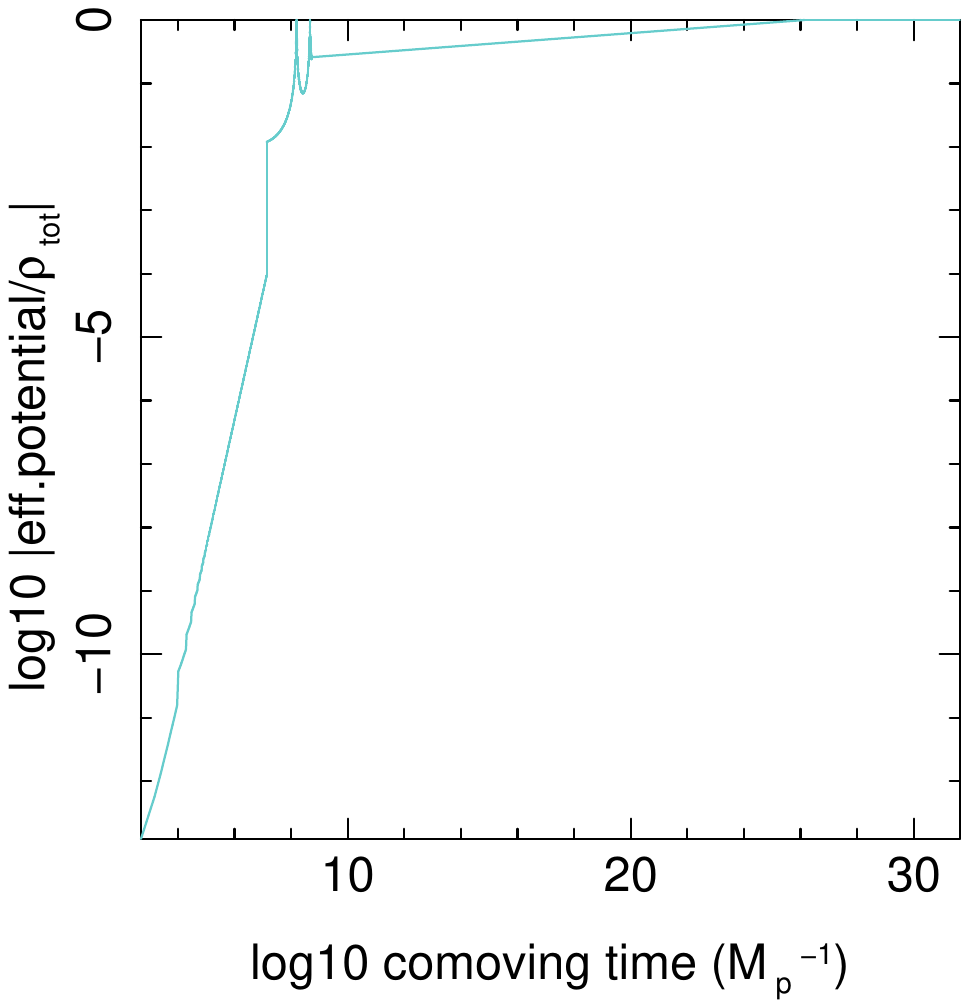} 
\end {tabular}
\end {center}
\vspace{-1cm}\caption{Evolution of classical and quantum components of effective energy density 
with time in Model 2: a) Classical potential of condensate; b) Effective quantum potential; c) 
Ratio of effective quantum potential to total energy density. \label{fig:dedensity}}
\end{figure}

Figs. \ref{fig:dedensity}-a, \ref{fig:dedensity}-b, and \ref{fig:dedensity}-c show classical 
potential of the condensate, quantum binding energy, and its fractional contribution to the total 
energy density in Model 2, respectively. Similar to the case of inflation, the contribution of 
classical potential is completely negligible and even the addition of quantum corrections in the 
effective potential (not shown here) does not make the contribution of condensate in the total 
energy density significant. However, the sharp increase in quantum corrections is certainly due 
to the low resolution of our simulations. Most of other conclusions which we discussed for 
inflation apply also to the late accelerating expansion and do not need to be repeated.

\section {Discussion} \label {sec:discuss}
We do not see an excess in longest modes, neither in inflation nor dark energy simulations. Therefore, 
there is no evidence of IR instability in this model. Up to precision of our simulations this result 
confirms the approximate analytical results obtained for de Sitter space 
in~\cite{desitterstable,infdesiter2pi,rgdesitter}. Our simulations show that, as expected, the light 
field acquires an effective time dependent mass. Moreover, after dissipation of initial instabilities, 
the hierarchy of masses is recovered and when inflation approaches to its end and when dark energy 
become dominant, the mass $\Phi$ and its condensate approaches their initial values. We should 
however remind that in our simulations this field is not massless\footnote{Masslessness of a field in 
interaction can be preserved only if symmetries prevent acquisition of mass, as it is the case for 
photons in SM}. Therefore, the above conclusion is not in contradiction with~\cite{desitterirmassless}, 
which finds IR instability only for massless fields.

In its simplest form considered here the model does not have internal symmetry, but it can be easily 
extended to such cases. Our simulations show that at initial stages of inflation a condensate 
form, and therefore internal symmetries break. However, the amplitude of the condensate would be 
very small, specially during inflation and late accelerating expansion. This means that there may 
be Goldstone boson due to symmetry breaking, but it would not be completely massless. This may 
explain the small but nonzero mass of $\Phi$.

The fast decay of condensate component during accelerating expansion is consistent with 
the results of~\cite{houricond}. However, the small contribution of condensate in the total energy 
density even before fast acceleration found here is in contradiction with conclusions 
of~\cite{houricond}, which predicts that a significant amount of condensate may survive the 
expansion. One of the reasons for this difference may be the lack of consistent evolution of 
expansion factor in approximative analytical method employed in~\cite{houricond}. Moreover, the latter  
analysis is mostly concentrated on finding time variation of condensate rather than comparing its 
contribution in the total density. In any case, both analytical approach in~\cite{houricond} and 
numerical simulations here are far from perfection and better analytical techniques and/or 
simulations are necessary to confirm or refute these findings.

Results of our simulations raise an important question: How can we discriminate 
between a quantum binding energy and an effective classical potential in cosmological observations ? 
Even for cases in which experiments can be performed in laboratory discrimination between quantum and 
classical correlations is not easy. For instance, although measurement of the spectrum of excited 
electrons in atoms can be relatively easily achieved, performing similar experiments for strongly 
coupled partons in hadrons or weakly coupled molecules is very difficult. In the former case the 
strong coupling makes isolation of one parton extremely difficult, and in the latter example thermal 
noise and strong interaction of a probe with atoms intertwine and influence the measurement of a weak 
molecular binding energy. In cosmological measurements correlation between causally decoupled modes 
may help to discriminate non-local quantum effects. But, due to the expansion of the Universe it is 
not trivial to discriminate between correlations induced by past causal interactions and inherently 
non-local quantum effects. The observation of non-Gaussianity may be a signature, but as we saw in 
the simulation of inflation, the amplitude of fluctuations of the quantum component, and thereby its 
non-Gaussianity, is very small. 

A point which is not addressed in this work is the effect of unobservable IR modes. They are suggested 
to be responsible for the late accelerating expansion~\cite{quinir}, but so far the issue is not 
investigated in a fully quantum setting. In the framework of 2PI formulation the incompleteness 
(openness) of cosmological observations is presented by a mixed density matrix, which is not 
considered here. Nonetheless, analysis for a toy model in de Sitter space at lowest quantum 
order~\cite{cosmosqmopen} shows that IR modes dissipate. This is consistent with our simulations of 
pure states. Another issue that low resolution of our simulations did not allow to investigate is a 
relation between inflation and dark energy. We showed that most probably inflaton condensate does not 
survive inflation and its remnant would be too diluted to generate another epoch of accelerating 
expansion at later times. However, as we demonstrated, the quantum binding energy of the same fields 
may induce late time accelerating expansion. We were not able to connect the two epochs. This needs 
a detail simulation of particle production at the end of inflation, which couldn't be followed with 
our code. 

There is also a need for exploring more extensively parameter space and extensions to this model, 
namely: larger coupling among the three fields, self-coupling of other fields, other combination of 
masses, other coupling models, internal symmetries, gauge symmetry, etc.

\section {Outlines} \label {sec:conclusion}
In conclusion, we studied a simple multi-component model for early and late accelerating expansion 
of the Universe in a fully quantum field theoretical framework. Through numerical simulation we 
investigated the process of formation of a quantum condensate from an initially null state and 
followed its evolution, as well as the evolution of quantum component of the fields and their effect 
on the geometry of the Universe. 

We assessed the reliability of simulations by changing resolution and break points of intermediate 
simulations. We also performed simulations with different masses and couplings. We observed that 
although small scale features in the evolution curves are not reliable, general behaviour of 
quantities do not significantly depend on the simulation setup. Interpretations and conclusions 
discussed in previous sections and summarized below are restricted to general aspects rather than 
features which may be artefacts. Evidently and as usual better simulations are necessary to confirm 
our conclusions.

Our simulations show that in realistic models containing multiple fields and hierarchy of masses 
and couplings, the non-local binding energy between interacting constituents, which depends on all 
fields, rather than the condensate may have important role in triggering inflation and generation of 
anisotropies and/or late accelerating expansion of the Universe. On the other hand, although 
energetically subdominant, numerical domination of light field quanta has crucial role in controlling 
the behaviour of heavier fields, and thereby the content and geometry of the Universe. This highlights 
the shortcomings of making conclusions about fundamental physics of early Universe by comparing 
cosmological data with predictions of classical or semi-classical models of inflation and dark 
energy. Additionally, it demonstrates that many quantum phenomena, which cannot be described by 
classical effective models, might have dominated processes leading to the Universe as we find it now.

\appendix
\section{Classical potential} \label{app:classpot}
Classical potential in (\ref{lagrange}) and its first and second derivatives are:
\bea
&& \mathcal{V} (\Phi, A, X) = \frac{1}{2}m_{\Phi}^2 {\Phi}^2 + \frac{\lambda}{4!}{\Phi}^4 + \frac{1}{2}m_X^2 X^2 + \frac{1}{2}m_A^2 A^2 - \mg \Phi X A \label{potentialphixa} \\
&& \frac {\partial \mathcal{V}}{\partial \Phi} = m_{\Phi}^2 \Phi + \frac{\lambda}{3!}{\Phi}^3 - \mg X A, \quad \quad \frac {\partial \mathcal{V}}{\partial X} = m_X^2 X - \mg \Phi A, \quad \quad \frac {\partial \mathcal{V}}{\partial A} = m_A^2 A - \mg \Phi X. \label{potentialphixad} \\
&& \frac {\partial^2 \mathcal{V}}{\partial \Phi^2} = m_{\Phi}^2 + \frac{\lambda}{2!}{\Phi}^2 \quad \quad 
\frac {\partial^2 \mathcal{V}}{\partial X^2} = m_X^2 \quad \quad \frac {\partial^2 \mathcal{V}}{\partial A^2} = m_A^2 \label{potentialphixad2} \\
&& \frac {\partial^2 \mathcal{V}}{\partial \Phi \partial X} = - \mg A \quad \quad \frac {\partial^2 \mathcal{V}}{\partial \Phi \partial A} = - \mg X \quad \quad \frac {\partial^2 \mathcal{V}}{\partial X \partial A} = - \mg \Phi \label{potentialphixad2p}
\eea
There are 3 extrema points indexed as $0, 1, \& 2$:
\bea
&& \Phi_0 = A_0 = X_0 = 0 \label{trivialext} \\
&& X_{1,2} = \pm \biggl (\frac{\lambda m_X^2 m_A^4}{3! \mg^4} + \frac{m_\Phi^2 m_A^2}{\mg^2} 
\biggr)^{\frac{1}{2}}, \quad \quad \Phi_{1,2} = \frac{m_X m_A}{\mg}, \quad \quad 
A_{1,2} = \frac{\mg \Phi_{1,2} X_{1,2}}{m_A^2} \label{ext12}
\eea
Second derivative (\ref{potentialphixad2}) and (\ref{potentialphixad2p}) show that the trivial minimum 
(\ref{trivialext}) is the true minima of the system. For both positive and negative $\mg$ other 
extrema are unstable. 

For the value of parameters used in the simulations 
$|X_{1,2}| = 10^{-6} / \sqrt{3!},~A_{1,2} = 10^{-4},~\Phi_{1,2} = 0.1$. Therefore, even at classical level 
the amplitude of condensates of $X$ and $A$ at quasi-equilibrium is much smaller than that of 
$\Phi$ and neglecting them in the simulations is justified.

\section {Propagators and decomposition of self-energy} \label{app:prop}
For a bosonic field $\psi$ propagators are defined as (for $\langle \psi \rangle = 0$):
\bea
G^F(x,y) & \equiv & \frac{1}{2} \langle \{\hat{\psi} (x), \hat{\psi}^\dagger(y)\} \rangle = 
\frac{i}{2} (G^> + G^<) \label{fprog} \\
G^\rho(x,y) & \equiv & i \langle [\hat{\psi} (x),\hat{\psi}^\dagger(y)] \rangle = 
- (G^> - G^<) \label{rhoprog} \\
iG^> (x,y) & \equiv & \langle \hat{\psi} (x) \hat{\psi}^\dagger (y) \rangle = 
\tr (\hat{\psi} (x)\hat{\psi}^\dagger(y)\hat{\varrho}) \label{gtprog} \\ 
iG^< (x,y) & \equiv & \langle \hat{\psi}^\dagger (y) \hat{\psi} (x) \rangle = 
\tr (\hat{\psi}^\dagger (y) \hat{\psi} (x) \hat{\varrho}) \label{lsprog} \\
G_{Fey} (x,y) & \equiv & -i \langle T\hat{\psi} (x) \hat{\psi}^\dagger (y) \rangle \nonumber \\
&=& G^> (x,y) \Theta (x^0-y^0) + G^< (x,y) \Theta (y^0-x^0) \nonumber \\
&=& G^F (x,y) - \frac{i}{2} \text{sign} (x^0 - y^0) G^\rho (x,y) \label{progfey} \\
\bar{G}_{Fey} (x,y) & \equiv & -i \langle \bar{T}\hat{\psi} (x) \hat{\psi}^\dagger (y)\rangle 
\nonumber \\
&=& G^> (x,y) \Theta (y^0-x^0) + G^< (x,y) \Theta (x^0-y^0) \nonumber \\
&=& G^F (x,y) + \frac{i}{2} \text{sign} (x^0 - y^0) G^\rho (x,y) \label{progfeybar}
\eea
where $T$ and $\bar{T}$ time ordering and inverse ordering operators, respectively. When 
$\langle \psi \rangle \neq 0$, $G^F(x,y) = G^F(x,y) 
\biggl |_{\langle \psi \rangle = 0} - ~\langle \psi (x) \rangle \langle \psi (y) \rangle$.

Properties of propagators can be summarized as the followings:
\bea
&& [iG^{>,<} (x,y)]^\dagger = iG^{>,<} (y,x), \quad \quad iG^> = G^F -\frac{i}{2} G^\rho , 
\quad \quad iG^< = G^F +\frac{i}{2} G^\rho \nonumber \\
&& G^{F~\dagger}(x,y) = G^F(x,y), \quad \quad G^{\rho~\dagger}(x,y) = -G^\rho(y,x) \label{progconjug}
\eea
Here $\hat{\varrho}$ is the density operator of the quantum state of the system. The advantage of 
using $G^F(x,y)$ and $G^\rho(x,y)$ is that they include both time paths and their evolution equations 
are explicitly causal and suitable for numerical simulations~\cite{2pirev}.

In a similar manner the self-energy $\Pi (x,y)$ can be decomposed to symmetric (F) and 
anti-symmetric ($\rho$) components~\cite{2pirev}. For this purpose we first separate local 
component of the self-energy, then we decompose non-local part in analogy with (\ref {progfey}):
\bea
\Pi (x,y) & \equiv & -i \Pi^0 (x) \delta^{(4)} (x-y) + \bar{\Pi} (x,y) \label{selfenerloc} \\
M^2 (x) & \equiv & m^2 + \Pi^0 (x) \label {effmassdef} \\
\bar{\Pi} (x,y) & \equiv & \Pi^F (x,y) - \frac{i}{2} \text{sign} (x^0 - y^0) \Pi^\rho (x,y) 
\label{selfenerdecomp}
\eea
By using the decomposition of Feynman propagator (\ref{progfey}) in self-energy diagrams we obtain 
the following expression for contribution of a diagram including $k$ propagators:
\bea
\Pi^F & \propto & \sum_{j=0}^{[k/2]} C^k_{k-2j} (-1)^j 2^{-2j} G_F^{k-2j} G_\rho^{2j} 
\label{selfenerf} \\
\Pi^\rho & \propto & \sum_{j=0}^{[k/2]} C^k_{k-2j-1} (-1)^j 2^{-2j} G_F^{k-2j-1} G_\rho^{2j+1} 
\label{selfenerrho}
\eea
where $[k]$ is the integer part of $k$. Here we have used proportionality sign rather than equality 
because couplings, number of degeneracies, and traces are not shown in (\ref{selfenerf}) and 
(\ref{selfenerrho}). These factors depend on the topology of corresponding Feynman diagram, order 
of interactions and 1-point expectation value. They are the same for both components.

\section{Density matrix of a many-particle state} \label {app:initparticle}
A pure quantum many-particle state can be decomposed as:
\bea
&& [a_{\beta_1}, a^{\dagger}_{\beta_2}] = \delta_{\beta_1 \beta_2} \quad 
[a_{\beta_1}, a_{\beta_2}] = 0 \quad [a^{\dagger}_{\beta_1}, a^{\dagger}_{\beta_2}] = 0\label{canon} \\
&& |\Psi \rangle \equiv \sum_{\beta_1 \beta_2 \cdots} \Psi_{\beta_1 \beta_2 \cdots} 
|\beta_1 \beta_2 \cdots \rangle = \sum_{\beta_1 \beta_2 \cdots} \Psi_{\beta_1 \beta_2 \cdots} 
a_{\beta_1}^\dagger a_{\beta_2}^\dagger \ldots 
|0\rangle, \quad \quad \hat{\varrho} \equiv |\Psi\rangle \langle \Psi | \label{genstate} \\
&& a_\beta |0 \rangle = 0~~\forall \beta \in \{\beta_1, \beta_2, \ldots \}, 
\quad \quad \sum_{\beta_1 \beta_2 \cdots} |\Psi_{\beta_1 \beta_2 \cdots}|^2 = 1. \nonumber
\eea
where $\beta$'s are a set of quantum numbers, including momentum mode $k$ and field identification 
indices. They define properties of a particle or mode at a given instance of time. The number of 
particles/modes in $|\Psi\rangle$ can be infinite. The absence (zero value) for some of 
$\Psi_{\beta_1 \beta_2 \cdots}$ coefficients presents an initial quantum entanglement between particles. 
In the simplest cases, such as the model studied here, fields are scalars without internal 
symmetries and only position or momentum modes are of physical interest. Thus, 
$\beta = \{\mathbf{x}, i \in \Phi,A,X\}$ or $\beta = \{k, i \in \Phi,A,X\}$ in coordinates or 
momentum representation, respectively. Note that all the operators in (\ref{canon}) and 
(\ref{genstate}) are defined at the same time coordinate, e.g. the initial time $t_0$. Therefore, the 
latter is not explicitly mentioned. Creation and annihilation operators in coordinate and momentum 
representations are related to each other:
\bea
&& \hat{\Phi}^-(x) \equiv \frac{1}{(2\pi)^3} \int d^3k~\um_k (t) a_k e^{-ik.\mathbf{x}}, \quad 
\hat{\Phi}^+(x) \equiv \frac{1}{(2\pi)^3} \int d^3k~ \um^*_k (t) a^\dagger_k e^{ik.\mathbf{x}}, \quad 
\hat\Phi (x) = \hat{\Phi}^-(x) + \hat{\Phi}^+(x)
 \nonumber \\
&& \label{fieldcomp} 
\eea
where $\um_k (t_0)$ is the spatial Fourier transform of a solution of the field equation at initial 
time $t_0$.

The density operator of pure states (\ref{genstate}) can be expanded as:
\bea
\hat{\varrho} &=& \int d^4 x_1 \sqrt{-g(x_1)} \delta (t_{x_1} - t_0) \ldots d^4 y_1 \sqrt{-g(y_1)} 
\delta (t_{y_1} - t_0) \ldots \Psi^*_{y_1,y_2,\cdots} \Psi_{x_1,x_2,\cdots} \nonumber \\
&& \hat{\Phi}^+ (x_1) \hat{\Phi}^+ (x_2) \ldots |0\rangle \langle 0| \ldots \hat{\Phi}^- (y_2) 
\hat{\Phi}^- (y_1) \label {rhoexpan}
\eea
From comparison of (\ref{rhoexpan}) with the expansion of $F[\Phi]$ in (\ref{fexpan}) it is 
straightforward to show that:
\bea
&& \alpha (x_1, x_2, \ldots, y_1, y_2, \ldots) = \Psi^*_{y_1,y_2,\cdots} \Psi_{x_1,x_2,\cdots} = \nonumber \\
&& \quad \quad \sum_{\{\beta\}\{beta'\}}\Psi^*_{\beta'_1 \beta'_2 \cdots} \Psi_{\beta'_1 \beta'_2 \cdots} \langle \beta'_1, \beta'_2 
\ldots | \Phi (x_1), \Phi (x_2) \ldots \rangle \langle \Phi (y_1), \Phi (y_2) \ldots | \beta_1, \beta_2 
\ldots \rangle \label{alphaexpan}
\eea


\section{Propagators of free scalar fields} \label{app:propag}
Applying (\ref{canon}) with $\alpha = k$ to the definition of propagators given in 
Appendix \ref{app:prop}, their decomposition in momentum space can be obtained as the followings:
\bea
iG_{Fey} (x,y) &\equiv &\langle\Psi|T\Phi(x)\Phi(y)|\Psi\rangle = \nonumber \\
&& \sum_k \sum_i \sum_{k_1 k_2 \ldots k_n} \delta_{kk_i} |\Psi_{k_1 k_2 
\ldots k_n}|^2 \biggl [\um_k^* (x)\um_k (y) \Theta (x_0 - y_0) + 
\um_k (x)\um_k^* (y) \Theta (y_0 - x_0)\biggr ] + \nonumber \\
&& \sum_k  \biggl [\um_k (x)\um_k^* (y) \Theta (x_0 - y_0) + \um_k^* (x)
\um_k (y) \Theta (y_0 - x_0)\biggr ] \label{propst}
\eea
From (\ref{propst}) we can extract the expression for advanced and retarded propagators:
\bea
iG^> (x,y) &\equiv& \langle\Psi|\Phi(x)\Phi(y)|\Psi\rangle = \sum_k \sum_i \sum_{k_1 k_2 \ldots k_n} 
\delta_{kk_i} |\Psi_{k_1 k_2 \ldots k_n}|^2 \um_k^* (x)\um_k (y) + \nonumber \\
&& \sum_k \biggl [1 + \sum_i \sum_{k_1 k_2 \ldots k_n} \delta_{kk_i} |\Psi_{k_1 k_2 \ldots k_n}|^2 \biggr ] 
\um_k (x)\um_k^* (y) \label{propgr} \\
iG^< (x,y) &\equiv& \langle\Psi|\Phi(x)\Phi(y)|\Psi\rangle = \sum_k \sum_i \sum_{k_1 k_2 
\ldots k_n} \delta_{kk_i} |\Psi_{k_1 k_2 \ldots k_n}|^2 \um_k (x) \um_k^* (y) + \nonumber \\
&& \sum_k \biggl [1 + \sum_i \sum_{k_1 k_2 \ldots k_n} \delta_{kk_i} 
|\Psi_{k_1 k_2 \ldots k_n}|^2 \biggr ] \um_k^* (x)\um_k (y) \label{propless}
\eea
Using (\ref{propgr}) and (\ref{propless}) we find (\ref{propfu1part}) and (\ref{proprhou1part}) 
expressions for $G^F$ and $G^\rho$, respectively.

For a gas of free particles the wave-function of the multi-particle initial state can be factorized 
to 1-particle functions:
\bea
|\Psi_{k_1 k_2 \ldots k_n}|^2 &=& \prod_{i \in \{1,2,\cdots\}} \sum_{k_i} |\psi_{k_i}|^2 \label {onepartstate}\\
G^F (\vec{x},\eta_0,\vec{y},\eta_0)& = & \sum_k \biggl [\biggl (\frac{1}{2} + 
f (k, \bar{x}, \eta_0) \biggr ) \biggl (\um_k (\vec{x},\eta_0)\um_k^* (\vec{y},\eta_0) + 
\um_k^* (\vec{x},\eta_0)\um_k (\vec{y},\eta_0) \biggr) \biggr ] \label{propfu1part} \\
G^\rho (\vec{x},\eta_0,\vec{y},\eta_0)& = & i\sum_k \biggl (\um_k (\vec{x},\eta_0)
\um_k^* (\vec{y},\eta_0) - \um_k^* (\vec{x},\eta_0)\um_k (\vec{y},\eta_0) \biggr) 
\label{proprhou1part} \\
f (k, \bar{x}, \eta_0) & \equiv & \sum_i |\psi_{k_i}(\bar{x}_i, \eta_0)|^2 = N (\bar{x},\eta_0) 
|\psi_k|^2 \label{1pdist} \\
\um_{k} (\vec{x},\eta_0) &\equiv& \um_{k} (\eta_0) e^{-i\vec{k}.\vec{x}}, \quad\quad \sum_k \equiv 
\frac{1}{(2\pi)^3} \int d^3k \label{sumdef}
\eea
where $\psi_k$ is the 1-particle wave-function and $N$ is a normalization factor. From (\ref{sumdef}) 
it is clear that the r.h.s. of (\ref{propfu1part}) is a Fourier transform with respect to 
$\vec{x} - \vec{y}$. Therefore, after a Wigner transformation, the amplitude of wave function 
$|\psi_k|^2$ and thereby 1-particle distribution $f$ will depend on the average coordinate 
$\bar{x} \equiv (\vec{x}+\vec{y}) / 2$. For a Gaussian distribution in a matter dominated Universe:
\be
N = \frac{3\pi \mathcal{H}^2 e^{\frac{M^2}{4\sigma^2}}}{2 \mathcal{G} a^2 \sigma^2 M^2 
K_1 (\frac{M^2}{4\sigma^2})} \label{distnormalizgauss}
\ee
where $M$ is the effective mass of particles. The function $K_1$ is modified Bessel function of 
second kind. The antisymmetric propagator $iG^\rho (\vec{x},\eta_0,\vec{y},\eta_0)$ does not depend 
on the initial state and its expression is (\ref{proprhou1part}) irrespective for any state. 
We remind that this expression is the normalization factor $N$ obtained in (\ref{distnormaliz}) 
after solving the constraint equation (\ref{xprogconstrint}). 

Classically, $f (k, \eta_0)$ is interpreted as statistical distribution of particles, for instance 
Boltzmann or Bose-Einstein distribution. Nonetheless, the expression (\ref{propfu1part}) can be easily 
extended to entangled particles. For instance, if the initial state consists of pair of particles 
entangled by their momentum, $f (k_1,k_2,\eta_0)$ presents the distribution of entangled pair with 
momenta $(k_1,k_2)$. Therefore, this formulation covers both single field and coherent oscillations 
studied in~\cite{infinitsuperpos}. We remind that if the scalar field has an internal symmetry, 
that is multiple flavors, $iG^F (x,y)$ and $f (k,\eta_0)$ will have implicit flavor indices. 

In the model studied here at initial 
time $G_i^F,~i=X,A,\Phi$ are free and proportional to $N+1/2$ where $N$ is the number of particles 
in the initial state, see (\ref{propfu1part}). Even without considering initial free fields, 
by taking Wigner transformation, it can be shown that $G_i^F$'s can be considered as distribution 
functions with respect to the center of mass coordinate and Fourier modes of relative coordinates.  
They evolve according to Boltzmann equations and quantum correction integrals in the r.h.s. of 
(\ref{evolgf}) play the role of collisional terms~\cite{inin,boltzlepto,2picurved}. The 
wave-function amplitude in (\ref{propfu1part}) at classical limit is replaced by a 1-particle distribution and 
its normalization is equal to (\ref{distnormalizgauss}).

\section {Distribution of remnants} \label{app:distphia}
We define the rest frame of $X$ particles as the frame in which the maximum of $f_X (k)$ is at 
$|\vec{k}| = 0$\footnote{Here we consider local inertial, called also normal coordinates, which is 
locally equivalent to a flat space.}. At lowest order decay of particles occurs locally. Thus, we 
use local inertial frame for calculating momentums of remnants. For interaction model (a) remnants 
$\Phi$ and $A$ have opposite 3-moment in the rest frame of decaying $X$ particle, that is 
$\vec{p_\Phi} = -\vec{p_A}$ and 
$|\vec{p_\Phi}| = |\vec{p_A}| = [(M_X^2 - M_\Phi^2 - M_A^2)^2 - 4 M^2_\Phi M^2_A ]^{1/2} / 2M_X$. In a 
frame in which $|\vec{k}| \neq 0$, the 4-momentum of $\Phi$ and $A$ are ($a_0 = 1$ is assumed):
\bea
p'_\Phi (k) = \begin {pmatrix} \gamma (E_\Phi + \vec{\beta}.\vec{p}_\Phi) \\ 
\vec{p}_\Phi + \biggl (\frac{\gamma - 1}{\beta^2} (\vec{\beta}.\vec{p}_\Phi) + \gamma E_\Phi \biggr ) 
\vec{\beta} \end {pmatrix} & \quad & 
p'_A (k) = \begin {pmatrix} \gamma (E_A + \vec{\beta}.\vec{p}_A) \\ 
\vec{p}_A + \biggl (\frac{\gamma - 1}{\beta^2} (\vec{\beta}.\vec{p}_A) + \gamma E_A \biggr ) 
\vec{\beta} \end {pmatrix} \label{kframe} \\
E_\Phi^2 = |\vec{p_\Phi}|^2 + M_\Phi^2 & \quad & E_A^2 = |\vec{p_A}|^2 + M_A^2 \label{cmenerphia}
\eea
where $\vec{\beta} = \vec{k} / \omega_k$, $\gamma = \omega_k/M_X$. Thus, there is a one-to-one 
relation between the momentum of a $X$ particle and those of its remnants. The inverse 
transformation, that is $k (p'_\Phi)$ can be written as:
\bea
\gamma & = & \frac{E'_\Phi (E'_\Phi + E_\Phi) - \mathcal {C}}{E_\Phi (E'_\Phi + E_\Phi) + \mathcal {C}} 
\label{gammak} \\
\mathcal {A} & \equiv & \vec{\beta}.\vec{p}_\Phi = \frac{E'_\Phi - \gamma E_\Phi}{\gamma} 
\label{betap} \\
\vec{\beta} & = & \frac{\mathcal {C}}{\frac{(\gamma - 1) \mathcal {A}}{\beta^2} + \gamma E_\Phi} 
\label {betavec}\\
\mathcal {C} & \equiv & \vec{p}'_\Phi.\vec{p}_\Phi - \vec{p}_\Phi^2 \label{cdef}
\eea
where $E'_\Phi \equiv p^{'0}_\Phi$. 

If the boundary condition (\ref{dirichletneumannijmom}) is chosen for $G^F_\Phi$, there is a direct 
relation between initial number density of $X$ and initial increase of number density of $\Phi$. 
For this reason, initial distribution of $\Phi$'s at $t_{+0}$ can be derived from momentum 
distribution of their parent $X$ particles:
\be
\omega_{p'_\Phi} f_\Phi (p'_\Phi (k)) \delta (p_\Phi^{'2} - M_\Phi^2) d^4 p'_\Phi = \mathcal{N} \omega_k 
f_X (k) \delta (k^2 - M_X^2) d^4k \label{phidist}
\ee
where $\omega_{p'_\Phi} = \sqrt {\vec {p'}_\Phi^2 + M_\Phi^2}$, $\omega_k = \sqrt {\vec {k}^2 + M_X^2}$, 
$\beta = |\vec {\beta}|$, and $\mathcal{N} = 1$ presents the multiplicity of $\Phi$ in the decay of 
$X$. The function $f_\Phi (p'_\Phi)$ determines the distribution of energy 
levels - partial condensates - in a generalized coherent state as described in 
Sec. \ref{sec:initcondcond}.

Using (\ref{phidist}) $f_\Phi (p'_\Phi (k))$ is determined as:
\bea
f_\Phi (p'_\Phi) d^3p'_\Phi & = & \mathcal{N} J (k (p'_\Phi)) f_X (k (p'_\Phi)) d^3p'_\Phi \\
J & = & \biggl |\frac{\mathcal{D} \beta + \frac{\mathcal{B}}{\gamma^2 \omega_k}}{|p'_\Phi|^4} 
\biggl [|p'_\Phi|^2 \biggl (\beta^2 \mathcal {B}^2 + (1-\beta)(\gamma - 1) 
\frac{(\vec{p}_\Phi.\vec{\beta})^2}{\beta^2} \biggr ) + \nonumber \\
&& \biggl (\frac{\vec{p}_\Phi.\vec{k}}{\beta |p_\Phi|} - 1 \biggr ) \biggl (|p_\Phi|^2 + 
(\gamma - 1) \frac{(\vec{p}_\Phi.\vec{k})^2}{\beta^2} - \frac{(\vec{p}_\Phi.\vec{\beta})^2 \gamma E_\Phi}
{\beta |p_\Phi|} \biggr ) \times \nonumber \\
&& \biggl (\gamma \beta^2 \mathcal{B}^2 + \mathcal {B} (\gamma + \beta (\gamma - 1))
(\vec{p}_\Phi.\vec{\beta}) + \frac{\gamma - 1}{\beta} (\vec{p}_\Phi.\vec{\beta})^2 \biggr ) \biggr ] 
\biggr | \label{initphidistexp} \\
\mathcal{D} & \equiv & \frac{(\vec{p}_\Phi.\vec{\beta})}{\omega_k \beta^2} \biggl (\beta \gamma - 
\frac{\gamma - 1}{\beta \gamma^2} \biggr ) + \frac{\beta E_\Phi}{M_X} \label {ddefine} \\
\mathcal {B} & \equiv & \frac{(\gamma - 1) (\vec{p}_\Phi.\vec{\beta})}{\beta^2} + \gamma E_\Phi 
\label {bdefine}
\eea
where $J$ is the Jacobian of volume transformation. In the case of (\ref{lagrangint}-b \& c) 
interactions, we have to multiply $f_X (k')$ by matrix element $|{\mathcal M} (k',p_\Phi)|^2$ 
of decay of $X$ particles with momentum $k'$ to $\Phi$ with momentum $p_\Phi$~\cite{houridmquin}. 

If the boundary condition (\ref{dirichletneumannijsrc}) is used for $G^F_\phi$, there would be no 
direct relation between the initial distribution of $X$ particles. In this case, initial variation 
of condensate modes must be put by hand. For instance, if $\Upsilon_\Phi (k)$ in 
(\ref{dirichletneumannijsrc}) is a Gaussian, based on properties of Gaussian distribution 
the best candidate for $f_\Phi (p'_\Phi)$ is a Gaussian.

\section{Einstein equations for linear perturbations} \label {app:connect}
We first calculate components of connection $\Gamma^\mu_{\nu\rho}$ for metric:
\be
ds^2 = a^2(\eta) (1 + 2 \psi (x,\eta) d\eta^2 - a^2(\eta) [(1-2\phi) \delta_{ij} + h_{ij}] dx^i dx^j 
\label{perturbmetricgen}
\ee
The metric defined in (\ref{perturbmetric}) is the special case of (\ref{perturbmetricgen}) with 
$\phi = \psi$. For metric (\ref{perturbmetricgen}) Christoffel coefficients of the connection at 
linear order of perturbations have following expressions:
\bea
&& \Gamma^0_{00} = \frac{a'}{a} + \psi', \quad \quad \Gamma^i_{00} = \delta^{ik} \psi_{,k}, \quad\quad 
\Gamma^0_{0i} = \psi_{,i}, \label{gamm000} \\
&& \Gamma^0_{ij} = \biggl [\frac{a'}{a} \biggl ((1 - 2\psi - 2\phi) \delta_{ij} + h_{ij} \biggr ) - 
\phi' + \frac{h'_{ij}}{2} \biggr ], \quad \quad \Gamma^i_{j0} = \Gamma^i_{0j} = 
\frac{a'}{a} \delta^i_j + \frac{h'^i_j}{2} - \phi' \delta^i_j, \label{gammaij0} \\
&& \Gamma^i_{jk} = \frac{1}{2} (h^i_{j,k} + h^i_{k,j} - {h_{jk,}}^i) - (\phi_{,k} \delta^i_j + 
\phi_{,j} \delta^i_k - {\phi_,}^i \delta_{jk}) \label{gammaijk}
\eea
We remind that at linear order $h_{ij} = h^{ij} = h^i_j$. Nonetheless, for the sake of consistency 
of notation in the description of $\Gamma^\mu_{\nu\rho}$ above and elsewhere we respect 
covariant/contravariant presentation of indices.

Riemann curvature tensor $R_{\mu\nu}$ can be expanded with respect to connection as:
\be
R_{\mu\nu} = \partial_\rho \Gamma^\rho_{\mu\nu} - \Gamma^\rho_{\mu\rho} \Gamma^\sigma_{\nu\sigma} - 
D_\nu\biggl (\partial_\mu (\ln \sqrt {-g}) \biggr ) \label{riemanncurve}
\ee 
Finally the semi-classical Einstein equations in this gauge are written as:
\bea
G_{00} &=& 3 \hm^2 - 6\hm \phi' + 2 \phi^i_i + \frac{3}{2} \hm' h' + \frac{1}{2} ({h^i_k}_{,i}^k - 
h^i_i) = 8\pi \gm \langle T_{00} \rangle \label{G00gauge} \\
G_{0i} &=& 2\phi'_{,i} + 2\hm \psi_{,i} + \frac{1}{2} ({h'}^k_{i,k} - {h'}_{,i}) = 
8\pi \gm \langle T_{0i} \rangle \label{G0igauge} \\
G_{ij} &=& -(2\hm' + \hm^2) \biggl [(1 - 2\psi - 2\phi)~\delta_{ij} + h_{ij} \biggr ] + 
2 \hm (\psi' + 2 \phi')~\delta_{ij} + \hm (\frac{3}{2} {h'}_{ij} - h'\delta_{ij}) + \nonumber \\
&& (2 \phi'' - \frac{1}{2}h'')~\delta_{ij} + (\psi_{,k}^k - \phi_{,k}^k)~\delta_{ij} + \phi_{,ij} - 
\psi_{,ij} + \nonumber \\
&& \frac{1}{2}{h_{ij}}'' + \frac{1}{2}(h^k_{i,jk} + {h_{jk,i}}^k - {h_{ij,k}}^k - h_{,ij}) + 
\frac{1}{2} (h_{,k}^k - {h_k^l}_{,l}^k)~\delta_{ij} = 8\pi \gm \langle T_{ij} 
\rangle \label{Gijgauge}
\eea
where $h \equiv h^i_i$. Due to diffeomorphism invariance only 6 of above equations are independent. 
This means that 2 of 8 metric components $\psi$, $\phi$ and $h_{ij}$ can be chosen arbitrarily. 
An interesting choice which simplifies Einstein equations is $h=0$ and $\phi = \psi$. In Newtonian 
gauge without tensor perturbations the latter relation is satisfied when the anisotropic shear is 
null. Here this choice does not impose any constraint on $T^{\mu\nu}$ because independent components 
of $h_{ij}$ can include the effect of an anisotropic shear.

\section{Solution of free field equation in homogeneous FLRW geometry} \label{app:solution}
To find solutions of free field equations in a homogeneous FLRW geometry similar to 
(\ref{evolinithomo}), it is better to perform a scaling similar to (\ref{chidef}) with 
${\boldsymbol \psi} = h = 0$. Additionally, we first ignore the spacetime dependence of $M^2$ and 
find exact or approximate solutions with a constant mass, and then use WKB approximation to take 
into account coordinate dependence of effective mass.

After the change of variable the homogeneous evolution equation (\ref{evolinithomo}) becomes:
\be
\Xi_k^{\chi''} + (k^2 + M^2 a^2 - \frac{a''}{a}) \Xi_k^\chi = 0 \label{homoevolk}
\ee
where $\Xi_k$ is the Fourier transform of propagators or fields and $\Xi_k^\chi = a \Xi_k$. Solutions 
of this equation depends on the explicit expression of $a (\eta)$, which in turn depends on the 
equation of state of dominant component of matter. Here we separately discuss the solutions for 
radiation dominated and matter dominated eras, and for a general FLRW cosmology.

\paragraph*{Radiation domination:}
\be
\frac{a}{a_0} = \biggl (\frac{t}{t_0} \biggr )^{\frac{1}{2}} = \frac{\eta}{\eta_0}, \quad \quad 
a'' = 0, \quad \frac{a'}{a} = \frac{1}{\eta} \label {expan}
\ee
For this case an exact solution is known~\cite{integbook,qmcurve}:
\bea
&& \Xi_k^\chi (\eta) = c_k U_k (z') + d_k V_k (z') \label{homosol} \\
&& U_k (z') = D_{-1/2 + i\alpha} (z' e^{i\frac{\pi}{4}}), \quad \quad 
V_k (z') = D_{-1/2 - i\alpha} (z' e^{-i\frac{\pi}{4}}) \label{parbcylv} \\
&& z' \equiv \theta \frac{\eta}{\eta_0}, \quad \theta \equiv \sqrt{2 a_0 \eta_0 M} = 
\sqrt{\frac{2M}{H_0}} \quad \alpha \equiv \frac{k^2\eta_0}{2 a_0 M} = 
\frac{k^2/a_0^2}{2M H_0} \label{homosolparam}
\eea
where $a_0$ and $H_0$ are the expansion factor and Hubble constant at initial 
conformal time $\eta_0$, respectively; and $D_\nu (x)$ is parabolic cylinder function of order $\nu$. Their 
derivatives, which are  necessary for determining integration constants, can be 
determined from the following recursive relation:
\be
\frac{dD_{-1/2 \pm i\alpha} (z)}{dz} = \frac{z}{2} D_{-1/2 \pm i\alpha} (z) + D_{1/2 \pm i\alpha} (z) 
\label{parbcylder}
\ee
When local quantum corrections are considered $M^2 \rightarrow M^2 + \Delta M_k^2 (\eta)$. In this 
case, no exact analytical solution is known. Thus, under the assumption that $\Delta M_k / M \ll 1$ 
and $\Delta M'_k / M \ll 1$, we use a WKB-like technique, and to obtain an approximate solution we 
perform the following replacement in (\ref{parbcylv}):
\be
z' \rightarrow \int dz' (1 + \frac{\Delta M_k^2}{M^2})^{1/4}
\ee

\paragraph*{Matter domination:}
In matter domination regime the expansion factor evolves as:
\be
\frac{a}{a_0} = \biggl( \frac{t}{t_0}\biggr )^{\frac{2}{3}} = \biggl( \frac{\eta}{\eta_0}\biggr )^2, 
\quad \quad \frac{a''}{a} = \frac{2}{\eta^2} \quad , \quad \frac{a'}{a} = \frac{2}{\eta} 
\label{amatter}
\ee
In this case analytical solution for (\ref{homoevolk}) is known only for $k = 0$ or $M = 0$:
\bea
&& U_k \& V_k
\begin{cases}
\sqrt{\frac{\eta}{\eta_0}} J_{\pm \frac{1}{2}} (\beta \frac{\eta^3}{\eta_0^3}), 
\quad \quad \beta \equiv \frac{a_0\eta_0 M}{3} = \frac{2M}{3H_0} & \text{For $k^2 = 0$}  \\
\sqrt{\frac{\eta}{\eta_0}} J_{\pm \frac{3}{2}} (k \eta) = \sqrt{\frac{\eta}{\eta_0}} J_{\pm \frac{3}{2}} 
(\frac{2k}{H_0 a (\eta)}\frac{\eta^3}{\eta_0^3}) & \text{For $M = 0$}
\end{cases}\label {solkmmatter} \\
&& \nonumber \\
&& J_{\frac{1}{2}} (x) = \sqrt{\frac{2}{\pi x}} \sin x, \quad 
J_{-\frac{1}{2}} (x) = \sqrt{\frac{2}{\pi x}} \cos x \label {besselonehalf} \\
&& J_{\frac{3}{2}} (x) = \sqrt{\frac{2}{\pi x}} (\frac{\sin x}{x} - \cos x), \quad 
J_{-\frac{3}{2}} (x) = \sqrt{\frac{2}{\pi x}} (-\sin x - \frac{\cos x}{x}) \label {besselthreehalf}
\eea
Explicit expansion of Bessel functions in these solutions shows that at lowest order in $\eta$ 
the solutions for the two cases with analytical solutions are equal. Moreover, $J_{\pm \frac{3}{2}}(x) \xrightarrow {x \gg 1}
J_{\mp \frac{1}{2}}(x)$ up to a constant factor. Therefore, if the mass term in (\ref{homoevolk}) is 
dominant, an approximate solution can be obtained by replacing the argument of solution for 
$k = 0$ with:
\be
\frac{2M}{3H_0} \frac{\eta^3}{\eta_0^3} \rightarrow \frac{2}{3H_0} \biggl (M^2 + \frac{k^2}{a^2} 
\biggr )^{\frac{1}{2}} ~ \frac{\eta^3}{\eta_0^3} \label{wkbu}
\ee
But this approximation does not converge to the exact solution when $M^2 \rightarrow 0$. A better 
approximation is an interpolation in $(M^2, k^2, \Xi)$ space:
\be
U_k \& V_k \approx \sqrt{\frac{\eta}{\eta_0}} \biggl [M^2 J^2_{\pm \frac{1}{2}} (y) + 
\frac {k^2}{a_0^2} J^2_{\mp \frac{3}{2}} (y) \biggr ]^{\frac{1}{2}}, \quad y \equiv 
\frac{2}{3H_0} \biggl (M^2 + \frac{k^2}{a^2} \biggr )^{\frac{1}{2}} ~ \frac{\eta^3}{\eta_0^3} 
\label{solmatterapp}
\ee
which approaches to exact solutions (\ref{solkmmatter}) for both $M^2 \rightarrow 0$ and 
$k^2 \rightarrow 0$. Moreover, to obtain a better approximate solution, coordinate or equivalently $k$ and time dependence of $M^2$ can be 
directly added to $y$ defined in (\ref{solmatterapp}).

\paragraph{Other cosmologies}
Equation (\ref{homoevolk}) does not have an exact analytical solution and the previous two cases 
are exceptional in having exact solutions, at least for some special values of 
parameters\footnote{Exact solutions exist also for de Sitter space, and are described extensively 
in literature about inflation. For this reason we do not repeat them here.}. Thus, when $a(\eta)$ is 
an arbitrary function, we have to use a general approach such as WKB to obtain approximate solutions expanded 
with respect to derivatives of the expansion factor. In addition to providing a solution for field equations, 
such an expansion is crucial for the adiabatic regularization, in which along with the 
evolution of bare propagators - usually performed numerically - the evolution of free vacuum is 
necessary for removing singularities~\cite{renormadiab,qmcurve,heatkernel}.

The second-order WKB approximate solutions have the following general form:
\bea
U_k, V_k & = & \frac{1}{\sqrt{2W}} e^{\pm i \int d\eta W (\eta)}, \label{wkbappoxsol} \\
W^2 (\eta) & = & \Omega^2 (\eta) + \frac{3W'^2}{4W^2} - \frac{W''}{2W}, \quad \quad 
\Omega^2 (\eta) = k^2 + a^2 (\eta) M_R^2 (\eta) + 6 (\zeta - \frac{1}{6}) \frac{a''}{a}\label{wkbappoxw}
\eea
where $\zeta = 0$ for FLRW and $\zeta = 1/6$ for conformal geometries. The function $W$ and amplitude 
of solutions $|U_k|^2 = |V_k|^2$ have the following expressions with respect to expansion rate and its 
derivatives up to second order, obtained from perturbative solution of (\ref{wkbappoxw}) for 
$\zeta = 0$:
\bea
W^2 & \approx & \omega_k^2 - \frac{1}{2} \biggl (\frac {C''}{C} - \frac{C^{'2}}{2C^2} \biggr ) + 
\frac{5C^{'2} M^4}{16 \omega_k^4} - \frac{C'' M^2}{4 \omega_k^2} \label {wkbappoxw} \\
|U_k|^2 = |V_k|^2 & = & \frac{1}{2 \omega_k} \biggl [1 + \frac{1}{4 \omega_k^2} 
\biggl (\frac {C''}{C} - \frac{C^{'2}}{2C^2} \biggr ) + \frac{C'' M^4}{8 \omega_k^4} - 
\frac{5 C^{'2} M^4}{32 \omega_k^6} \biggr ] \label {wkbappoxampl}
\eea
where $C \equiv a^2$. The integral in the phase term in (\ref{wkbappoxsol}) can be approximated as:
\be
\int d\eta W (\eta) \approx \omega_k \eta - \frac{C'}{8 \omega_k C} + \frac{5CC' M^4}{32 \omega_k^5} - 
\frac{C' M^2}{8 \omega_k^3} + \ldots \label{wkbphaseint}
\ee
The perturbative expansion in (\ref{wkbappoxw}) is with respect to derivatives of the expansion 
factor and corresponds to adiabatic orders defined in Sec. \ref{sec:renorm}. Introducing a time scale 
$T$, derivatives of $C$ can be written as $C^{(n)} = T^{-n} d^nC/d\upeta^n,~\upeta \equiv \eta/T$. For 
adiabatic time scale $T \rightarrow \infty$, derivatives $C^{(n)} \rightarrow 0$ and approximate WKB 
solution approaches exact solution of wave equation in Minkowski space.

Higher order adiabatic solution of (\ref{wkbappoxw}) can be determined from the following recursive 
relation~\cite{quinadiabir}:
\be
W_{(n)}^2 = \Omega_k - \frac{1}{2}\biggl [\frac{W''_{(n-2)}}{W_{(n-2)}} - \frac{3}{2}
\biggl (\frac{W'_{(n-2)}}{W_{(n-2)}} \biggr ) \biggr ], \quad \quad W_{(0)} = C \omega_k \label{adiabaticn}
\ee

\section {Solution of constraint equations} \label {app:gensolution}
Equations (\ref{dirichletneumannij}) and (\ref{derivcond}) provide a system of linear equations 
with respect to $|c_k|^2$, $|d_k|^2$, $c_kd^*_k = |c_kd_k|e^{i\Delta\theta}$, and 
$c^*_kd_k = |c_kd_k|e^{-i\Delta\theta}$, which must be solved to determine these integration constants. 
The coefficients in these equations depend on renormalized values of mass and couplings fixed at 
the initial time $\eta_0$\footnote{Because the value of physical momentum $k/a$ changes with time, 
$1/\eta_0$ can be considered as the energy scale for defining renormalized 
quantities~\cite{renormadiab0}.}.

For $X$ field the constraints (\ref{dirichletneumannij}) and (\ref{derivcond}) are expanded as: 
\bea
A_0 |c^X_k|^2 + A_1 d^X_k c^{X*}_k + A_2 c^X_k d^{X*}_k + A_3 |d^X_k|^2 & = & -i \label {initeqxqm} \\
B_0 |c^X_k|^2 + B_1 d^X_k c^{X*}_k + B_2 c^X_k d^{X*}_k + B_3 |d^X_k|^2 & = & 0 \label {initeqxdecay}
\label {initeqxhubble} 
\eea
and for $\Phi$ and $A$ as:
\bea
A_0 |c^i_k|^2 + A_1 d^i_k c^{i*}_k + A_2 c^i_k d^{i*}_k + A_3 |d^i_k|^2 & = & -i \label {initeqiqm} \\
E_0 |c^i_k|^2 + E_1 d^i_k c^{i*}_k + E_2 c^i_k d^{i*}_k + E_3 |d^i_k|^2 & = & \mathcal {D}_i \quad \quad 
i = \Phi, A \label {initeqidecay}
\eea
where:
\be
\mathcal {D}_i \equiv a_0 \Gamma_X G^X_{p(k)} (\eta_0), \quad \quad i \in \Phi,~A \label {didef}
\ee
and $p(k)$ is determined from relation between momentum of $A$ and $\Phi$ particles in the 
decay of $X$ discussed in Appendix \ref {app:distphia}. Alternatively, if the boundary condition 
(\ref{dirichletneumannijsrc}) is assumed, in equation (\ref{initeqidecay}) one has to replace 
$\mathcal {D}_i (k)$ with $\Upsilon_i (k)$. 

Coefficients $A_i$, $B_i$, and $E_i$ have following expressions:
\bea
A_0 & \equiv & U'_i (k) U^*_i (k) - U_i (k) U^{*'}_i (k) \nonumber \\
A_1 & \equiv & V'_i (k) U^*_i (k) - V_i (k) U^{*'}_i (k) \nonumber \\
A_2 & \equiv & U'_i (k) V^*_i (k) - U_i (k) V^{*'}_i (k) \nonumber \\
A_3 & \equiv & V'_i (k) V^*_i (k) - V_i (k) V^{*'}_i (k), \quad \quad i \in X,~\Phi,~A \label{a0def}
\eea
\bea
B_0 & \equiv & U'_X (k) U^*_X (k) + U_X (k) U^{*'}_X (k) - 
U_X (k) U^*_X (k) (2\mathcal{H} + a_0{\mathcal K}_X - a_0 \Gamma_X) \nonumber \\ 
B_1 & \equiv & V'_X (k) U^*_X (k) + V_X (k) U^{*'}_X (k) - 
V_X (k) U^*_X (k) (2\mathcal{H} + a_0{\mathcal K}_X - a_0 \Gamma_X) \nonumber \\
B_2 & \equiv & U'_X (k) V^*_X (k) + U_X (k) V^{*'}_X (k) - 
U_X (k) V^*_X (k) (2\mathcal{H} + a_0{\mathcal K}_X - a_0 \Gamma_X) \nonumber \\
B_3 & \equiv & V'_X (k) V^*_X (k) + V_X (k) V^{*'}_X (k) - 
V_X (k) V^*_X (k) (2\mathcal{H} + a_0{\mathcal K}_X - a_0 \Gamma_X) \label{b0def}
\eea
\bea
E_0 & \equiv & U'_i (k) U^*_i (k) + U_i (k) U^{*'}_i (k) - U_i (k) U^*_i (k) 
(2\mathcal{H} + a_0{\mathcal K}_i) \nonumber \\ 
E_1 & \equiv & V'_i (k) U^*_i (k) + V_i (k) U^{*'}_i (k) - V_i (k) U^*_i (k) 
(2\mathcal{H} + a_0{\mathcal K}_i) \nonumber \\
E_2 & \equiv & U'_i (k) V^*_i (k) + U_i (k) V^{*'}_i (k) - U_i (k) V^*_i (k) 
(2\mathcal{H} + a_0{\mathcal K}_i) \nonumber \\
E_3 & \equiv & V'_i (k) V^*_i (k) + V_i (k) V^{*'}_i (k) - V_i (k) V^*_i (k) 
(2\mathcal{H} + a_0{\mathcal K}_i), \quad \quad i \in \Phi,~A \label{e0defi}
\eea
For the sake of notation simplicity the species index of $A_i, B_i, \& E_i$ are dropped. 
The $A_i$ coefficients satisfy the following properties:
\be
A^*_i = -A_i~~ i \in \{0,~3\}, \quad A^*_1 = -A_2 \label{acoefprop}
\ee
Therefore, real and imaginary part of constraints (\ref{initeqxqm}) and (\ref{initeqiqm}) are not 
independent and there is no degeneracy or over-constraining in the model. Equalities in 
(\ref{acoefprop}) are valid for any $U_k$ and $V_k$ solutions. In spacial cases, e.g. $V_k = U^*_k$ 
or when both solutions are real, there may be new relations between coefficients, but they do not 
induce additional degeneracies.

\subsection*{Acknowledgment} The author thanks Dominik Schwarz and Bj\"orn Garbrecht for their support and helpful discussions.


\begin{thebibliography}{99}
\bibitem {infrev} A.H. Guth, \Journal {\PRD}{23}{1981}{347}; A.D. Linde, \Journal {\PLB}{108}{1982}{389}; J. Martin, C. Ringeval, V. Vennin, \Journal {\PDU}{5-6}{2014}{75} \href{http://arxiv.org/abs/1303.3787}{[arXiv:1303.3787]} (review)
\bibitem {derev} E.J. Copeland, M. Sami, S. Tsujikawa, \Journal {\IMD}{15}{2006}{1753} \href{http://arxiv.org/abs/hep-th/0603057}{[hep-th/0603057]}(review); A. Silvestri, M. Trodden, \Journal {\RPP}{72}{2009}{096901} \href{http://arxiv.org/abs/arXiv:0904.0024}{[arXiv:0904.0024]}(review).
\bibitem {modgrrev} A. Joyce, B. Jain, J. Khoury, M. Trodden, \Journal {\PRE}{568}{2015}{1} \href{http://arxiv.org/abs/1407.0059}{[arXiv:1407.0059]}.
\bibitem {quin} B. Ratra, P.J.E. Peebles, \Journal {\PRD}{37}{1988}{3406}; C. Wetterich, \Journal {\NPB}{302}{1988}{668}.
\bibitem {deeft2pi} C. Wetterich, \Journal {\PRD}{92}{2015}{083507} \href{http://arxiv.org/abs/1503.07860}{[arXiv:1503.07860]}.
\bibitem {quininfunify} M. Wali Hossain, R. Myrzakulov, M. Sami, Emmanuel N. Saridakis, \Journal{\IMD}{24}{2015}{1530014} \href{http://arxiv.org/abs/1410.6100}{[arXiv:1410.6100]}.
\bibitem {hubbleanisoconstr} B. Kalus, D.J. Schwarz, M. Seikel, A. Wiegand, \Journal {\AA}{553}{2013}{A56} \href{http://arxiv.org/abs/1212.3691}{[arXiv:1212.3691]}.
\bibitem {infplanck} Planck Collaboration, \Journal{\AA}{594}{2016}{A20} \href{http://arxiv.org/abs/1502.02114}{[arXiv:1502.02114]}.
\bibitem {higgssymmbrrev} A. Djouadi \Journal {\PRE}{457}{2008}{1} [hep-ph/0503172], \href{http://arxiv.org/abs/1505.07950}{[arXiv:1505.07950]}.
\bibitem {reheatrev} L. Kofman, A. Linde, A. Starobinsky, \Journal {\PRD}{56}{1997}{3258} \href{http://arxiv.org/abs/hep-ph/9704452}{[hep-ph/9704452]}; P. Greene, L. Kofman, A. Linde, A. Starobinsky, \Journal {\PRD}{56}{1997}{6175} \href{http://arxiv.org/abs/hep-ph/9705347}{[hep-ph/9705347]}, L. Kofman, \Journal{\LNP}{738}{2008}{55} (review).
\bibitem {renormadiab0} S.A. Ramsey, B.L. Hu, \Journal {\PRD}{56}{1997}{678}; Erratum \Journal {\PRD}{57}{1998}{3798} \href{http://arxiv.org/abs/hep-ph/9706207}{[hep-ph/9706207]}. 
\bibitem {renormadiab1} A. Tranberg, \Journal {\JHE}{0811}{2008}{037} \href{http://arxiv.org/abs/0806.3158}{[arXiv:0806.3158]}.
\bibitem {infsecular} D. Boyanovsky, H.J De Vega, \Journal {\PRD}{70}{2004}{063508} \href{http://arxiv.org/abs/astro-ph/0406287}{[astro-ph/0406287]}.
\bibitem {infsecularir} D. Boyanovsky, H.J De Vega, N. Sanchez, \Journal {\PRD}{72}{2005}{103006} \href{http://arxiv.org/abs/astro-ph/0507596}{[astro-ph/0507596]}.
\bibitem {infqftnongauss} S. Weinberg, \Journal{\PRD}{72}{2005}{043514} [hep-th/0506236]; S. Weinberg, \Journal {\PRD}{74}{2006}{023508} \href{http://arxiv.org/abs/hep-th/0605244}{[hep-th/0605244]}; C. Burrage, R.H. Ribeiro, D. Seery, \Journal {\JCA}{07}{2011}{032} \href{http://arxiv.org/abs/1103.4126}{[arXiv:1103.4126]}.
\bibitem {desitterinstab} A.D. Dolgov, M. B. Einhorn, V. I. Zakharov, \Journal {\em Acta Phys.Polon. B}{26}{1995}{65} \href{http://arxiv.org/abs/gr-qc/9405026}{[gr-qc/9405026]}.
\bibitem {desitcorrmass} A.M. Polyakov, \Journal {\NPB}{797}{2008}{199} \href{http://arxiv.org/abs/0709.2899}{[arXiv:0709.2899]}; D.Krotov \& A.M. Polyakov, \Journal {\NPB}{849}{2011}{410} \href{http://arxiv.org/abs/1012.2107]}{[arXiv:1012.2107]}.
\bibitem {desitterirmassless} S. Hollands, \Journal {\em Annales Henri Poincar\'e}{13}{2012}{1039} \href{http://arxiv.org/abs/1105.1996}{[arXiv:1105.1996]}.
\bibitem {desitterirmassless0} R. Verch, in the proceedings of "Quantum Field Theory and Gravity", Regensburg, Germany, Sep 28 - Oct 1, 2010, F. Finster \etal. (eds.), Birkhaeuser, Basel, (2011) \href{http://arxiv.org/abs/1105.6249}{[arXiv:1105.6249]}.
\bibitem {desittersymmbreak} D. Boyanovsky, R. Holman, \Journal {\JHE}{05}{2011}{047} \href{http://arxiv.org/abs/1103.4648}{[arXiv:1103.4648]}; D. Boyanovsky, \Journal {\PRD}{85}{2012}{123525} \href{http://arxiv.org/abs/1203.3903}{[arXiv:1203.3903]}; D. Boyanovsky, \Journal {\PRD}{86}{2012}{023509} \href{http://arxiv.org/abs/1205.3761}{[arXiv:1205.3761]}; A. Albrecht, R. Holman, B.J, Richard, \Journal {\PRL}{114}{2015}{171301} \href{http://arxiv.org/abs/1412.6879}{[arXiv:1412.6879]}. 
\bibitem {wwmethod} V. Weisskopf, E. Wigner, \Journal {\ZPH}{63}{1930}{54}.
\bibitem {desitterinstab1} P.R. Anderson, E. Mottola, \Journal {\PRD}{89}{2014}{104038} \href{http://arxiv.org/abs/1310.0030}{[arXiv:1310.0030]}; \Journal {\PRD}{89}{2014}{104039} \href{http://arxiv.org/abs/1310.1963}{[arXiv:1310.1963]}.
\bibitem {desitterir} L. Lello, D. Boyanovsky, R. Holman, \Journal {\PRD}{89}{2014}{063533} \href{http://arxiv.org/abs/1307.4066}{[arXiv:1307.4066]}. 
\bibitem {deultralightir} L. Parker, A. Raval, \Journal {\PRD}{60}{1999}{063512}; \Journal{Erratum-ibid.D}{67}{2003}{029901} \href{http://arxiv.org/abs/gr-qc/9905031}{[gr-qc/9905031]}; \Journal {\PRD}{60}{1999}{123502}; \Journal {Erratum-ibid.D}{67}{2003}{029902} \href{http://arxiv.org/abs/gr-qc/9908013}{[gr-qc/9908013]}. 
\bibitem {desitterstable} D. Marlof, I.A. Morrison, \Journal {\PRD}{84}{2011}{044040} \href{http://arxiv.org/abs/1010.5327}{[arXiv:1010.5327]}; S. Hollands, \href{http://arxiv.org/abs/1010.5367}{[arXiv:1010.5367]}.
\bibitem {infdesiter2pi} B. Garbrecht, G. Rigopoulos, \Journal {\PRD}{84}{2011}{063516} \href{http://arxiv.org/abs/1105.0418}{[arXiv:1105.0418]}; F. Gautier, J. Serreau, \Journal {\PLB}{727}{2013}{541} \href{http://arxiv.org/abs/1305.5705}{[arXiv:1305.5705]}.
\bibitem {inf2piremormcurv} M. Herranen, T. Markkanen, A. Traunberg, \Journal {\JHE}{05}{2014}{026} \href{http://arxiv.org/abs/1311.5532}{[arXiv:1311.5532]}; S. Park, T. Prokopec, R.P. Woodard, \Journal {01}{2016}{074} \href{http://arxiv.org/abs/1510.03352}{[arXiv:1510.03352]}. 
\bibitem {rgdesitter} A. Kaya, \Journal {\PRD}{87}{2013}{123501} \href{http://arxiv.org/abs/1303.5459}{[arXiv:1303.5459]}; J. Serreau, \Journal {\PLB}{730}{2014}{271} \href{http://arxiv.org/abs/1306.3846}{[arXiv:1306.3846]}.
\bibitem {infstochas} A.A. Starobinsky, J. Yokoyama, \Journal {\PRD}{50}{1994}{6357}; S. Winitzki, A. Vilenkin, \Journal {\PRD}{61}{2000}{084008} \href{http://arxiv.org/abs/gr-qc/9911029}{[gr-qc/9911029]}; F. K\"uhnel, D.J. Schwarz, \Journal {\PRD}{78}{2008}{103501} \href{http://arxiv.org/abs/0805.1998}{[arXiv:0805.1998]}; F. K\"uhnel, D.J. Schwarz, \Journal {\PRL}{105}{2010}{211302} \href{http://arxiv.org/abs/1003.3014}{[arXiv:1003.3014]}.
\bibitem {infstochas0} J. Martin, M. Musso \Journal {\PRD}{73}{2006}{043516} \href{http://arxiv.org/hep-th/0511214}{[hep-th/0511214]}, \Journal {\PRD}{73}{2006}{043517} \href{http://arxiv.org/hep-th/0511292}{[hep-th/0511292]}.
\bibitem {infqftstoch} G. Lazzari, T. Prokopec, (2013) \href{http://arxiv.org/abs/1304.0404}{[arXiv:1304.0404]}. 
\bibitem {infstochasirequiv} B. Garbrecht, G. Rigopoulos, Y. Zhu, \Journal {\PRD}{89}{2014}{063506} \href{http://arxiv.org/abs/}{[arXiv:1310.0367]}; B. Garbrecht, G. Rigopoulos, Y. Zhu, \Journal {\PRD}{91}{2015}{063520} \href{http://arxiv.org/abs/}{[arXiv:1412.4893]}.
\bibitem {cond2piexpand} B. Garbrecht, P. Millington, \Journal {\NPB}{906}{2016}{105} \href{http://arxiv.org/abs/1509.07847}{[arXiv:1509.07847]}.
\bibitem {quinvactunnel} J. Yokoyama, \Journal {\PRL}{88}{2002}{151302} \href{http://arxiv.org/abs/hep-th/0110137}{[hep-th/0110137]}; \Journal {\IMD}{11}{2002}{1603} \href{http://arxiv.org/abs/gr-qc/0205104}{[gr-qc/0205104]}. 
\bibitem {houricond} H. Ziaeepour, \Journal {\PRD}{81}{2010}{103526} \href{http://arxiv.org/abs/1003.2996}{[arXiv:1003.2996]}; in ``Advances in Dark Energy Research'', Ed. M. L. Ortiz, Nova Science Inc. New York (2015).
\bibitem {houridmquin} H. Ziaeepour, \Journal {\PRD}{69}{2004}{063512} \href{http://arxiv.org/abs/astro-ph/0308515}{[astro-ph/0308515]}.
\bibitem {quinqft0} E. Elizalde, J.E. Lidsey, S. Nojiri, S.D. Odintsov, \Journal {\PLB}{574}{2003}{1} \href{http://arxiv.org/abs/hep-th/0307177}{[hep-th/0307177]}.
\bibitem {dediscrimscale} E. Bertschinger, P. Zukin, \Journal {\PRD}{78}{2008}{024015} \href{http://arxiv.org/abs/0801.2431}{[arXiv:0801.2431]}.
\bibitem {houridephono} H. Ziaeepour, \Journal {\PRD}{86}{2012}{043503} \href{http://arxiv.org/abs/1112.6025}{[arXiv:1112.6025]}; C. Wetterich, in "Modifications of Einstein's Theory of Gravity at Large Distances'', Page 57, E. Papantonopoulos (ed.), Springer, (2015) \href{http://arxiv.org/abs/1402.5031}{[arXiv:1402.5031]}.
\bibitem {dediscrim} H. Steigerwald, J. Bel, C. Marinoni, \Journal {JCA}{05}{2014}{042} \href{http://arxiv.org/abs/1403.0898}{[arXiv:1403.0898]}; L. Taddei, L. Amendola, \Journal {JCA}{02}{2015}{001} \href{http://arxiv.org/abs/1408.3520}{[arXiv:1408.3520]}; D. Shi, C.M. Baugh, \Journal {\MRA}{459}{2016}{3540} \href{http://arxiv.org/abs/1511.00692}{[arXiv:1511.00692]}.
\bibitem {inin} E. Calzzetta, B.L. Hu, \Journal {\PRD}{37}{1988}{2878}.
\bibitem {2piinitcond} S.A. Fulling, S.N.M. Ruijsenaars, \Journal {\PRE}{152}{1987}{135}; J. Burges, J. Cox, \Journal {\PLB}{517}{2001}{369} \href{http://arxiv.org/abs/hep-ph/0006160}{[hep-ph/0006160]}; M. Garny, M.M. M\"uller, \Journal {\PRD}{80}{2009}{085011} \href{http://arxiv.org/abs/0904.3600}{[arXiv:0904.3600]}.
\bibitem {scalereffpot} Y.K.E. Cheung, M. Drewes, J.U. Kang, J.C. Kim, \Journal{\JHE}{08}{2015}{059} \href{http://arxiv.org/abs/1504.04444}{arXiv:1504.04444}.
\bibitem {2pirev} K. Chou, Z. Su, B. Hao L. Yu, \Journal {\PRE}{118}{1985}{1}, J. Berges, \Journal {\AIP}{739}{2005}{3} \href{http://arxiv.org/abs/hep-ph/0409233}{[hep-ph/0409233]}, update \href{http://arxiv.org/abs/1503.02907}{[arXiv:1503.02907]} 
\bibitem {infnongauss} D. Seery, J.E. Lidsey, \Journal{\JCA}{0509}{2005}{011} \href{http://arxiv.org/abs/astro-ph/0506056}{[astro-ph/0506056]}.
\bibitem {houriustate} H. Ziaeepour, (2000) \href{http://arxiv.org/abs/astro-ph/0002400}{[astro-ph/0002400]}; in ``Progress in Dark Matter Research'', Nova Science Inc. New York (2005) p. 175, \href{http://arxiv.org/abs/astro-ph/0406079}{[astro-ph/0406079]}.
\bibitem {infmassivecmb} N. Kaloper, M. Kleban, A. Lawrence, S. Shenker, \Journal {\PRD}{66}{2002}{123510} \href{http://arxiv.org/abs/hep-th/0201158}{[hep-th/0201158]}; R. Allahverdi, M. Drees, \Journal {\PRL}{89}{2002}{091302} \href{http://arxiv.org/abs/hep-ph/0203118}{[hep-ph/0203118]}; N. Barnaby, Z. Huang, \Journal {\PRD}{80}{2009}{126018} \href{http://arxiv.org/abs/0909.0751}{[arXiv:0909.0751]}.
\bibitem {shiftsymm} K. Freese, J.A. Frieman, A.V. Olinto, \Journal{\PRL}{65}{1990}{3233}.
\bibitem {2piintro} R. Jackiw \Journal {\PRD}{9}{1974}{1686}; L. Dalon \& R. Jackiw, \Journal {\PRD}{9}{1974}{3320}; H. Schnitzer \Journal {\PRD}{10}{1974}{1800}; J.M. Cornwall, R. Jackiw, E. Tomboulis, \Journal {\PRD}{10}{1974}{2428}.
\bibitem {schwingerkeldyshpath} J. Schwinger, \Journal {\JMP}{2}{1961}{407}; L.V. Keldysh, \Journal {\em Zh. Eksp. Teor. Fiz}{47}{1964}{1515}.
\bibitem {kadanoffbaym} G. Baym, L.P. Kadanoff, \Journal {\PRV}{124}{1961}{287}.
\bibitem {heatkernel} S.A. Ramsey, B.L. Hu, \Journal {\PRD}{56}{1997}{661} \href{http://arxiv.org/abs/gr-qc/9706001}{[gr-qc/9706001]}.
\bibitem {2picurved} A. Hohenegger, A. Kartavtsev, M. Lindner, \Journal {\PRD}{78}{2008}{085027} \href{http://arxiv.org/abs/0807.4551}{[arXiv:0807.4551]}.
\bibitem {qmcurve} N.D. Birrell, P.C.W. Davies, ``Quantum fields in curved space'', Cambridge University Press (1982), L. Parker, D. Toms, ``Quantum Field Theory in Curved Spacetime'', Cambridge Univ. Press, Cambridg, UK (2009).
\bibitem {hourivacuum}  H. Ziaeepour, \Journal{\MPL}{27}{2012}{1250154} \href{http://arxiv.org/abs/1205.3304}{[arXiv:1205.3304]}.
\bibitem {infinitcond0} K. Schalm, G. Shiu, J.P. van der Schaar, \Journal {\JHE}{0404}{2004}{076} \href{http://arxiv.org/abs/hep-th/0401164}{[hep-th/0401164]}.
\bibitem {oneparticledens} P. Danielewicz \Journal {\APH}{152}{1984}{239}.
\bibitem {renormflat} H. van Hees, J. Knoll, \Journal {\PRD}{65}{2002}{025010} \href{http://arxiv.org/abs/hep-ph/0107200}{[hep-ph/0107200]}; \href{http://arxiv.org/abs/hep-ph/0202263}{[hep-ph/0202263]}; \href{http://arxiv.org/abs/hep-ph/0203008}{[hep-ph/0203008]}; F. Cooper, B. Mihaila, J.F. Dawson, \Journal {\PRD}{70}{2004}{105008} \href{http://arxiv.org/abs/hep-ph/0407119}{[hep-ph/0407119]}, J. Berges, S. Borsanyi, U. Reinosa, J. Serreau, \Journal{\APH}{320}{2005}{344} \href{http://arxiv.org/abs/hep-ph0503240}{[hep-ph0503240]}.
\bibitem {renormflat0} S. Borsanyi, U. Reinosa, \Journal {\PRD}{80}{2009}{125029} \href{http://arxiv.org/abs/0809.0496}{[arXiv:0809.0496]}. 
\bibitem {renormflat1} U. Reinosa, J. Serreau, \Journal {\JHE}{0607}{2006}{028} \href{http://arxiv.org/abs/hep-th/0605023}{[hep-th/0605023]}. 
\bibitem {renormbphz} N.N. Bogoliubov, O.S. Parasiuk, \Journal {\AMT}{97}{1957}{97}; K. Hepp, \Journal {\CMP}{2}{1966}{301}, W. Zimmermann, in ``Lectures in Elementary Particles and Quantum Field Theory'', Proc. 1970 Brandeis Summer Institute, Ed. S. Deser \etal, MIT Press, Cambridge, Massachusetts (1970).
\bibitem {nprge} C. Wetterich, \Journal {\PLB}{301}{1993}{90}; J. Berges, N. Tetradis, C. Wetterich, \Journal {\PRE}{363}{2002}{223} \href{http://arxiv.org/abs/hep-ph/0005122}{[hep-ph/0005122]}.
\bibitem {rgoptimalregul} D.F. Litim, \Journal {\PRD}{64}{2001}{105007} \href{http://arxiv.org/abs/hep-th/0103195}{[hep-th/0103195]}.
\bibitem {renormflatsimul} F. Lyonnet, I. Schienbein, F. Staub, A. Wingerter, \Journal {\CPC}{185}{2014}{1130} \href{http://arxiv.org/abs/1309.7030}{[arXiv:1309.7030]}; M.E. Carrington, Wei-Jie Fu, D. Pickering, J.W. Pulver, \Journal {\PRD}{91}{2015}{025003} \href{http://arxiv.org/abs/1404.0710}{[arXiv:1404.0710]}. 
\bibitem {renormadiab} L. Parker,  S.A. Fulling, \Journal {\PRD}{9}{1974}{341}; S.A. Fulling, L. Parker, and B.L. Hu, \Journal {\PRD}{10}{1974}{3905}; N.D. Birrell, \Journal {\PRLA}{361}{1978}{513}, T.S. Bunch, L. Parker, \Journal {\PRD}{20}{1979}{2499}.
\bibitem {qftdergflow} L. Parker, D.A.T. Vanzella, \Journal {\PRD}{69}{2004}{104009} \href{http://arxiv.org/abs/gr-qc/0312108}{[gr-qc/0312108]}. 
\bibitem {renorminf2pi} T. Markkanen, A. Tranberg, \Journal {\JCA}{08}{2013}{045} \href{http://arxiv.org/abs/1303.0180}{[arXiv:1303.0180]}; M. Herranen, T. Markkanen, A. Tranberg, \Journal {\JHE}{05}{2014}{026} \href{http://arxiv.org/abs/1311.5532}{[arXiv:1311.5532]}; F. Gautier, J. Serreau, \Journal {\PRD}{92}{2015}{105035} \href{http://arxiv.org/abs/1509.05546}{[arXiv:1509.05546]}. 
\bibitem {infinitcond1} P.R. Anderson, C. Molina-Paris, E. Mottola, \Journal {\PRD}{72}{2005}{043515}  \href{http://arxiv.org/abs/hep-th/0504134}{[hep-th/0504134]}.
\bibitem {renormsingular} S.A. Fulling, M.S. Sweeny, R.M. Wald, \Journal {\CMP}{63}{1978}{257}; A. del R\'io, J. Navaro-Salas, \href{http://arxiv.org/abs/1412.7570}{[arXiv:1412.7570]}.
\bibitem {adiaborgin} L. Parker, \Journal {\PRL}{21}{1968}{562}.
\bibitem {qftcurvreno} B.S. De Witt, \Journal {\PRE}{19}{1975}{295}.
\bibitem {adabaticinf} S. Weinberg, \Journal {\PRD}{67}{2003}{123504} \href{http://arxiv.org/abs/astro-ph/0302326}{[astro-ph/0302326]}.
\bibitem {infqftsimul} D. Cormier, R. Holman, \Journal {\PRD}{62}{2000}{023520} \href{http://arxiv.org/abs/hep-ph9912483}{[hep-ph9912483]}.
\bibitem {infinitsuperpos} A. Albrecht, N. Bolis, R. Holman, \Journal {\JHE}{11}{2014}{093} \href{http://arxiv.org/abs/1408.6859}{[arXiv:1408.6859]}.
\bibitem {bec} M.H. Anderson, J.R. Ensher, M.R. Mathews, C.E. Wieman, E.A. Cornell, \Journal {\SCI}{269}{1995}{198}.
\bibitem {coherglauber} R.J. Glauber, \Journal {\PRV}{131}{1963}{2766}.
\bibitem {coherstaterev} J.P. Gazeau, ``Coherent States in Quantum Physics'', Wiley-VCH, Verlag, Weinheim (2009).
\bibitem {bunchdavis} T.S. Bunch \& P.C.W. Davis, \Journal {\PRLA}{360}{1978}{117}.
\bibitem {infinitcond} N. Kaloper, M. Kleban, A. Lawrence, S. Shenker, L. Susskind, \Journal {\JHE}{0211}{2002}{037} \href{http://arxiv.org/abs/hep-th/0209231}{[hep-th/0209231]}; B.R. Greene, K.Schalm, G. Shiu, J.P. van der Schaar, Journal {\JCA}{0502}{2005}{001} \href{http://arxiv.org/abs/hep-th/0411217}{[hep-th/0411217]}.
\bibitem {infrenorm} H. Collins, R. Holman, \Journal {\PRD}{71}{2005}{085009} \href{http://arxiv.org/abs/hep-th/0501158}{[hep-th/0501158]}; \href{http://arxiv.org/abs/hep-th/0507081}{[hep-th/0507081]}.
\bibitem {boltzlepto} T. Prokopec, M.G. Schmidt, S. Weinstock, \Journal {\APH}{314}{2004}{208} \href{http://arxiv.org/abs/hep-ph/0312110}{[hep-ph/0312110]}; \Journal {\APH}{314}{2004}{267} \href{http://arxiv.org/abs/hep-ph/0406140}{[hep-ph/0406140]}; A. de Simone, A. Riotto, \Journal {\JCA}{0708}{2007}{002} \href{http://arxiv.org/abs/hep-ph/0703175}{[hep-ph/0703175]}; B. Garbrecht, M. Herranen \Journal {\NPB}{861}{2012}{17} \href{http://arxiv.org/abs/1112.5954}{[arXiv:1112.5954]}.
\bibitem {covarthermal} J. Ehlers, in "General Relativity and Cosmology", ed. B.K. Sachs, Academic Press NewYork (1971).
\bibitem {covarthermal0} G. Chacon-Acosta, L. Dagdug, H.A. Morales-Tecotl, \Journal {\PREE}{81}{2009}{021126} \href{http://arxiv.org/abs/0910.1625}{[arXiv:0910.1625]}.
\bibitem {houriqmgr} H. Ziaeepour, ``And what if gravity is intrinsically quantic ?'', \Journal{\JPC}{174}{2009}{012027}, \href{http://arxiv.org/abs/0901.4634}{[arXiv:0901.4634]}.
\bibitem {infmfieldsimul} M.C. David Marsh, L. McAllister, E. Pajer, T. Wrase, \Journal {\JCA} {11}{2013}{040} \href{http://arxiv.org/abs/1307.3559}{[arXiv:1307.3559]}; J. Elliston, S. Orani, D.J. Mulryne, \Journal {\PRD}{89}{2014}{103532} \href{http://arxiv.org/abs/1402.4800}{[arXiv:1402.4800]}.
\bibitem {hdmmassunitarity} K. Griest, M. Kamionkowski, \Journal {\PRL}{64}{1990}{615}.
\bibitem {infscale} Z.K. Guo, D.J. Schwarz, Y.-Z. Zhang, \Journal {\PRD}{83}{2011}{083522} \href{http://arxiv.org/abs/1008.5258}{[arXiv:1008.5258]}; M. Kleban, M. Mirbabayi, M. Porrati, \Journal {\JCA}{01}{2016}{017} \href{http://arxiv.org/abs/1508.01527}{[arXiv:1508.01527]}.
\bibitem {infpowerlaw} F. Lucchin, S. Matarrese, \Journal {\PRD}{32}{1985}{1316}; J. Yokoyama, \Journal {\PLB}{207}{1988}{31}, Andrew R. Liddle, \Journal {\PLB}{220}{1989}{502}.
\bibitem {infparam} D.J. Schwarz, C.A. Terrero-Escalante, A.A. Garcia, \Journal {\PLB}{517}{2001}{243} \href{http://arxiv.org/abs/astro-ph/0106020}{[astro-ph/0106020]}; J. Martin, D.J. Schwarz, \Journal {\PRD}{67}{2003}{083512} \href{http://arxiv.org/abs/astro-ph/0210090}{[astro-ph/0210090]}.
\bibitem {cmbinf} Z.K. Guo, D.J. Schwarz, Y.-Z. Zhang, \Journal {\JCA}{08}{2011}{031} \href{http://arxiv.org/abs/1105.5916}{[arXiv:1105.5916]}.
\bibitem {qftfixedpoint} J. Berges, A. Rothkopf, J. Schmidt, \Journal {\PRL}{101}{2008}{041603} \href{http://arxiv.org/abs/0803.0131}{[arXiv:0803.0131]}.
\bibitem {adiabaticdef} D. Langlois, \Journal{\CRP}{4}{2003}{953}.
\bibitem {infparticleprod} L. Parker, \Journal {\PRL}{21}{1968}{562}; \Journal {\PRV}{183}{1969}{1057}; \Journal {\PRD}{3}{1971}{346}; A. Enea Romano, M. Sasaki, \Journal {\PRD}{78}{2008}{103522} \href{http://arxiv.org/abs/0809.5142}{[arXiv:0809.5142]}; G. D'Amico, R. Gobbetti, M. Kleban, M. Schillo, \Journal {\JCA}{11}{2013}{013} \href{http://arxiv.org/abs/1306.6872}{[arXiv:1306.6872]}. 
\bibitem {preinfnongauss} N. Barnaby, \Journal {\PRD}{82}{2010}{106009} \href{http://arxiv.org/abs/1006.4615}{[arXiv:1006.4615]}; I. Agullo, L. Parker, \Journal {\GRG}{43}{2011}{2541} \href{http://arxiv.org/abs/1106.4240}{[arXiv:1106.4240]}.
\bibitem {infdecohere} R. Brandenberger, R. Laflamme, M. Miji\'c, \Journal {\MPL}{05}{1990}{2311}; F. Lombardo, F. Mazzitelli, \Journal {\PRD}{53}{1996}{2001} \href{http://arxiv.org/abs/hep-th/9508052}{[hep-th/9508052]}; C.P. Burgess, R. Holman, D. Hoover, \Journal {\PRD}{77}{2008}{063534} \href{http://arxiv.org/abs/astro-ph/0601646}{[astro-ph/0601646]}.
\bibitem {quinir} G. Geshnizjani, R. Brandenberger, \Journal {\PRD}{66}{2002}{123507} \href{http://arxiv.org/abs/gr-qc/0204074}{[gr-qc/0204074]}; \Journal {\JCA}{0504}{2005}{006} \href{http://arxiv.org/abs/hep-th/0310265}{[hep-th/0310265]}.
\bibitem {cosmosqmopen} S. Shandera, N. Agarwal, A. Kamal \href{http://arxiv.org/abs/1708.00493}{[arXiv:1708.00493]}.
\bibitem {integbook} I.S. Gradshteyn, I.M. Ryzhik ``Table of integrals, series, and products'', Accademic Press, INC. (1980).
\bibitem {quinadiabir} J.D. Bates, P.R. Anderson, \Journal {\PRD}{82}{2010}{024018} \href{http://arxiv.org/abs/1004.4620}{[arXiv:1004.4620]}.
\end{thebibliography}
\end{document}